\documentclass[a4paper,11pt]{article}
\pdfoutput=1

\usepackage{Class/jcappub}
\usepackage{amsmath,float,amssymb,bm, Misc/Macro,mathrsfs,tabularx,url,amsfonts,scalerel}
\usepackage{threeparttable}
\usepackage{xcolor}
\usepackage{fontawesome5}
\usepackage{multirow}
\usepackage{hhline}

\usepackage{subcaption}

\DeclareMathAlphabet\mathbfcal{OMS}{cmsy}{b}{n}
\newlength\bshft
\def\fakebold#1{\ThisStyle{\ooalign{$\SavedStyle#1$\cr%
  \kern-\bshft$\SavedStyle#1$\cr%
  \kern\bshft$\SavedStyle#1$}}}
  
\author[1,2]{M.\,Ruiz-Granda,}
\author[3]{P.\,Diego-Palazuelos,}
\author[1]{C.\,Gimeno-Amo,}
\author[1]{P.\,Vielva,}
\author[4]{A.\,I.\,Lonappan,}
\author[5]{T.\,Namikawa,}
\author[6,7]{R.\,T.\,Génova-Santos,}
\author[8,9]{M.\,Lembo,}
\author[10]{R.\,Nagata,}
\author[1]{M.\,Remazeilles,}
% ...
\affiliation[1]{Instituto de Fisica de Cantabria (IFCA, CSIC-UC), Avenida los Castros SN, 39005, Santander, Spain}
\affiliation[2]{Dpto. de Física Moderna, Universidad de Cantabria, Avda. los Castros s/n, E-39005 Santander, Spain}
\affiliation[3]{Max Planck Institute for Astrophysics, Karl-Schwarzschild-Str. 1, D-85748 Garching, Germany}
\affiliation[4]{University of California, San Diego, Department of Physics, San Diego, CA 92093-0424, USA}
\affiliation[5]{Kavli Institute for the Physics and Mathematics of the Universe (Kavli IPMU, WPI), UTIAS, The University of Tokyo, Kashiwa, Chiba 277-8583, Japan}
\affiliation[6]{Instituto de Astrofísica de Canarias, E-38200 La Laguna, Tenerife, Canary Islands, Spain}
\affiliation[7]{Departamento de Astrofísica, Universidad de La Laguna (ULL), E-38206, La Laguna, Tenerife, Spain}
\affiliation[8]{Institut d'Astrophysique de Paris, CNRS/Sorbonne Université, Paris, France}
\affiliation[9]{Dipartimento di Fisica e Scienze della Terra, Università di Ferrara, Via Saragat 1, 44122 Ferrara, Italy}
\affiliation[10]{Japan Aerospace Exploration Agency (JAXA), Institute of Space and Astronautical Science (ISAS), Sagamihara, Kanagawa 252-5210, Japan}
% ...

\emailAdd{ruizm@ifca.unican.es}

\abstract{
Cosmic microwave background (CMB) photons are deflected by large-scale structure through gravitational lensing. This secondary effect introduces higher-order correlations in CMB anisotropies, which are used to reconstruct lensing deflections. This allows mapping of the integrated matter distribution along the line of sight, probing the growth of structure, and recovering an undistorted view of the last-scattering surface. Gravitational lensing has been measured by previous CMB experiments, with \textit{Planck}'s $42\,\sigma$ detection being the current best full-sky lensing map. We present an enhanced \textit{LiteBIRD} lensing map by extending the CMB multipole range and including the minimum-variance estimation, leading to a $49$ to $58\,\sigma$ detection over $80\,\%$ of the sky, depending on the final complexity of polarized Galactic emission. The combination of \textit{Planck} and \textit{LiteBIRD} will be the best full-sky lensing map in the 2030s, providing a $72$ to $78\,\sigma$ detection over $80\,\%$ of the sky, almost doubling \textit{Planck}'s sensitivity. Finally, we explore different applications of the lensing map, including cosmological parameter estimation using a lensing-only likelihood and internal delensing, showing that the combination of both experiments leads to improved constraints. The combination of \textit{Planck} + \textit{LiteBIRD} will improve the $S_8$ constraint by a factor of 2 compared to \textit{Planck}, and \textit{Planck} + \textit{LiteBIRD} internal delensing will improve \textit{LiteBIRD}'s tensor-to-scalar ratio constraint by $6\,\%$. We have tested the robustness of our results against foreground models of different complexity, showing that improvements remain even for the most complex foregrounds.
} 

\title{\textit{LiteBIRD} science goals and forecasts: improved full-sky reconstruction of the gravitational lensing potential through the combination of \textit{Planck} and \textit{LiteBIRD} data}

\begin{document}
\maketitle
\flushbottom

\section{Introduction}

Gravitational lensing deflects the trajectory of cosmic microwave background (CMB) photons as they traverse the large-scale structure of the universe~\cite{Blanchard:1987AA, Lewis:2006:review}. This collection of numerous arcminute-level deflections smooths the acoustic oscillation peaks observed in the CMB angular power spectra and introduces secondary $B$-mode polarization signals and higher-order correlations in CMB temperature and polarization anisotropies. 

Reconstructing these lensing deflections allows us to map the integrated matter distribution along the line of sight, known as the lensing potential $\phi$, and probe the growth of structure. In particular, the lensing angular power spectrum $C_L^{\phi\phi}$ is sensitive to a combination of the late-time amplitude of density fluctuations $\sigma_8$, the matter density $\Omega_{\rm m}$, and the Hubble constant $H_0$~\cite{2021Baxter}, and, when cross-correlated with galaxy surveys, it offers a tomographic view into the amplitude of structure as a function of redshift $\sigma_8(z)$~\cite{Krolewski:2021, 2022White, 2023_DES_SPT}. $C_L^{\phi\phi}$ also encapsulates any suppression of the matter power spectrum due to (dark and ordinary) light relics~\cite{Lewis:2006:review}, currently providing tighter upper limits on the sum of neutrino masses than laboratory experiments~\cite{2006Lesgourgues, 2022Lokhov}. The lensing map also helps calibrate the mass of galaxy clusters~\cite{2000Seljak, 2007Hu, 2015Melin, 2015Baxter} and, in combination with the thermal and kinetic Sunyaev-Zeldovich (SZ) effects and X-ray measurements, allows the study of the thermodynamics of galaxy formation and evolution~\cite{2017Battaglia, 2023Bolliet}.

If known, the deflections produced by lensing can be undone to recover an undistorted view of the last-scattering surface. This process, known as delensing~\cite{ComparisonDelensing}, sharpens the acoustic peaks~\cite{P18:phi} and reduces the lensing contribution to covariance matrices~\cite{2007Li, 2012Benoit-Levy, Peloton:2016:RDN0, Namikawa_Nagata_2015JCAP}, thus improving constraints on primordial gravitational waves~\cite{2017Manzotti, 2020Adachi, 2021_BICEPKeck_SPT} (parametrized by the tensor-to-scalar ratio, $r$), primordial non-Gaussianity~\cite{2020Coulton}, and other cosmological parameters, including those beyond the standard model physics~\cite{2017Green, 2022Hotinli, 2023Ange}. Hence, accurate lensing reconstruction is of high interest for many astrophysical and cosmological studies.

The lensing potential can be reconstructed from the correlations between off-diagonal angular multipoles that it induces in CMB temperature and polarization anisotropies~\cite{OkamotoHu:quad, Flat_sky_QE_2002}. CMB experiments such as \textit{Planck}~\cite{P13:phi, P15:phi, P18:phi, Planck_PR4_lensing}, BICEP/Keck~\cite{BKVIII, BICEPKeck:2022:los-dist},  the Atacama Cosmology Telescope (ACT)~\cite{Das:2011, ACT16:phi, ACT_DR6_lensing, 2024Madhavacheril}, POLARBEAR~\cite{PB:phi:2013, PB:phi:2019}, and the South Pole Telescope (SPT)~\cite{2012vanEngelen, 2013Holder, 2015Story, 2017Omori, 2019Wu, 2020Bianchini, 2021Millea, 2023Pan, 2024Ge} have measured lensing and its angular power spectrum, with the analysis of ACT DR6 temperature and polarization data over $9400$~deg$^2$ of the sky providing the highest signal-to-noise measurement to date at $43\,\sigma$~\cite{ACT_DR6_lensing}. A comparable signal-to-noise of $42\,\sigma$ is reached by the full-sky analysis of \textit{Planck} PR4 temperature and polarization data~\cite{Planck_PR4_lensing}. However, these state-of-the-art measurements are still mainly driven by temperature data, with polarization only contributing up to $20\,\sigma$ and $13\,\sigma$ for ACT and \textit{Planck}, respectively~\cite{ACT_DR6_lensing}. Recently, the SPT Collaboration obtained the first competitive polarization-only measurement by providing a $38\,\sigma$ detection from the analysis of the 1500~deg$^2$ observed with the SPT-3G camera during the 2019 and 2020 winter seasons~\cite{2024Ge}. Ongoing and future experiments like SPT, the Simons Observatory~\cite{2019_SO,2022_SO_delensing}, and the Ali CMB Polarization Telescope~\cite{AliCPT:phi} plan to continue exploiting the untapped potential of CMB polarization to improve lensing reconstruction from the ground.

After \textit{COBE}~\cite{1992ApJ...397..420B}, \textit{WMAP}~\cite{WMAP:2003ogi}, and \textit{Planck}~\cite{Planck_overview_2020}, \textit{LiteBIRD}~\cite{LiteBIRD_PTEP_2023} is the next CMB space-based mission aiming to measure the polarized signal at large angular scales with unprecedented precision. This work is part of a series of papers presenting the science achievable by the \textit{LiteBIRD} space mission and expanding on the overview published in ref.~\cite{LiteBIRD_PTEP_2023}. In particular, we focus on forecasting the advantages that the combination of \textit{Planck}'s high-resolution temperature anisotropies ($T$) and \textit{LiteBIRD}'s high-precision measurements of large angular scale polarization ($E$ and $B$ modes) will provide in the reconstruction of the gravitational lensing potential. Beyond the combination of independent measurements suffering from different systematics, the high complementarity of these data sets also allows the extension of the frequency range available for component separation and the access to angular scales otherwise unreachable to \textit{LiteBIRD} only. Hence, their combination will provide the highest signal-to-noise full-sky lensing map to date, reaching a $72$ to $78\,\sigma$ detection compared to the $43$ to $44\,\sigma$ and $49$ to $58\,\sigma$ detections that \textit{Planck} and \textit{LiteBIRD} can respectively achieve on their own. In addition, this work also extends the results presented in ref.~\cite{Lonappan_lensing} by increasing the CMB multipole range and by including the use of $TT$, $TE$, $TB$, $EE$ quadratic estimators, and their minimum-variance combination, to estimate the lensing potential from \textit{LiteBIRD} temperature and polarization data.

The paper is organized as follows. We describe the simulations used for our study in section~\ref{sec:simulation}. Section~\ref{sec:reconst} details the different steps of the analysis pipeline. We report our results for the lensing maps (section~\ref{sec: lensing maps}), band powers (section~\ref{sec: band powers}), and reconstruction signal-to-noise (section~\ref{sec: SNR}) obtained with the different quadratic estimators in section~\ref{sec:results}, focusing on the improvement that the \textit{Planck} + \textit{LiteBIRD} combination provides compared to \textit{LiteBIRD} alone. To contextualize these results, section~\ref{sec:applications} shows the improvement provided by the \textit{Planck} + \textit{LiteBIRD} combination in common lensing science cases like cosmological parameter estimation (section~\ref{sec: cosmo params}) or $B$-mode delensing in tensor-to-scalar ratio constraints (section~\ref{sec: delensing}). We finish by summarizing the main conclusions of our work in section~\ref{sec:summary}. The software and data used for this study are publicly available at \url{https://github.com/litebird/PlanckBIRD-lens}.

\section{Simulations} \label{sec:simulation}

In this work, we simulate 400 observations from the \textit{Planck} and \textit{LiteBIRD} missions. Our mock data includes \textit{LiteBIRD}'s 22 channels, which redundantly sample 15 unique frequencies ranging between $40$ to $402$\,GHz~\cite{LiteBIRD_PTEP_2023}, and the $9$ ($7$) \textit{Planck} temperature (polarization) channels, covering frequencies from $30$ to $857$\,GHz ($30$ to $353$\,GHz)~\cite{Planck_overview_2020}.

We start by generating random Gaussian realizations of the spherical harmonic coefficients for the lensing potential, $\phi_{\ell m}$, and the unlensed CMB temperature, $\tilde{t}_{\ell m}$, and $E$-mode polarization, $\tilde{e}_{\ell m}$. To enforce the proper correlations between them~\cite{Cholesky_lensing}, we use the Cholesky decomposition of the covariance matrix of the three fields~\cite{Linear_filter},
\begin{equation}\label{corrCholesky}
\begin{pmatrix}
\phi_{\ell m}\\
\tilde{t}_{\ell m} \\
\tilde{e}_{\ell m}
\end{pmatrix}=
\begin{pmatrix}
L_{11} & 0 & 0\\
L_{21} & L_{22} & 0\\
L_{31} & L_{32} & L_{33}
\end{pmatrix}
\begin{pmatrix}
h_{\ell m} \\
j_{\ell m}\\
k_{\ell m}
\end{pmatrix},
\end{equation}
where the matrix elements are
\begin{equation}
\begin{aligned}
L_{11}&=\sqrt{C_\ell^{\phi\phi}},\quad & L_{22}&=\sqrt{\widetilde{C}_\ell^{TT}-L_{21}^2},\\
L_{21}&=\frac{C_\ell^{T\phi}}{\sqrt{C_\ell^{\phi\phi}}},\quad & L_{32}&=\frac{C_\ell^{T\phi}-L_{21}L_{31}}{L_{22}},\\
L_{31}&=\frac{C_\ell^{E\phi}}{\sqrt{C_\ell^{\phi\phi}}},\quad & L_{33}&=\sqrt{\widetilde{C}_\ell^{EE}-L_{31}^2-L_{32}^2},
\end{aligned}
\end{equation}
and the coefficients $h_{\ell m}$, $j_{\ell m}$, and $k_{\ell m}$ are uncorrelated complex Gaussian variables of zero mean and unit variance. Theoretical unlensed angular power spectra are calculated using \texttt{CLASS}\footnote{\url{https://github.com/lesgourg/class_public}}~\cite{CLASS} for the \textit{Planck} 2018 $TT$, $TE$, $EE$ + low$E$ + lensing cosmological parameters in table 2 of ref.~\cite{P18:cosmological-parameters} and tensor-to-scalar ratio $r=0$. Additionally, as $\phi_{\ell m}$, $\tilde{t}_{\ell m}$, and $\tilde{e}_{\ell m}$ are the complex harmonic coefficients of a real field, $h_{\ell m}$, $j_{\ell m}$, and $k_{\ell m}$ have to verify the reality condition $a_{\ell m}^{\ast}= (-1)^m a_{\ell -m}$, where $a\in\{h, j, k\}$ \cite{Statistics_CMB}. We then lens the CMB maps using \texttt{lenspyx}\footnote{\url{https://github.com/carronj/lenspyx}}~\cite{Lenspyx_citation}.

Next, we produce $31$ frequency maps ($9$ for \textit{Planck} and $22$ for \textit{LiteBIRD}) of synchrotron, thermal dust, free-free, anomalous microwave emission (AME), and molecular CO Galactic emission using \texttt{pysm3}\footnote{\url{https://github.com/galsci/pysm}}~\cite{Thorne_2017, Zonca_2021, pysm3_2025}. We consider the following three different scenarios,
\begin{itemize}
    \item \textbf{No foregrounds}. These simulations contain only CMB and the instrumental noise and beam of each experiment. They provide a useful benchmark to evaluate the efficiency of component separation by comparison with the ideal scenario.
    \item \textbf{Simple foregrounds}. These simulations are based on the \texttt{s1}, \texttt{d1}, \texttt{a1}, \texttt{f1}, and \texttt{co1} \texttt{pysm3} foreground models. \texttt{s1} models the spectral energy distribution (SED) of synchrotron emission through a power-law scaling with a spatially varying spectral index derived from the analysis of Haslam 408~MHz and \textit{WMAP} 23~GHz 7-year data~\cite{WMAP_7years}. Synchrotron amplitude templates come from the reprocessed Haslam 408~MHz~\cite{Haslam_reproc} and \textit{WMAP} 9-year 23~GHz $Q$/$U$ maps~\cite{WMAP_Bennet_2013}. \texttt{d1} models thermal dust through a one-component modified blackbody SED, with amplitudes and spatially varying spectral parameters from the \texttt{Commander} intensity and polarization \textit{Planck}-2015 analysis~\cite{Commander_Planck_2015}. The AME model \texttt{a1} is based on the sum of two unpolarized spinning dust populations derived from the same \texttt{Commander} analysis~\cite{Commander_Planck_2015}. \texttt{f1} describes the SED of the unpolarized free-free emission with a constant power-law index of $-2.14$ using the amplitudes from the 30 GHz \texttt{Commander} fit to \textit{Planck}-2015 data~\cite{Commander_Planck_2015,Draine_free_free}. \texttt{co1} includes the unpolarized emission from the first three CO rotational lines using templates from the \texttt{MILCA} analysis of \textit{Planck} data~\cite{CO_MILCA_2014}. 
    \item \textbf{Complex foregrounds}. These simulations are based on the \texttt{s5}, \texttt{d10}, \texttt{a1}, \texttt{f1}, and \texttt{co3} \texttt{pysm3} foreground models. Compared to \texttt{s1}, in \texttt{s5} the dispersion of the spectral index's spatial variability is rescaled to match that observed in S-PASS~\cite{SPASS}. \texttt{d10} uses templates from the \texttt{GNILC} needlet-based analysis of \textit{Planck} data~\cite{GNILC-Planck2016, Planck_diffuse_2020}, which reduce the contamination from point sources and the cosmic infrared background (CIB) compared to \texttt{Commander} products. Compared to \texttt{co1}, \texttt{co3} is polarized at the level of $0.1\,\%$ and includes mock CO clouds up to $20^\circ$ outside of the Galactic plane simulated with \texttt{MCMole3D}~\cite{2017Puglisi}.
\end{itemize}
These simulations extend our previous work in ref.~\cite{Lonappan_lensing} through the addition of temperature maps, and hence the free-free, AME, and CO components, and through the use of more complex synchrotron and dust models in polarization. We do not include here extragalactic foregrounds such as point sources, the CIB, or the SZ effect. The study of the impact of extragalactic contaminants is left for future work~\cite{Mishra:2019, Sailer:2020, 2021Fabbian, 2022Lembo, 2022Cai, 2023Darwish}.

Finally, we model the instrument response by adding white Gaussian noise and a symmetric Gaussian beam to each observed frequency band. The specifications for \textit{Planck} channels are provided in table 4 of ref.~\cite{Planck_overview_2020}, and for \textit{LiteBIRD}, in table 13 of ref.~\cite{LiteBIRD_PTEP_2023}. For temperature observations, we scaled down by a $\sqrt{2}$ factor the polarization sensitivities tabulated for \textit{LiteBIRD}. We assume delta bandpasses centered at the nominal frequency. Our maps are defined on a \texttt{HEALPix}\footnote{\url{https://healpix.sourceforge.io/}}~\cite{gorski} grid of $N_\mathrm{side}=512$ for \textit{LiteBIRD} and $N_\mathrm{side}=2048$ for \textit{Planck}.\footnote{Although \texttt{pysm3} developed several strategies to add small-scale information to foreground models~\cite{Thorne_2017, Zonca_2021}, the \texttt{s1}, \texttt{d1}, \texttt{a1}, and \texttt{f1} templates were derived at $N_\mathrm{side}=512$ and then upsampled to higher resolutions using the \texttt{healpy}~\cite{healpy} function \texttt{ud\_grade}. This produces some negligible numerical artifacts around $\ell\sim 1500$ in the angular power spectra of polarized foregrounds in our \textit{Planck} simulations. Nevertheless, the impact of these artifacts is negligible because they are several orders of magnitude below the noise. The issue is resolved for \texttt{s5} and \texttt{d10} templates because they were natively generated at higher resolution.}

\section{Methodology} \label{sec:reconst}

In this section, we detail the different steps of the analysis pipeline. First, we specify how CMB maps are obtained through component separation (section~\ref{sec:HILC}), the combination of \textit{Planck} and \textit{LiteBIRD} data (section~\ref{subsec:apply HILC}), and C-inverse filtering (section~\ref{sec: filtering}). Secondly, we present the quadratic estimators used to reconstruct the lensing potential (section~\ref{sec: qe recons}), the characterization and subtraction of the main biases (section~\ref{sec: lensing ps}), and the calculation of the minimum-variance (MV) lensing estimate (section~\ref{sec: MV lensing}). 
 
\subsection{Component separation: harmonic internal linear combination} \label{sec:HILC}

In our forecast, we separate the different components of the observed signal using a harmonic internal linear combination (HILC)~\cite{Tegmark_HILC, HILC_derivation_2008}. In harmonic space, the observed signal in CMB thermodynamic units at a frequency $\nu$ is 
\begin{equation}
    a_{\ell m}^{\nu} = (a_{\ell m}^\mathrm{s} + a_{\ell m}^{\mathrm{fg}, \nu})B_\ell^{\nu}P_\ell^\mathrm{w} + a_{\ell m}^{\mathrm{n}, \nu},
\end{equation}
where $a_{\ell m}^\mathrm{s}$ is the CMB signal, and $a_{\ell m}^{\mathrm{fg}, \nu}$ and $a_{\ell m}^{\mathrm{n}, \nu}$ are, respectively, the foreground and noise signals at a frequency $\nu$. We assume the instrumental beam at a frequency $\nu$, $B_\ell^{\nu}$, to be Gaussian and symmetric, thus completely characterized by its full width at half maximum (FWHM). The pixel window function $P_\ell^\mathrm{w}$ depends on the map resolution and reflects the binning of time-ordered data into a discrete pixel grid.

These multifrequency observations are convolved with different beams. In order to combine them, first we deconvolve each one by its corresponding beam. We then define the cleaned map as the weighted sum of the deconvolved coefficients, $\hat{a}_{\ell m}^{\nu}$:
\begin{equation}
    a_{\ell m}^\mathrm{clean} = \sum_{\nu}w_\ell^\nu\frac{a_{\ell m}^\nu}{B_\ell^\nu P_\ell^\mathrm{w}} = \sum_{\nu}w_\ell^\nu \hat{a}_{\ell m}^{\nu}.
\end{equation}

The $w_\ell^\nu$ weights for each multipole and frequency are calculated by minimizing the variance for each multipole $\ell$ of the observed signal, 
\begin{equation}\label{eq: minimise variance HILC}
\text{Var}\left[ \sum_{\nu} w_\ell^\nu\hat{a}_{\ell m}^\nu\right]=\sum_{\nu\nu'}w_\ell^\nu w_\ell^{\nu'}\hat{C}_\ell^{\nu\nu'}
\end{equation}
under the restriction that $\sum_\nu w_\ell^\nu =1$. In eq.~(\ref{eq: minimise variance HILC}), $\hat{C}_\ell^{\nu\nu'}$ is the frequency-frequency covariance matrix calculated as the cross-power spectrum between the $\nu$ and $\nu'$ channels,
\begin{equation}
    \bR{\hat{C}}_\ell = \hat{C}_\ell^{\nu\nu'} = \frac{1}{2\ell+1}\sum_{m=-\ell}^{\ell}\hat{a}_{\ell m}^\nu\hat{a}_{\ell m}^{\nu'\ast}.
\end{equation}
Using the Lagrange multiplier theorem, the optimal weights are calculated, in matrix notation, as
\begin{equation}
    \bm{w}_\ell = \frac{\bm{1}^{\sf T}\bR{\hat{C}}_\ell^{-1}}{\bm{1}^{\sf T}\bR{\hat{C}}_\ell^{-1}\bm{1}},
\end{equation}
where $\bm{1}$ denotes a column vector of ones with length equal to the number of observed frequencies.

Since the CMB signal is frequency-independent in thermodynamic units, the $\sum_\nu w_\ell^\nu =1$ restriction guarantees that the CMB power is preserved. This claim is only partially true. The existing chance correlations, mainly between the signal and noise fluctuations, can bias the recovered power spectra at low multipoles and lead to a loss of signal at large angular scales~\cite{HILC_derivation_2008, Delabrouille_HILC_derivation, NILC}. This topic will be discussed in more detail in appendix~\ref{Appendix Bias HILC}.

Although HILC is a powerful component separation tool, it has some limitations. HILC assumes that the CMB, foregrounds, and noise are Gaussian, with all their information contained at the power spectrum level. It also assumes the isotropy of the signal in the weight derivation, which prevents us from optimally capturing the spatial variability of the foreground emission. In contrast, other blind component-separation methods such as the needlet ILC~\cite{NILC,2023Carones} or parametric methods (e.g., \texttt{Commander}~\cite{2023Galloway}, \texttt{FGBuster}~\cite{2009Stompor}, or \texttt{B-SeCRET}~\cite{2020delaHoz}) do not require this assumption and can provide a more efficient foreground cleaning at low $\ell$. Nevertheless, here we decided to keep HILC as our component-separation algorithm because it is computationally cheap, simple, and still provides a lensing signal-to-noise compatible with the official \textit{Planck} results. We leave it to future work to test the impact of using different component-separation methods on the lensing reconstruction.

\subsection{Optimal combination of \textit{Planck} and \textit{LiteBIRD} data} \label{subsec:apply HILC}

When analyzed separately, we run the HILC using multipoles $2\leq\ell\leq 1000$ for \textit{LiteBIRD} and $2\leq\ell\leq 2048$ for \textit{Planck}.\footnote{\textit{Planck} analyses only used $100\leq\ell\leq 2048$ multipoles for their lensing estimation~\cite{P18:phi, Planck_PR4_lensing}.} Having a different number of temperature and polarization channels for \textit{Planck} is not a problem, as the HILC treats $T$, $E$, and $B$ modes independently. 

\begin{figure}[t]
    \centering
    \includegraphics[width=1\textwidth]{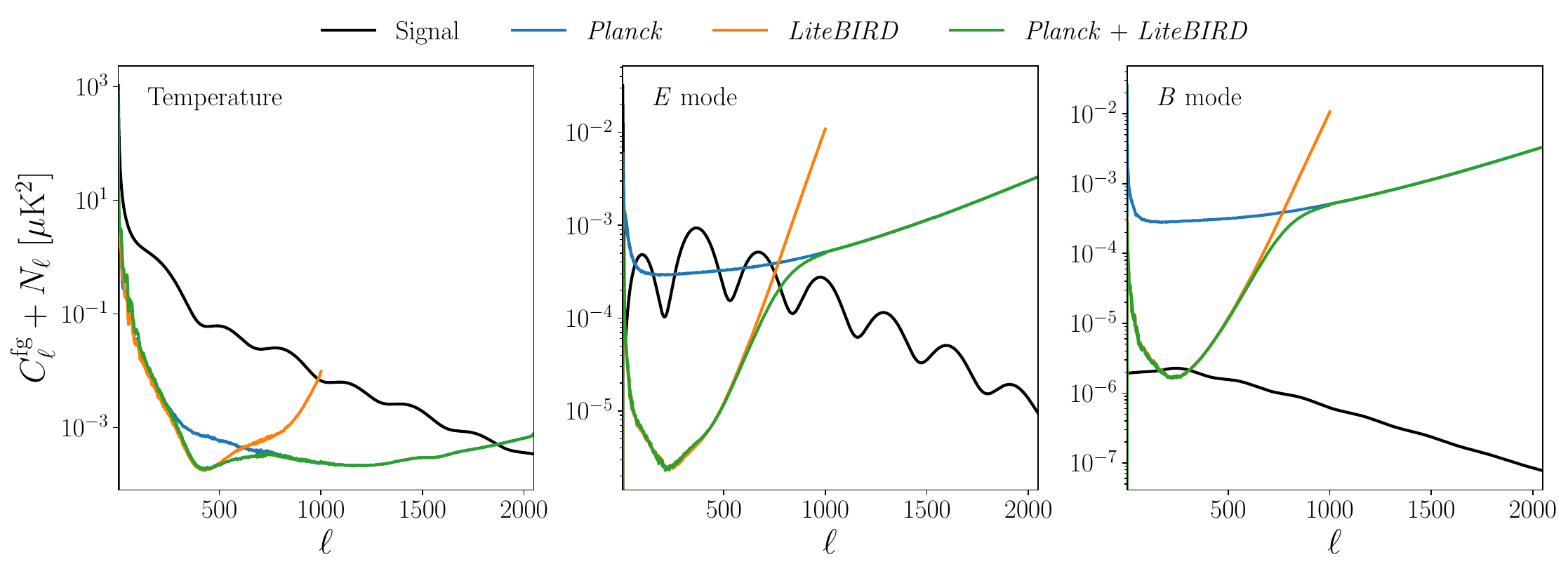}
    \caption{
    Foreground and noise residuals computed from $400$ component-separated temperature and polarization simulations in the simple-foregrounds case. Solid black lines correspond to the input signal (from left to right, $T$, $E$, and $B$). Residuals are shown for \textit{LiteBIRD} (blue), \textit{Planck} (orange) and \textit{Planck} + \textit{LiteBIRD} (green). Temperature power spectra are calculated with the $97\,\%$ \textit{Planck} Galactic mask, while polarization results are for full sky.
    }
    \label{fig:res_s1_d1}
\end{figure}

The combination of \textit{Planck} and \textit{LiteBIRD} data is straightforward in harmonic space since the HILC also treats each multipole independently. This allows us to divide the multipole range into two regions:
\begin{itemize}
    \item[i)] we use the full set of \textit{Planck} + \textit{LiteBIRD} bands in the multipole range $2 \leq \ell\leq 1000$, which includes 31 (29) frequency maps in temperature (polarization);
    \item[ii)] we rely completely on \textit{Planck} data to recover multipoles $1001 \leq \ell\leq 2048$, with only 9 (7) frequency maps being available in temperature (polarization) at these scales. 
\end{itemize}
Choosing $\ell=1000$ as the transition multipole allows us to have a smooth transition between both multipole regions, guaranteeing that we are taking the best of both experiments as can be seen in figure~\ref{fig:res_s1_d1}. \textit{LiteBIRD}’s noise explodes well before $\ell=1000$, leaving a buffer of a couple of hundred multipoles where the HILC weights smoothly combine the data from both experiments.

We estimate the frequency-frequency covariance matrix from the frequency maps of each simulation. Exceptionally, in the no-foreground scenario, we build the covariance matrix directly from the white-noise maps since the minimization of the HILC reduces to an inverse-variance weighting problem in this case.

We apply the $97\,\%$ \textit{Planck} Galactic mask\footnote{All \textit{Planck} Galactic masks can be found in the \texttt{HFI\_Mask\_GalPlane-apo0\_2048\_R2.00.fits} file available at \url{https://pla.esac.esa.int/\#maps}.} to the temperature frequency maps of both experiments when calculating the frequency-frequency covariance matrix. To mitigate the mode coupling between multipoles, the mask is apodized using a $2^\circ$ C1 apodization by \texttt{NaMaster}\footnote{\url{https://github.com/LSSTDESC/NaMaster}} \cite{pymaster}. We decided to mask temperature maps because, since the HILC weights minimize the total power, including CMB, foregrounds, and noise, the very bright foreground signal in the Galactic center dominates over the full-sky signal and leads to a high level of noise residuals at small scales. In particular, the noise residuals are higher than those obtained in official \textit{Planck} analyses, leading to a significantly lower signal-to-noise ratio (SNR) of the lensing detection than what was reported by the \textit{Planck} collaboration. Note that we only mask temperature maps when calculating the HILC weights, which are then applied to the full-sky data to produce a full-sky component-separated map. This leads to full-sky temperature maps with significant leftover foreground contamination in the small region obscured by the $97\,\%$ mask, but overall lower noise residuals at small scales. Although it can also reduce small-scale noise residuals, no masking is applied in polarization, since the improvement is not that significant compared to temperature.

\begin{figure}[t]
    \centering
    \includegraphics[width=1\textwidth]{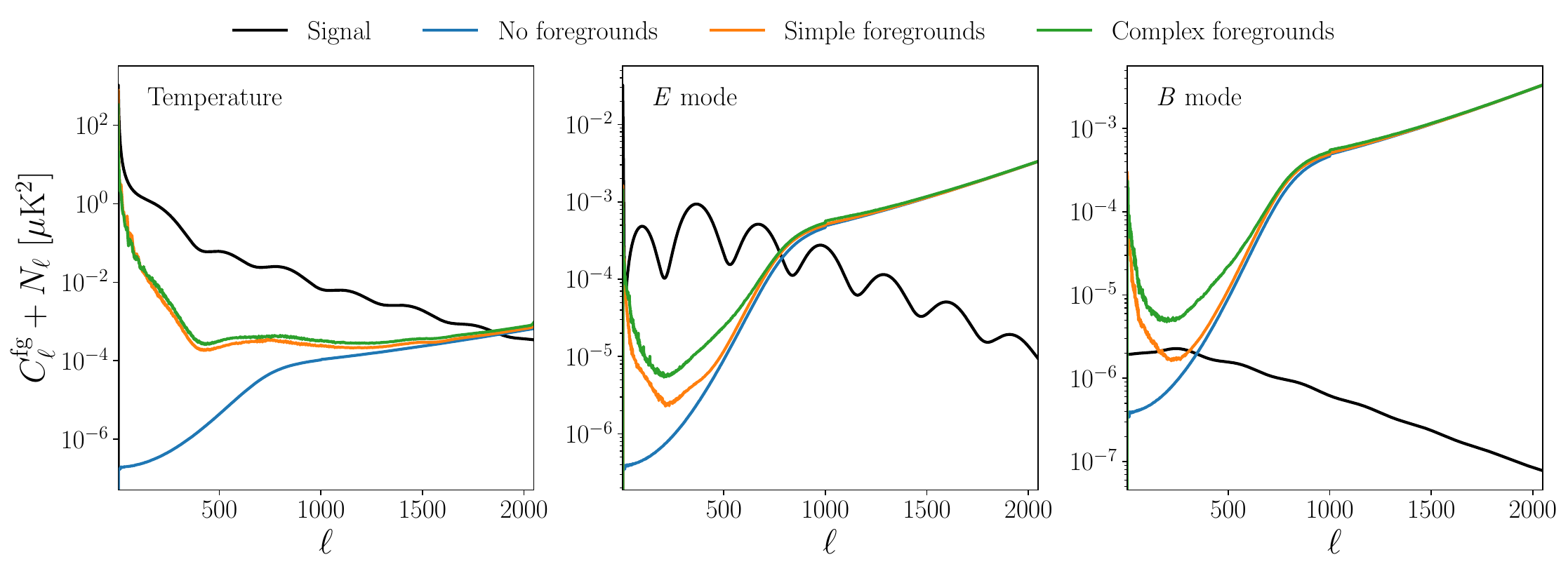}
    \caption{
    Foreground and noise residuals computed from $400$ component-separated temperature and polarization simulations of \textit{Planck} + \textit{LiteBIRD}. Solid black lines correspond to the input signal (from left to right, $T$, $E$, and $B$). Residuals are shown for the no-foregrounds (blue), simple-foregrounds (orange), and complex-foreground (green) cases. Temperature power spectra are calculated with the $97\,\%$ \textit{Planck} Galactic mask, while polarization results are for full sky. 
    }
    \label{fig:res_LiteBIRD_Planck}
\end{figure}

Figure~\ref{fig:res_s1_d1} shows the noise and foreground residuals for the simple-foreground simulations, illustrating the inner workings of the combination of \textit{Planck} and \textit{LiteBIRD} data sets. \textit{LiteBIRD} dominates in both temperature and polarization at large angular scales thanks to its lower noise. \textit{Planck} is the only contribution at small scales due to its higher angular resolution. At intermediate scales, the combination of both data sets achieves lower residuals than each individual experiment.

Figure~\ref{fig:res_LiteBIRD_Planck} focuses instead on how the \textit{Planck}+\textit{LiteBIRD} combination performs in different foreground scenarios. Results from the no-foreground simulations are included to illustrate the potential of the combination in an ideal scenario and provide a benchmark to evaluate the efficiency of component separation. A significant increase of residuals can be seen at intermediate polarization scales, especially for $B$ modes, when going from simple to complex foregrounds. Only a small difference is observed in temperature, since the increase in complexity of the \texttt{pysm3} models used here focuses mainly on the polarization signal. The increase on residuals mainly points out that HILC is not a sufficiently flexible component-separation method as it struggles to deal with the spatial variability of the spectral parameters in more complex foreground scenarios. The difference to the no-foreground case shows that there is still an ample margin for improvement at large angular scales by the use of a more efficient component-separation algorithm. Nevertheless, as previously mentioned, we will stick to HILC, since the scope of this work is to prove the advantages of a combined analysis of \textit{Planck} and \textit{LiteBIRD} data for the lensing reconstruction irrespective of the foreground models and component-separation technique considered. 

Bear in mind that no additional masking beyond the $97\,\%$ temperature mask for the calculation of HILC weights and power spectrum was applied to derive the results presented in figures \ref{fig:res_s1_d1} and \ref{fig:res_LiteBIRD_Planck}. Hence, the contribution from foreground residuals still needs to be reduced through masking before any cosmological analysis. In particular, we apply the $70\,\%$ \textit{Planck} Galactic mask to \textit{Planck}-only runs and the $80\,\%$ \textit{Planck} Galactic mask to \textit{LiteBIRD}-only and \textit{Planck}+\textit{LiteBIRD} runs. The former mask is chosen for consistency with \textit{Planck}'s lensing analyses~\cite{P18:phi, Planck_PR4_lensing} and the latter to be consistent with our previous work in \cite{Lonappan_lensing}. A $2^\circ$ C1 apodization is applied to these masks to mitigate mode coupling in the filtering step.

\subsection{Filtering of CMB anisotropies}\label{sec: filtering}

The clean CMB maps estimated in the previous section need to be filtered to remove masked regions and down-weight noisy areas, a required step before lensing reconstruction. We use an inverse-variance filter (or C-inverse filter) for our lensing analysis. Its definition can be derived from that of the well-known Wiener filter (WF), which in the case of Gaussian noise and signal, provides the optimal maximum a posteriori reconstruction of the signal~\cite{MAP_Carron_2017}. Wiener-filtered maps are given by
\begin{equation}
\bm{s}^\mathrm{WF}=\bm{S}(\bm{S}+\bm{N})^{-1}\bm{s}^\mathrm{data},
\end{equation}
where $\bm{s}^\mathrm{WF}=(T^\mathrm{WF}, Q^\mathrm{WF}, U^\mathrm{WF})^{\sf T}$ are the Stokes parameters of the filtered maps, $\bm{S}$ and $\bm{N}$ are respectively the pixel-space signal and noise covariance matrices, and the $\bm{s}^\mathrm{data}=(T, Q, U)^{\sf T}$ input maps are in our case the beam- and pixel-window-deconvolved maps produced by the HILC.

Starting in pixel space, we want to transform the filtering operation to harmonic space. To do so, we introduce the $\mathbfcal{Y}$ and $\mathbfcal{Y}^{-1}$ spherical harmonic matrix operators, which perform direct and inverse spherical harmonic transforms, respectively:
\begin{equation}\label{direct SHT}
    \bm{s}^\mathrm{data} = \mathbfcal{Y}\bm{s}_{\ell m}^\mathrm{data};
\end{equation}
\begin{equation}\label{inverse SHT}
    \bm{s}_{\ell m}^\mathrm{data} \cong w \mathbfcal{Y}^\dagger \bm{s}^\mathrm{data}=\mathbfcal{Y}^{-1}\bm{s}^\mathrm{data}.
\end{equation}
Here $\bm{s}_{\ell m}^\mathrm{data}=(t_{\ell m}, e_{\ell m}, b_{\ell m})^{\sf T}$ is a column vector of dimension $3L$  ($L=\sum_{\ell=2}^{\ell_\mathrm{max}} (2\ell+1)$), $\dagger$ is the conjugate transpose, and $w=4\pi/N_\mathrm{pix}$ is the pixel solid angle, which is the same for all pixels in the HEALPix pixelation scheme~\cite{gorski}. The approximation symbol appears because, when working with pixelated maps, the integral over the solid angle is approximated by a sum. The matrix $\mathbfcal{Y}$ has dimensions $3N_\mathrm{pix}\times 3L$ and its elements are  
\begin{equation}
    \mathbfcal{Y} = \begin{pmatrix}
\bm{Y} & \phantom{-}\bm{0} & \phantom{-}\bm{0}\\
\bm{0} & \ -\bm{X}_1 & -i\bm{X}_2\\
\bm{0} & \phantom{-}i\bm{X}_2 & \ -\bm{X}_1
\end{pmatrix},
\end{equation}
where $\bm{Y}$, $\bm{X}_1$, and $\bm{X}_2$ are $N_\mathrm{pix} \times L$ matrices that contain spin-weighted spherical harmonics. The elements of the  $\bm{Y}$ matrix are the scalar ($s=0$) spherical harmonics, and the elements of the $\bm{X}_1$ and $\bm{X}_2$ matrices are the linear combinations of $s=2$ spin-weighted spherical harmonics defined in eq.~(4) of ref.~\cite{Healpix_convention}. The matrix $\mathbfcal{Y}$ verifies that $\mathbfcal{Y}\mathbfcal{Y}^{-1}=w\mathbfcal{Y}\mathbfcal{Y}^\dagger=\bm{I}_{3N_{pix}}$ and $\mathbfcal{Y}^{-1}\mathbfcal{Y}=w\mathbfcal{Y}^\dagger \mathbfcal{Y}=\bm{I}_{3L}$, where $\bm{I}_n$ is the identity matrix of dimension $n\times n$. The matrix $\mathbfcal{Y}^{-1}$ is the pseudoinverse of $\mathbfcal{Y}$. 

The pixel covariance matrix of the signal can be expressed as $\bm{S}=\mathbfcal{Y}\mathbb{C}^{\rm s}\mathbfcal{Y}^\dagger$, where $\protect\fakebold{\mathbb{C}}^{\rm s}$ is the signal covariance matrix in harmonic space with elements 
\begin{equation}
\protect\fakebold{\mathbb{C}}^{\rm s}_{\ell\ell'mm'}=\delta_{\ell\ell'}\delta_{mm'}
\begin{pmatrix}
C_\ell^{TT} & C_\ell^{TE} & 0\\
C_\ell^{TE} & C_\ell^{EE} & 0\\
0 & 0 & C_\ell^{BB}
\end{pmatrix}.
\end{equation}

At this point, we have all the tools to calculate the harmonic transform of the Wiener filter:
\begin{equation}
\begin{aligned}
    \protect\bm{s}_{\ell m}^\mathrm{WF}&=w\mathbfcal{Y}^\dagger \bm{s}^\mathrm{WF} = w\mathbfcal{Y}^\dagger \bm{S}(\bm{S}+\bm{N})^{-1}\bm{s}^\mathrm{data}=w\mathbfcal{Y}^\dagger \mathbfcal{Y} \fakebold{\mathbb{C}}^{\rm s} \mathbfcal{Y}^\dagger (\mathbfcal{Y} \fakebold{\mathbb{C}}^{\rm s}\mathbfcal{Y}^\dagger+\bm{N})^{-1}\bm{s}^\mathrm{data}=\\
    &=\fakebold{\mathbb{C}}^{\rm s}\mathbfcal{Y}^\dagger (\mathbfcal{Y} \fakebold{\mathbb{C}}^{\rm s}\mathbfcal{Y}^\dagger+\bm{N})^{-1}\bm{s}^\mathrm{data}.
\end{aligned}
\end{equation}

Now, multiplying by $w\mathbfcal{Y} \mathbfcal{Y}^\dagger=\bm{I}_{3N_{\rm pix}}$, we obtain the final expression:
\begin{equation}\label{harmonic WF}
\protect \bm{s}_{\ell m}^\mathrm{WF} = \fakebold{\mathbb{C}}^{\rm s}\mathbfcal{Y}^\dagger (\mathbfcal{Y}\fakebold{\mathbb{C}}^{\rm s}  \mathbfcal{Y}^\dagger+\bm{N})^{-1}\mathbfcal{Y} w\mathbfcal{Y}^\dagger \bm{s}^\mathrm{data}=\fakebold{\mathbb{C}}^{\rm s}(\fakebold{\mathbb{C}}^{\rm s}+\fakebold{\mathbb{N}})^{-1}\bm{s}_{\ell m}^\mathrm{data},
\end{equation}
where $\protect\fakebold{\mathbb{N}}=w^2\mathbfcal{Y}^\dagger \bm{N} \mathbfcal{Y}$ is the harmonic-space noise covariance matrix. 

For optimal filtering in lensing reconstruction with quadratic estimators~\cite{Optimal_filtering_2019}, CMB observations must be further weighted by an additional $\protect\fakebold{\mathbb{C}}^{\rm s}$ factor. Hence the C-inverse filtered maps, $\bm{s}_{\ell m}^\mathrm{CI}$, are defined as
\begin{equation}\label{diag harmonic WF}
\protect \bm{s}_{\ell m}^\mathrm{CI}=(\fakebold{\mathbb{C}}^{\rm s}+\fakebold{\mathbb{N}})^{-1}\bm{s}_{\ell m}^\mathrm{data}.
\end{equation}
Computing the C-inverse filter in either pixel or harmonic space involves the inversion of a big covariance matrix. This operation can be computationally challenging and expensive, often requiring the use of conjugate gradient inversion methods~\cite{Eriksen:2004:wiener, 2022_SO_delensing}. This approach was used in our previous \textit{LiteBIRD} lensing analysis \cite{Lonappan_lensing}. 

In this paper, we implement instead the diagonal harmonic-space filtering of ref.~\cite{ACT_DR6_lensing}, which is an approximation to eq.~\eqref{diag harmonic WF}. First, we approximate $\protect\fakebold{\mathbb{C}}^{\rm s}$ as diagonal, neglecting the $TE$ correlation and thus filtering independently temperature and polarization maps. This diagonal approximation also neglects the mode mixing due to masking and only works if the noise is isotropic (or close to isotropic), which is the situation of this paper, since we do not include the scanning strategy or other systematics effects from \textit{Planck} and \textit{LiteBIRD}. Under those approximations, the matrix $\protect\fakebold{\mathbb{C}}^{\rm s}+\fakebold{\mathbb{N}}$ in eq.~\eqref{diag harmonic WF} can be approximated by the angular power spectra and
\begin{equation}\label{eq: diagonal harmonic filtering}
    x_{\ell m}^\mathrm{CI} \approx \frac{x^\mathrm{data}_{\ell m}}{C_\ell^{XX}+\langle N_\ell^{XX}\rangle},
\end{equation}
where $x\in \{t, e, b\}$, $C_\ell^{XX}$ are the fiducial lensed power spectra, and $\langle N_\ell^{XX}\rangle$ are the mean full-sky noise residuals after component separation calculated from the average of all simulations. We exclude foreground residuals from the calculation of $\langle N_\ell^{XX}\rangle$, assuming that the majority of leftover foreground emission would be covered by the mask.  

For the diagonal harmonic-space C-inverse filter, data can be masked either before or after filtering. In their DR6 lensing analysis, the ACT collaboration applied the filter to masked data~\cite{ACT_DR6_lensing}. We find this approach appropriate for an accurate recovery of the intermediate to small scales accessible by ACT. However, \textit{LiteBIRD} and \textit{Planck} are full-sky surveys and the largest scales must be considered. To obtain $t_{\ell m}^\mathrm{CI}$ and $e_{\ell m}^\mathrm{CI}$, the approach of starting from masked data and then filtering works well. Nevertheless, the mode coupling between different scales induced by masking produces a spurious increase of the $BB$ power at low $\ell$ due to $E$-to-$B$ leakage~\cite{pymaster}. To mitigate this effect, we estimate $b_{\ell m}^\mathrm{CI}$ by filtering first the observed $e_{\ell m}$ and $b_{\ell m}$, and obtaining from them filtered $Q$ and $U$ maps. Afterwards, we mask those $Q$ and $U$ maps and calculate the $b_{\ell m}^\mathrm{CI}$ coefficients through the standard spherical harmonic transform.

\begin{figure}[t]
    \centering
    \includegraphics[width=1\textwidth]{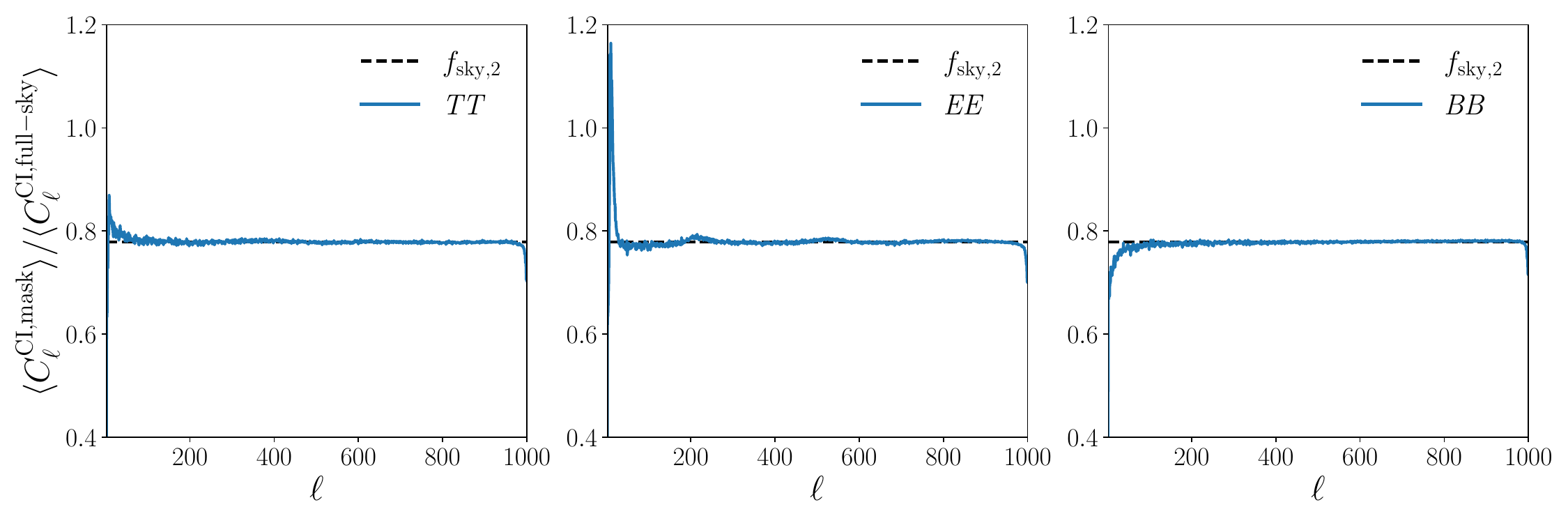}
    \caption{
    Ratio between the diagonal harmonic-space C-inverse filtered data of masked and full-sky angular power spectra for no-foreground \textit{LiteBIRD} simulations. Results are averaged over 400 simulations. Dashed black lines correspond to the unmasked sky fraction, $f_{\mathrm{sky},2}=\sum_i w_i^2/N_{\rm pix}$.
    }
    \label{fig:Harmonic_filtering_LiteBIRD_no_fg}
\end{figure}

In figure~\ref{fig:Harmonic_filtering_LiteBIRD_no_fg}, we show the ratio between the masked and full-sky filtered CMB power spectra. Results are only shown for no-foreground \textit{LiteBIRD} simulations, since this is the case where we can perform a fair comparison with the full-sky scenario because the signal and noise are homogeneous and isotropic. The diagonal harmonic filtering of eq.~(\ref{eq: diagonal harmonic filtering}) is exact in the full-sky limit, thus by comparing it with the partial-sky case, we can identify mask-induced effects by finding deviations of the ratio between the masked and full-sky filtered spectra from the effective sky fraction, $f_{\mathrm{sky},2}=\sum_i w_i^2/N_{\rm pix}$ (calculated from the value the mask takes at each pixel, $w_i$), that one would recover with an exact filtering. No obvious $E$-to-$B$ leakage is observed, validating our mitigation strategy for mask-induced couplings in $BB$ through filtering in harmonic space first and masking in real space second. The power-spectra ratio reasonably matches the effective sky fraction. We only observe significant deviations at large angular scales. Comparable differences were also observed in our previous analysis~\cite{Lonappan_lensing}, showing that the diagonal filtering approximation has a similar accuracy compared to the pixel-based approach. Furthermore, the diagonal harmonic-space C-inverse filter is $\mathcal{O}(100)$ times faster than its pixel space counterpart, making it a much more convenient option for forecasting.

\subsection{Lensing potential reconstruction}\label{sec: qe recons}

This section presents the quadratic estimators (QEs) used to reconstruct the lensing potential map. We first present QEs for the full-sky, and then explain how we correct for the bias that masking introduces.

The unlensed CMB spherical harmonic coefficients $\widetilde{X}_{\ell m}$ are assumed to be Gaussian and statistically isotropic, so that their statistical properties are fully characterized by the angular power spectra,
\begin{equation}
\langle \widetilde{X}_{\ell m}\widetilde{Y}_{\ell' m'}^{\ast} \rangle = \delta_{\ell\ell'}\delta_{mm'}\widetilde{C}_\ell^{XY},
\end{equation}
where $X,Y\in\{T,E,B \}$ and $\ast$ denotes the complex conjugate. When lensed by a fixed deflection field, the covariance acquires off-diagonal terms and becomes
\begin{equation}\label{eq: lensing-off-diag}
\langle X_{\ell m}Y_{\ell' m'} \rangle |_\phi =
C_\ell^{XY}\delta_{\ell\ell'}\delta_{m-m'}(-1)^m+
\sum_{LM}(-1)^M
\begin{pmatrix}
\ell\phantom{'} & \ell' &\phantom{-}L\\
m\phantom{'} & m' & -M
\end{pmatrix}
f_{\ell L\ell'}^{XY}\phi_{LM},
\end{equation}
where the matrix corresponds to the Wigner 3-$j$ symbols and $f_{\ell L\ell'}^{XY}$ are weights for the different quadratic $XY$ pairs, which depend on the lensed power spectra. The expressions for the weights can be found in table~1 of ref.~\cite{OkamotoHu:quad}. Note that the lensed angular power spectrum, $C_\ell^{XY}$, used in the first term of eq.~\eqref{eq: lensing-off-diag}, although not explicitly shown, already captures the convolution with the lensing potential.

We can build an estimator for the lensing potential, $\hat{\phi}_{LM}^{XY}$, using the correlations produced by lensing. It is expressed as a weighted sum over multipole pairs, where the weights minimize the variance of the estimator, leading to the following expression:
\begin{equation}\label{QE eq}
\hat{\phi}_{LM}^{XY} =
A_L^{XY}
\sum_{\ell_1 m_1}\sum_{\ell_2 m_2}(-1)^M
\begin{pmatrix}
\ell_1 & \ell_2 & \phantom{-}L\\
m_1 & m_2 & -M
\end{pmatrix}
f_{\ell_1 L\ell_2}^{XY}h^{XY}
X^\mathrm{CI}_{\ell_1 m_1}Y^\mathrm{CI}_{\ell_2 m_2},
\end{equation}
where $A_L^{XY}$ is a normalization factor, and $X^\mathrm{CI}_{\ell m}$ and $Y^\mathrm{CI}_{\ell m}$ are the C-inverse filtered data from section~\ref{sec: filtering}. Finally, $h^{XY}=1/2$ if $X=Y$ or 1 otherwise. The normalization in the ideal full-sky case is given by
\begin{equation}\label{QE norm}
A_L^{XY} = (2L+1)\left\{h^{XY} \sum_{\ell_1\ell_2}\frac{\left |f_{\ell_1 L\ell_2}^{XY}\right |^2}{\hat{C}^{XX}_{\ell_1}\hat{C}^{YY}_{\ell_2}}\right\}^{-1},
\end{equation}
where $\hat{C}^{XX}_{\ell}=C_{\ell}^{XX}+\langle N_{\ell}^{XX}\rangle$ from eq.~(\ref{eq: diagonal harmonic filtering}).

Throughout this paper, we use the $TT$, $EE$, $TE$, $TB$, and $EB$ QEs implemented in \texttt{lensQUEST}\footnote{\url{https://github.com/doicbek/lensquest}}~\cite{Beck_lensquest_2018, Beck_Thesis}. We ignore the $BB$ estimator since its contribution to the SNR is expected to be small~\cite{Okamoto_Hu_QE}. Although maximum a posteriori estimators provide the optimal lensing reconstruction~\cite{2015Anderes, MAP_Carron_2017, 2019Millea, 2020Millea}, here we stick to the standard QEs because not much signal-to-noise is gained by iteratively delensing \textit{LiteBIRD}~\cite{Lonappan_lensing,ComparisonDelensing}.

When averaged over simulations, the mean full-sky cross-spectrum between two different lensing reconstructions is given by:
\begin{equation}
\left \langle C_L^{\hat{\phi}^{XY} \hat{\phi}^{WZ}} \right \rangle = 
C_L^{\phi\phi}+N_L^{(0),XYWZ}+N_L^{(1),XYWZ}+\mathcal{O}([C_L^{\phi\phi}]^2),
\end{equation}
where $C_L^{\phi\phi}$ is the true lensing potential power spectrum and $N_L^{(a),XYWZ}$ are the reconstruction-noise biases of order $\mathcal{O}([C_L^{\phi\phi}]^a)$, henceforth denoted as N$a$ bias. The cross-spectrum is defined as
\begin{equation}
    C_L^{\hat{\phi}^{XY} \hat{\phi}^{WZ}} = \frac{1}{2L+1}\sum_{M=-L}^{L}\hat{\phi}_{LM}^{XY}\hat{\phi}_{LM}^{WZ\ast}.
\end{equation}
The major contribution to the reconstruction noise is the N0 bias, or Gaussian reconstruction noise, which originates from the disconnected part of the lensed CMB four-point function. The N1 bias captures secondary trispectrum contractions and is important on small scales, where it dominates over the lensing signal. The N2 bias has been mitigated by substituting the unlensed power spectra with the lensed power spectra in the $f_{\ell L\ell'}^{XY}$ weights~\cite{Hanson_2011}. Higher-order reconstruction biases have a negligible impact on the lensing reconstruction, and we are not considering them in this paper. The calculation and subtraction of all of these biases will be detailed in section~\ref{sec: lensing ps}. 

In a realistic scenario, working on the partial sky is necessary because of the presence of foreground contamination. In that situation, the normalization factor in eq.~\eqref{QE eq} becomes position-dependent and thus non-diagonal. However, lensing reconstruction is a very local operation, and keeping the normalization given by eq.~\eqref{QE norm} is correct away from the mask boundaries and across the majority of the sky~\cite{Carron_lensing_mask}. Additionally, masking introduces a source of anisotropies and off-diagonal couplings that are captured by QEs. Thus, the QE in eq.~\eqref{QE eq} contains the lensing potential signal, $\hat{\phi}_{LM}^{XY,\mathrm{lens}}$, and all the non-lensing sources of anisotropy, which can be included in the mean-field (MF) term, $\hat{\phi}_{LM}^{XY, \mathrm{MF}}$:
\begin{equation}
    \hat{\phi}_{LM}^{XY} = \hat{\phi}_{LM}^{XY, \mathrm{lens}} + \hat{\phi}_{LM}^{XY, \mathrm{MF}}.
\end{equation}

The MF can be estimated as the average of the QE over simulations and subtracted from the lensing reconstruction at the spherical harmonic coefficients level. In addition, we apply an apodized mask to reduce off-diagonal couplings and partially mitigate its impact~\cite{Benoit_Levy_2013}.

Calculating the MF as an average over simulations, $\hat{\phi}_{LM}^{XY, \mathrm{MF}}=\langle\hat{\phi}_{LM}^{XY}\rangle$, leads to a residual Monte Carlo (MC) noise equal to $N_L^{(0)}/N_{\rm sims}$, where $N_{\rm sims}$ is the number of simulations used to calculate the average. Instead, we can use a cross-spectra estimator to obtain the MF-free spectra without such MC noise~\cite{P18:phi}. The cross-spectra estimator for simulation $i$ is given by
\begin{equation}\label{MF free PS}
    C_L^{\hat{\phi}_i^{XY} \hat{\phi}_i^{WZ}} = \frac{1}{2L+1}\sum_{M=-L}^{L}\frac{N_{\rm sims}}{N_{\rm sims}-2}\left (\hat{\phi}_{i,LM}^{XY}-\hat{\phi}_{LM}^{\mathrm{MF1}, XY}\right) \left (\hat{\phi}_{i,LM}^{WZ}-\hat{\phi}_{LM}^{\mathrm{MF2}, XY}\right )^\ast,
\end{equation}
where we divide the 400 simulations into two splits named $S_1=\{1,\ldots, 200\}$ and $S_2=\{201,\ldots, 400\}$. For each split, we compute the MF as the average over simulations: $\hat{\phi}_{LM}^{\mathrm{MF}1,XY} = \langle \hat{\phi}_{LM}^{i,XY}\rangle_{S_{1}}$ and $\hat{\phi}_{LM}^{\mathrm{MF}2,XY} = \langle \hat{\phi}_{LM}^{i,XY}\rangle_{S_{2}}$. The factor $N_{\rm sims}/(N_{\rm sims}-2)$ excludes the $i$th-simulation from either $\hat{\phi}_{LM}^{\mathrm{MF}1,XY}$ or $\hat{\phi}_{LM}^{\mathrm{MF}2,XY}$. A detailed derivation can be found in appendix~\ref{Appendix MF cross}.  

To evaluate the importance of the MF bias over the different QEs, we compute the MF power spectra without MC noise using the cross-spectra estimator~\cite{CMBS4_Internal_Delensing},
\begin{equation}
    C_L^{\mathrm{MF},XYWZ} = \frac{1}{2f_{\rm{sky},4}}\left[C_L^{\hat{\phi}^{\mathrm{MF}1,XY}\hat{\phi}^{\mathrm{MF}2,WZ}} + C_L^{\hat{\phi}^{\mathrm{MF}1,WZ}\hat{\phi}^{\mathrm{MF}2,XY}}\right],
\end{equation}
where $f_{\mathrm{sky},4}=\sum_i w_i^4/N_{\rm pix}$ is the mean fourth-power of the mask \cite{Benoit_Levy_2013, Namikawa_fsky4_2013}.

\begin{figure}[t]
    \centering
    \includegraphics[width=1\textwidth]{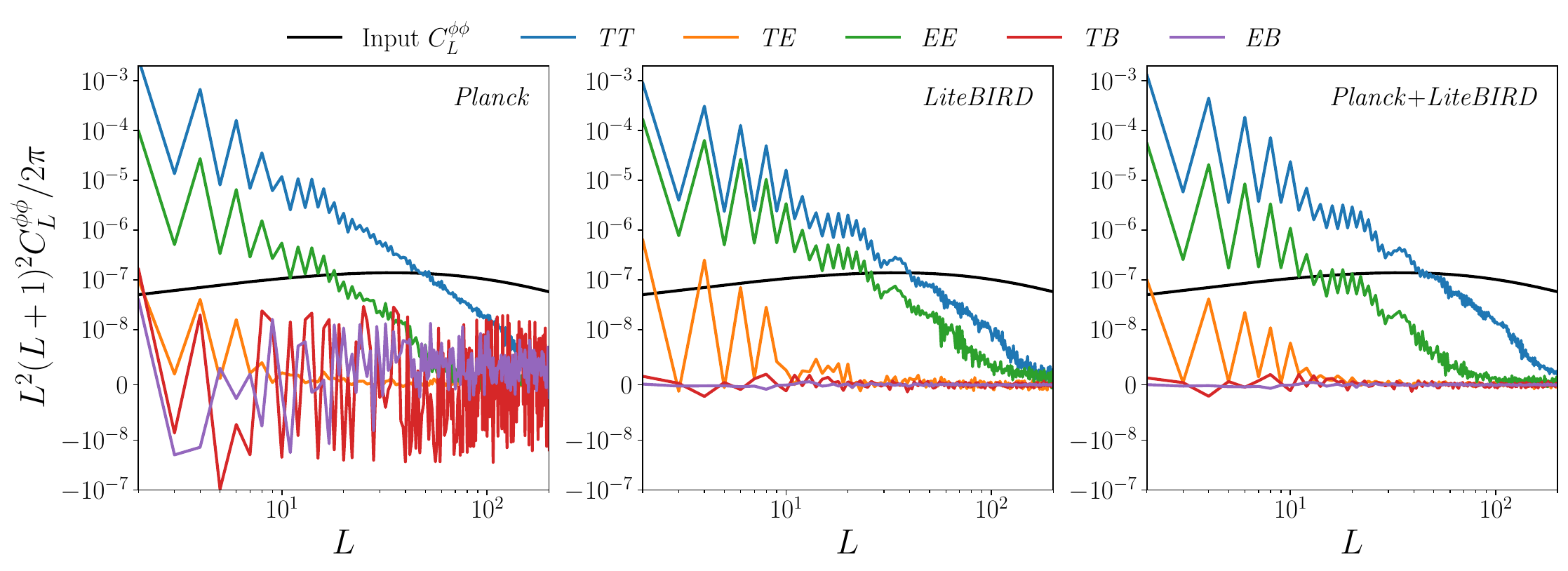}
    \caption{
    Power spectrum of the mean field (MF) calculated for \textit{Planck}, and for \textit{LiteBIRD} and \textit{Planck} + \textit{LiteBIRD} with our simple-foreground simulations. The black line shows the input lensing power spectrum used to generate the simulations.
    }
    \label{fig:MF_low_complexity}
\end{figure}

Figure~\ref{fig:MF_low_complexity} shows the MF for the simple-foregrounds case and the different experiments. We show only this case since the MF does not strongly depend on the foregrounds considered. This is because the main source of anisotropy is the mask, since the residual foregrounds are subdominant outside the mask. The MF is calculated for the 70\,\% Galactic mask in the case of \textit{Planck}, and the 80\,\% Galactic mask for the \textit{LiteBIRD} and \textit{Planck} + \textit{LiteBIRD} cases. From this plot, we see that while the $TT$ and $EE$ MF are very high at low $\ell$, the $EB$ MF is negligible~\cite{Namikawa_fsky4_2013, Namikawa_pol_2014}. \textit{LiteBIRD} relies mainly on the $EB$ QE for the lensing reconstruction, which allows an accurate measurement of lensing at large scales. In practice, \textit{LiteBIRD} MF subtraction is not critical for large-scale lensing reconstruction and is less susceptible than \textit{Planck} to biases due to MF misestimation.

The average partial-sky cross-spectrum between two different reconstructions of the lensing potential is
\begin{equation}\label{lensing ps}
\frac{1}{f_{\mathrm{sky},4}}\left \langle C_L^{\hat{\phi}^{XY} \hat{\phi}^{WZ}} \right \rangle = 
R_L^{\rm MC}C_L^{\phi\phi}+N_L^{(0),XYWZ}+N_L^{(1),XYWZ},
\end{equation}
where $R_L^{\rm MC}$ is a MC correction of the analytical normalization.

\subsection{Bias subtraction}\label{sec: lensing ps}

Solving eq.~\eqref{lensing ps}, the debiased power spectrum of the lensing potential for each simulation $i$ is given by:
\begin{equation}\label{cl_pp_debiased}
\hat{C}_{L,i}^{\phi^{XY}\phi^{WZ}}=\frac{1}{R_{L}^{\rm MC}}\left(\frac{1}{f_{\rm sky,4}}C_L^{\hat{\phi}_i^{XY} \hat{\phi}_i^{WZ}}-N_{L,i}^{(0),XYWZ}-N_L^{(1),XYWZ}\right),
\end{equation}
where the subscript $i$ in the N0 bias comes from its realization dependence, explained in section \ref{sec: N0 bias}. In the rest of this section, we will explain how to accurately estimate the different biases and the MC correction present in this expression.

\subsubsection{N0 bias} \label{sec: N0 bias}

The analytical full-sky expression for the N0 bias is given by
\begin{equation}\label{Analytical N0 bias}
\begin{aligned}
AN\text{-}N_L^{(0),XYWZ}=&\frac{A_L^{XY}A_L^{WZ}}{2L+1}
\sum_{\ell_1\ell_2} g^{XY\ast}_{\ell_1\ell_2}(L)\left [\bar{C}_{\ell_1}^{XW}\bar{C}_{\ell_2}^{YZ}
g^{WZ}_{\ell_1\ell_2}(L)\right.\\&+(-1)^{L+\ell_1+\ell_2}\bar{C}_{\ell_1}^{XZ}\bar{C}_{\ell_2}^{YW}\left. g^{WZ}_{\ell_2\ell_1}(L)
 \right ], 
\end{aligned}
\end{equation}
where $g^{XY}_{\ell_1\ell_2}(L)$ are the optimal weights derived for the QEs (they appear in eq.~\eqref{QE eq}),
\begin{equation}
    g^{XY}_{\ell_1\ell_2}(L) = \frac{h^{XY}f_{\ell_1 L\ell_2}^{XY\ast}}{\hat{C}^{XX}_{\ell_1}\hat{C}^{YY}_{\ell_2}},
\end{equation}
and $\bar{C}_{l}^{XY}$ is the power spectrum between two masked fields $X$ and $Y$ corrected by the unmasked pixel fraction $f_{\mathrm{sky},2}$, which includes both foreground and noise residuals in the case $X=Y$. The $\hat{C}^{XX}_{\ell}$ power spectra, which appear in both normalization and weights, are estimated considering only the signal and the mean noise of the HILC. For estimating a mean N0 bias, $\bar{C}_{\ell}^{XY}$ should be the mean observed power spectrum corrected by $f_{\rm{sky},2}$, $\langle\bar{C}_{\ell}^{XY}\rangle$, calculated using
\begin{equation} \label{mean_observed_PS}
    \bar{C}_{\ell}^{XY} = \frac{\hat{C}^{XX}_{\ell}\hat{C}^{YY}_{\ell}}{f_{\mathrm{sky,}2}}C_{\ell}^{XY, \mathrm{CI}},
\end{equation}
where $C_{\ell}^{XY, \mathrm{CI}}$ is the power spectrum of the C-inverse filtered maps.

The analytical full-sky expression is accurate up to the $1\,\%$ level, which is sufficient for large to intermediate scales, but biases the reconstructed lensing power spectrum at small angular scales. The mask-induced couplings are not considered in the analytical expression, for which an MC approach (named MCN0) is needed. We build the MCN0 estimator using QEs in which each of the input maps comes from a different simulation. In this case, the MF vanishes because the two maps are independent. We use all the 400 simulations to compute MCN0,
\begin{equation}
    \text{MC-}N_L^{(0), XYWZ} = \frac{1}{2f_{\rm{sky},4}}\left\langle C_L^{\hat{\phi}^{XY,i} \hat{\phi}^{WZ,i}} \right \rangle_{S}\,, \label{eq:mcn0}
\end{equation}
where
\begin{equation}
    \hat{\phi}_{LM}^{XY,i} = \hat{\phi}_{LM}^{X_iY_{i+1}} + \hat{\phi}_{LM}^{X_{i+1}Y_{i}},
\end{equation}
and $i$ is the simulation index and $S=\{1, 3, 5,\ldots, 399\}$. The factor $1/2$ comes from the fact that the QEs are built from two independent maps so each cross-power spectrum contains only half of the Gaussian noise \cite{P18:phi}. Since $C_L^{\hat{\phi}^{XY,i}\hat{\phi}^{WZ,i}}$ can be decomposed in the sum of four angular power spectra, twice the Gaussian noise is obtained.

\begin{figure}[t]
    \centering
    \includegraphics[width=1\textwidth]{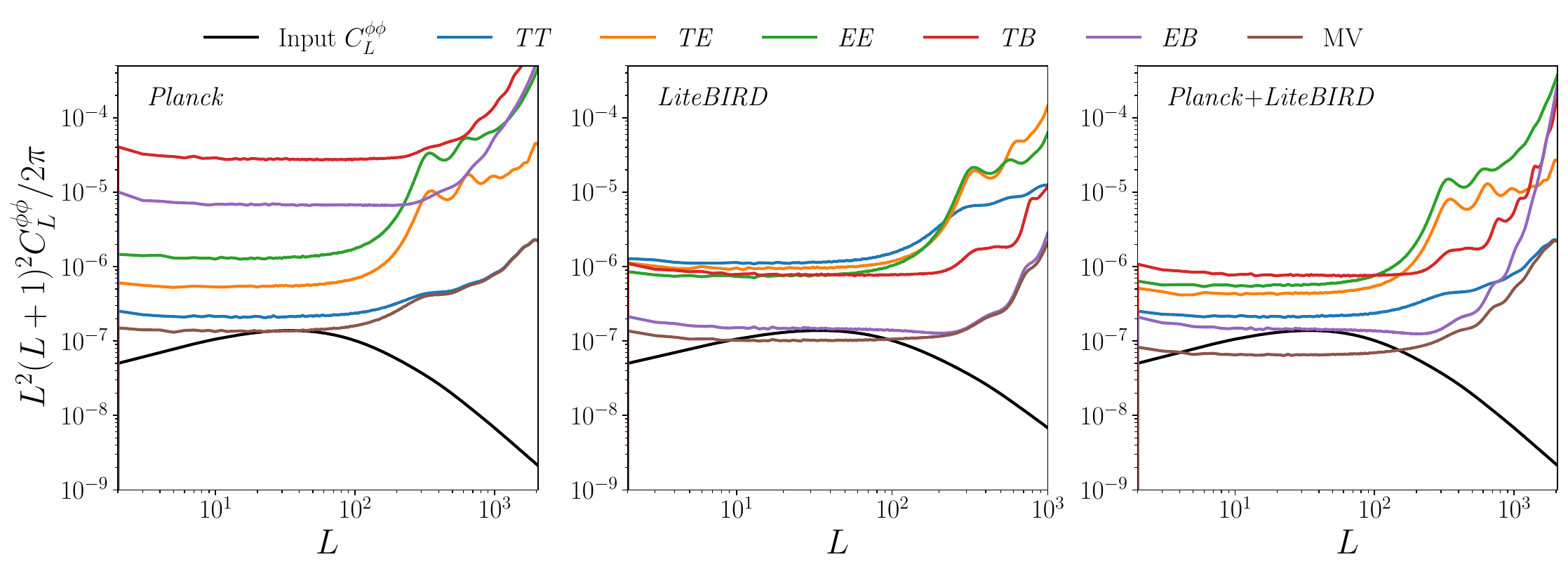}
    \caption{
    Lensing reconstruction MCN0 noise for \textit{Planck}, \textit{LiteBIRD} and \textit{Planck} + \textit{LiteBIRD} for the simple-foregrounds case. The MCN0 of the minimum variance (MV) estimator is also plotted. Black lines show the input lensing power spectrum used to generate the simulations.
    }
    \label{fig:MCN0_low_comp}
\end{figure}

Figure~\ref{fig:MCN0_low_comp} shows the MCN0 noise for the different QEs and the different experiments for our simple-foreground simulations. The $TT$ QE from \textit{Planck} and $EB$ from \textit{LiteBIRD} are the least noisy estimators for the combination of \textit{Planck} and \textit{LiteBIRD}. Additionally, using the MV estimator (further explained in section~\ref{sec: MV lensing}), \textit{LiteBIRD}'s N0 bias can be reduced compared to our previous analysis which was based on the $EB$ estimator only~\cite{Lonappan_lensing}. In conclusion, \textit{Planck} + \textit{LiteBIRD} MV will give the best full-sky lensing reconstruction.

\begin{figure}[t]
    \centering
    \includegraphics[width=1.0\textwidth]{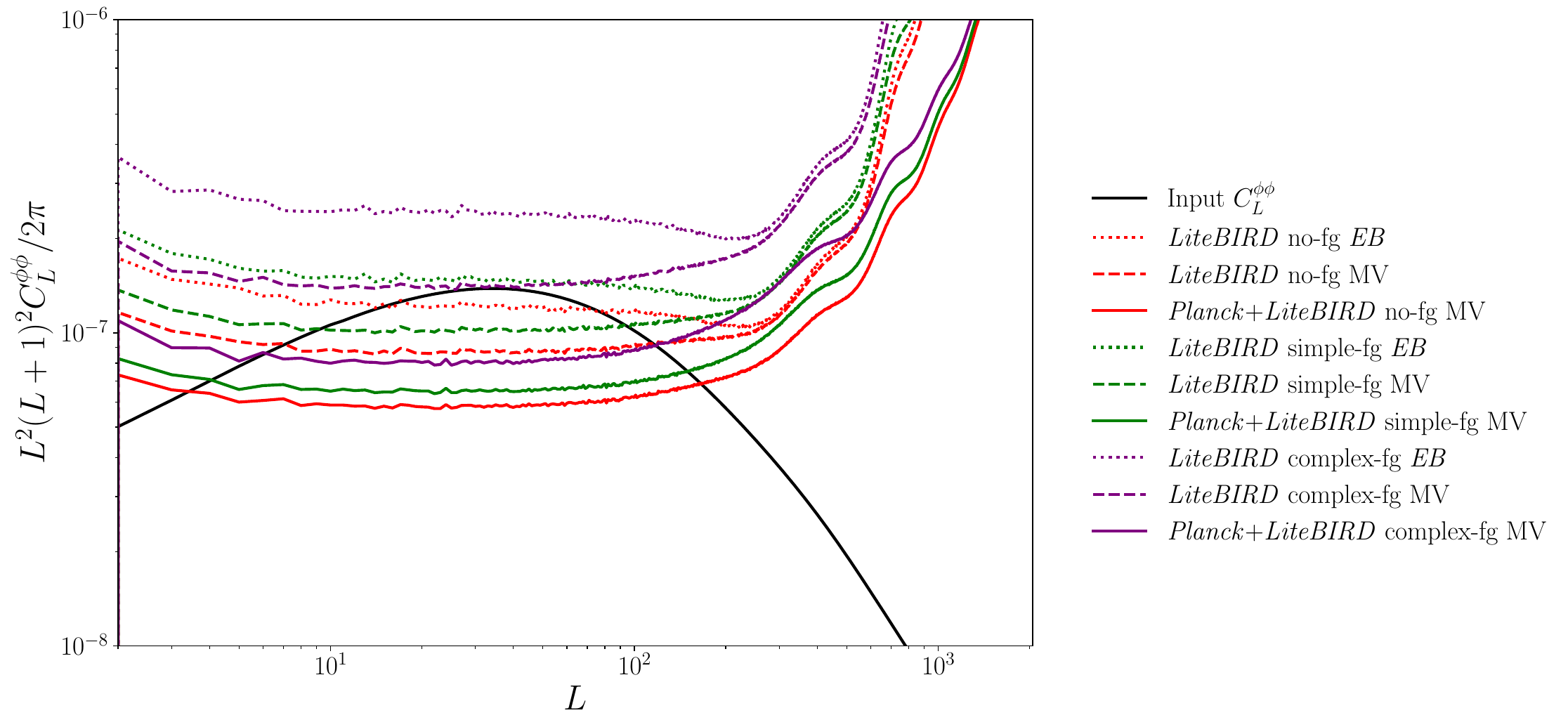}
    \caption{
    Comparison of the MCN0 noise for the \textit{LiteBIRD} $EB$ (dot-dashed lines), \textit{LiteBIRD} MV (dashed lined) and \textit{Planck} + \textit{LiteBIRD} MV (solid lines) lensing reconstructions from our no-foreground (in red), simple-foreground (in green), and complex-foreground (in purple) simulations. The black line shows the input lensing power spectrum used to generate the simulations.
    }
    \label{fig:comparison_MCN0}
\end{figure}

In figure~\ref{fig:comparison_MCN0}, we compare \textit{LiteBIRD} $EB$, \textit{LiteBIRD} MV and \textit{Planck} + \textit{LiteBIRD} MV for the different cases under analysis. Irrespective of the foregrounds considered, \textit{LiteBIRD} MV shows some improvement over the $EB$-only estimator, and the combination of \textit{Planck} and \textit{LiteBIRD} improves significantly the sensitivity that \textit{LiteBIRD} achieves alone. For all the cases studied, there is always a multipole range for which the \textit{Planck} + \textit{LiteBIRD} MCN0 MV is below the lensing signal.

The MCN0 estimator gives the correct estimation of the mean N0 bias, but it does not consider the variations of the N0 across simulations. When these variations are not considered, the covariance matrix of the lensing reconstruction exhibits correlations between off-diagonal multipoles. The deviations of the N0 bias with respect to the MCN0 are modeled using the realization-dependent N0 estimator (hereafter RDN0)~\cite{Hanson_2011}. In our previous \textit{LiteBIRD} lensing forecast~\cite{Lonappan_lensing}, an MC approach was used to estimate the N0 term accurately, however, due to its computational cost we decided to use another approach here. In this paper, we implement a mixed analytical- and simulation-based approach consisting of a first-order Taylor expansion of the N0 bias for simulation $i$ at the mean observed power spectra:
%
\begin{align}
N_{L,i}^{(0),XYWZ}(\bar{\boldsymbol{C}}_{\ell}^i)&\approx N_L^{(0),XYWZ}(\langle\bar{\boldsymbol{C}}_{\ell}\rangle)+\sum_{AB}\sum_{\ell'}\frac{\partial N_L^{(0),XYWZ}(\langle\bar{\boldsymbol{C}}_{\ell}\rangle)}{\partial \bar{C}_{\ell'}^{AB}}\left(\bar{C}_{\ell'}^{AB, i}-\langle \bar{C}_{\ell'}^{AB} \rangle\right)\nonumber\\
&=MC\text{-}N_L^{(0), XYWZ} + RD\text{-}N_{L,i}^{(0), XYWZ}, 
\end{align}
%
where $\bar{\boldsymbol{C}}_{\ell}^i$ is a vector containing the observed power spectra at simulation $i$, and $\langle\bar{\boldsymbol{C}}_{\ell}\rangle$ is a vector containing the average observed power spectra over the 400 simulations. The equation verifies that $\langle N_{L,i}^{(0),XYWZ}(\bar{\boldsymbol{C}}_{\ell}^i)\rangle =MC\text{-}N_L^{(0), XYWZ}$ and $\langle RD\text{-}N_{L,i}^{(0), XYWZ}\rangle =0$, which is the expected behavior for the small RDN0 perturbations over the mean MCN0. For calculating the RDN0 term, we use the analytical N0 expression from eq.~\eqref{Analytical N0 bias} because it is accurate enough for this correction~\cite{Peloton:2016:RDN0, ACT_DR6_lensing}:
\begin{equation}
\begin{aligned}
        RD\text{-}N_{L,i}^{(0), XYWZ} =&
        \frac{A_L^{XY}A_L^{WZ}}{2L+1}
\sum_{\ell_1\ell_2} g^{XY\ast}_{\ell_1\ell_2}(L)\Big[ \left(\bar{C}_{\ell_1}^{XW,i}\langle\bar{C}_{\ell_2}^{YZ}\rangle+\langle\bar{C}_{\ell_1}^{XW}\rangle\bar{C}_{\ell_2}^{YZ,i}\right)
g^{WZ}_{\ell_1\ell_2}(L)\\
&+(-1)^{L+\ell_1+\ell_2}\left(\bar{C}_{\ell_1}^{XZ,i}\langle\bar{C}_{\ell_2}^{YW}\rangle + \langle\bar{C}_{\ell_1}^{XZ}\rangle\bar{C}_{\ell_2}^{YW,i}\right) g^{WZ}_{\ell_2\ell_1}(L) \Big]\\
&- 2AN\text{-}N_L^{(0),XYWZ}.
\end{aligned}
\end{equation}
We use \texttt{lensQUEST} to compute the N0 bias, and modified the code to implement our calculation of the RDN0 bias.

\subsubsection{N1 bias}

\begin{figure}[t]
    \centering
    \includegraphics[width=0.7\textwidth]{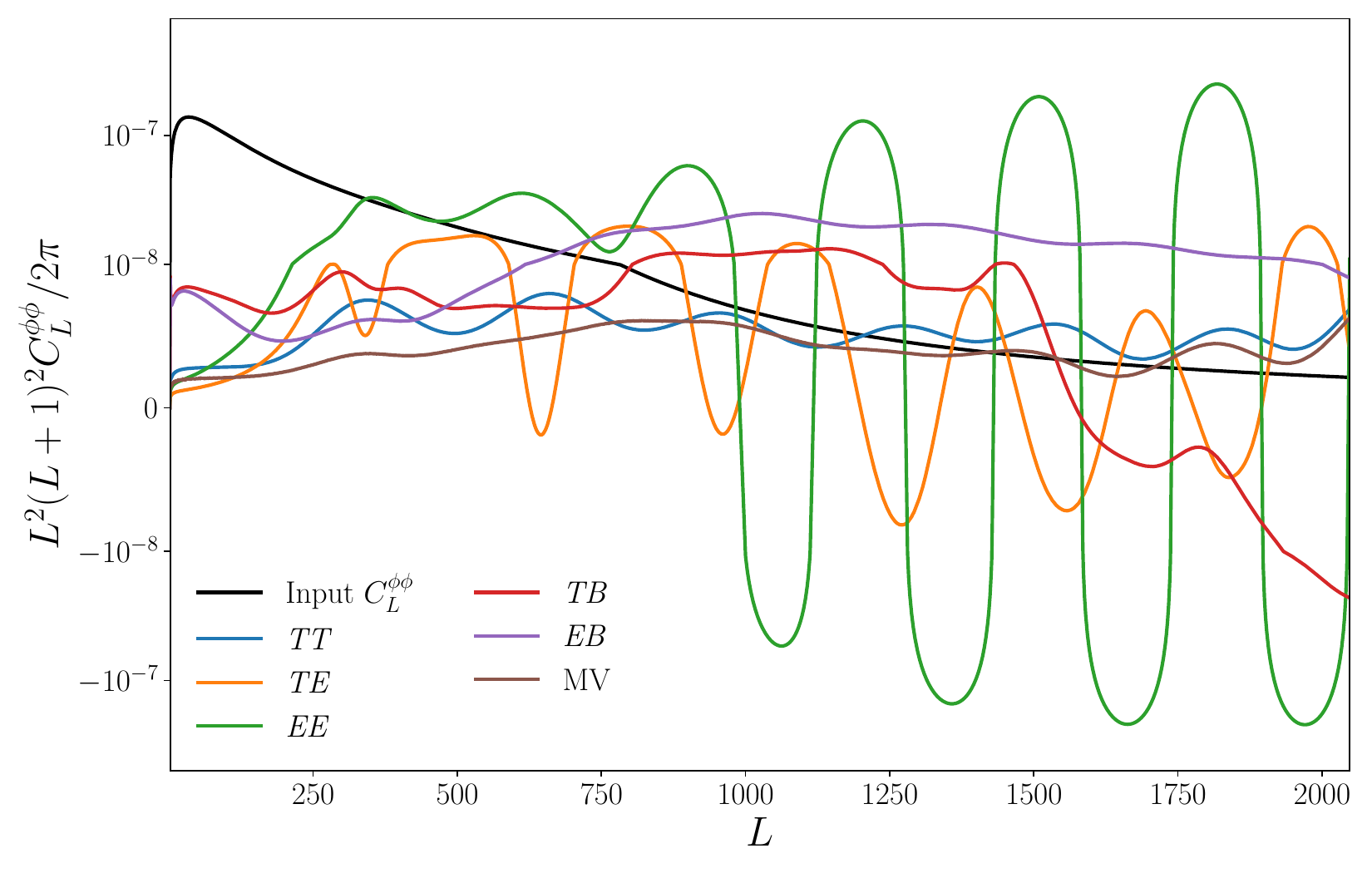}
    \caption{
    Analytical N1 bias for the \textit{Planck} + \textit{LiteBIRD} lensing reconstruction for our simple-foreground simulations. The minimum-variance (MV) N1 is also plotted. The black line shows the input lensing power spectrum used to generate the simulations.
    }
    \label{fig:N1_Planck_LiteBIRD}
\end{figure}

We use a modified version of \texttt{LensingBiases}\footnote{\url{https://github.com/JulienPeloton/lensingbiases}} to calculate the N1 noise in the flat-sky approximation. The N1 noise is important at small scales, $\ell>100$, for which the flat-sky approximation is sufficiently accurate. The analytical flat-sky  N1 estimator is
%
\begin{equation}
\begin{aligned}
    AN\text{-}N_L^{(1),XYWZ}=&A_L^{XY}A_L^{WZ}\int\frac{d^2\boldsymbol{\ell}_1}{(2\pi^2)}\int\frac{d^2\boldsymbol{\ell'}_1}{(2\pi^2)}G^{XY}_{{\boldsymbol{\ell}_1\boldsymbol{\ell}_2}}G^{WZ}_{{\boldsymbol{\ell'}_1\boldsymbol{\ell'}_2}}\\
   & \times\left[C_{|\boldsymbol{\ell}_1-\boldsymbol{\ell'}_1|}^{\phi\phi}f^{XW}_{-\boldsymbol{\ell}_1\boldsymbol{\ell'}_1}f^{YZ}_{-\boldsymbol{\ell}_2\boldsymbol{\ell'}_2} + C_{|\boldsymbol{\ell}_1-\boldsymbol{\ell'}_2|}^{\phi\phi}f^{XZ}_{-\boldsymbol{\ell}_1\boldsymbol{\ell'}_2}f^{YW}_{-\boldsymbol{\ell}_2\boldsymbol{\ell'}_1}\right],
\end{aligned}
\end{equation}
where $\boldsymbol{\ell}_1+\boldsymbol{\ell}_2=\boldsymbol{\ell'}_1+\boldsymbol{\ell'}_2=\boldsymbol{L}$, $A_L^{XY}$ corresponds to the full-sky normalization factor from eq.~\eqref{QE norm}, $f^{XY}_{\boldsymbol{\ell}\boldsymbol{\ell'}}$ are the flat-sky lensing weight functions that can be found in table~1 of ref.~\cite{Flat_sky_QE_2002}, and the optimal QE weights are given by\footnote{For the $TE$ N1, eq.~(\ref{eq: G weights}) is an approximation to the exact definition of the weights that is implemented in \texttt{LensingBiases} (see ref.~\cite{Flat_sky_QE_2002} for the exact expression). The differences between the exact expression and the approximation in eq.~(\ref{eq: G weights}) are negligible for the N1 computation.}
\begin{equation}\label{eq: G weights}
    G^{XY}_{{\boldsymbol{\ell}\boldsymbol{\ell'}}} = \frac{h^{XY}f^{XY}_{\boldsymbol{\ell}\boldsymbol{\ell'}}}{\hat{C}^{XX}_{\ell}\hat{C}^{YY}_{\ell'}}.
\end{equation}
This expression was derived in ref.~\cite{Kesden_N1}, and applied to the \textit{Planck} 2015 lensing analysis in ref.~\cite{P15:phi}.

Figure~\ref{fig:N1_Planck_LiteBIRD} shows the N1 bias for the combination of \textit{Planck} and \textit{LiteBIRD} in the simple-foreground simulations. The N1 noise is much lower than the N0 bias and only biases the lensing power spectrum at high $\ell$ ($\ell > 500$ for the $TT$, $EB$ and MV QE). Taking into account the N1 bias is required in order to recover an unbiased reconstruction of the lensing power spectrum at small scales.

\subsubsection{Monte Carlo normalization correction}

We estimate the MC correction to the QE normalization using the MV estimation of the lensing potential (see details in section~\ref{sec: MV lensing}). We do that because we observe the MC correction to be similar between the different QEs, and the MV estimate leads to the least noisy estimation of $R^{\rm MC}_L$. The MC normalization correction is calculated as 
\begin{equation}
    R^{\rm MC}_L = \frac{f_{\mathrm{sky}, 2}}{f_{\mathrm{sky}, 4}}\left\langle\frac{C_L^{\phi^{\rm in}\hat{\phi}^{\rm MV}}}{C_L^{\phi^{\rm in}\phi^{\rm in}}}\right\rangle,
\end{equation}
where $\phi^{\rm in}$ is the input lensing map masked with the same apodized mask used for the CMB maps, and $\hat{\phi}^{\rm MV}$ is the MV lensing reconstruction with the MF subtracted. Instead of directly masking the lensing potential $\phi$, we mask the lensing convergence $\kappa$\footnote{The lensing convergence is defined as $\kappa=\nabla^2\phi/2$, or $\kappa_{LM}=L(L+1)\phi_{LM}/2$ in harmonic space.}, because it has a much whiter power spectrum, especially on large angular scales. By masking $\kappa$ we significantly reduce the mask-induced leakage~\cite{P15:phi}. The $f_{{\rm sky},n}$ corrections take into account the sky coverage of the reconstructed power spectrum from the QE and of the masked input lensing maps.

\begin{figure}[t]
    \centering
    \includegraphics[width=1\textwidth]{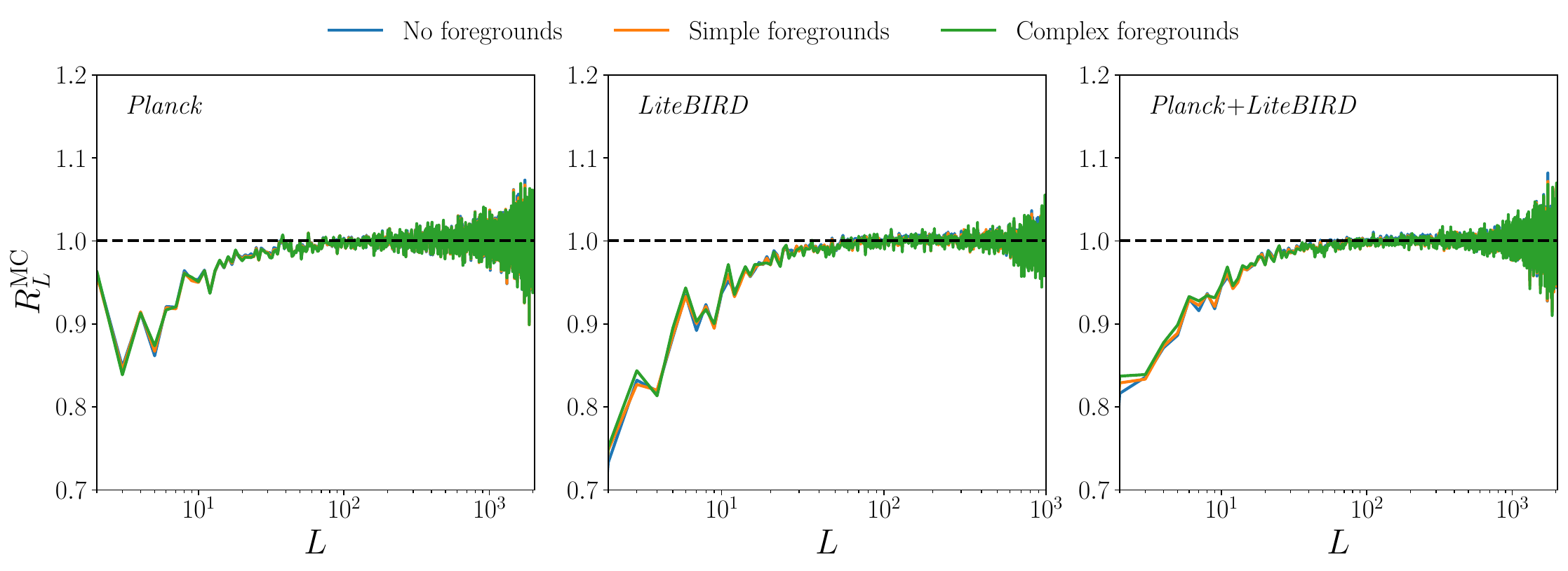}
    \caption{
    Minimum-variance Monte Carlo normalization correction to the analytical normalization, $R_L^{\rm MC}$, for the \textit{Planck}, \textit{LiteBIRD} and \textit{Planck} + \textit{LiteBIRD} experiments and our three different simulation sets. Dashed black lines correspond to the ideal situation in which no correction is needed, $R_L^{\rm MC}=1$.
    }
    \label{fig:R_MC_MV_all}
\end{figure}

Figure~\ref{fig:R_MC_MV_all} shows the MV MC normalization correction for all the experiments and types of simulations considered. We see that the MC correction does not change when the foreground complexity is varied and that a similar behaviour is observed for the three experiments. The correction is confined to the low-$\ell$ region, mainly affecting the large scales. The intermediate and small scales are not significantly affected by the MC normalization correction.

\subsection{Minimum-variance lensing estimator}\label{sec: MV lensing}

In this section, we combine the different QEs $\hat{\phi}_{LM}^{XY}$, with $XY\in\{TT,EE,TE,TB,EB\}$, to minimize the reconstruction noise.\footnote{The optimal way to combine QEs is through the generalized minimum-variance (GMV) estimator \cite{GMV_2021}. We performed an analytical comparison of the N0 biases obtained by the MV and the GMV using \texttt{GlobalLensQuest} (\url{https://github.com/abhimaniyar/GlobalLensQuest}) and found that the improvement is around $1\,\sigma$ for the \textit{Planck}+\textit{LiteBIRD} combination (details about the SNR calculation can be found in section \ref{sec: SNR}). Finding such a small improvement, we decided to stick to the MV formalism.} The objective is to obtain the QE of minimum variance, $\hat{\phi}_{LM}^{\rm MV}$, expressed as the weighted sum of the QEs from the different channels:
\begin{equation}\label{weightedSum}
\hat{\phi}_{\ell m}^{\rm MV}=\sum_{XY} w_L^{XY}\hat{\phi}_{LM}^{XY}.
\end{equation} 

Calculating the weights $w_L^{XY}$ to build the MV estimator is a quadratic optimization problem. It can be solved using the Lagrange multiplier theorem under the constraint that the sum of the weights is equal to one. The optimal values of the $\ell$-dependent weights can be derived using~\cite{Okamoto_Hu_QE}:
\begin{equation}\label{MVOkamotoHu}
w_L^{XY}=\frac{\sum_{WZ}(N_L^{-1})^{XYWZ}}{\sum_{XY}\sum_{WZ}(N_L^{-1})^{XYWZ}},
\end{equation}
where 
\begin{equation}\label{biases MV}
    N_L^{XYWZ} = MC\text{-}N_{L}^{(0),XYWZ}+N_L^{(1),XYWZ}.
\end{equation}

\begin{figure}[t]
    \centering
    \includegraphics[width=1\textwidth]{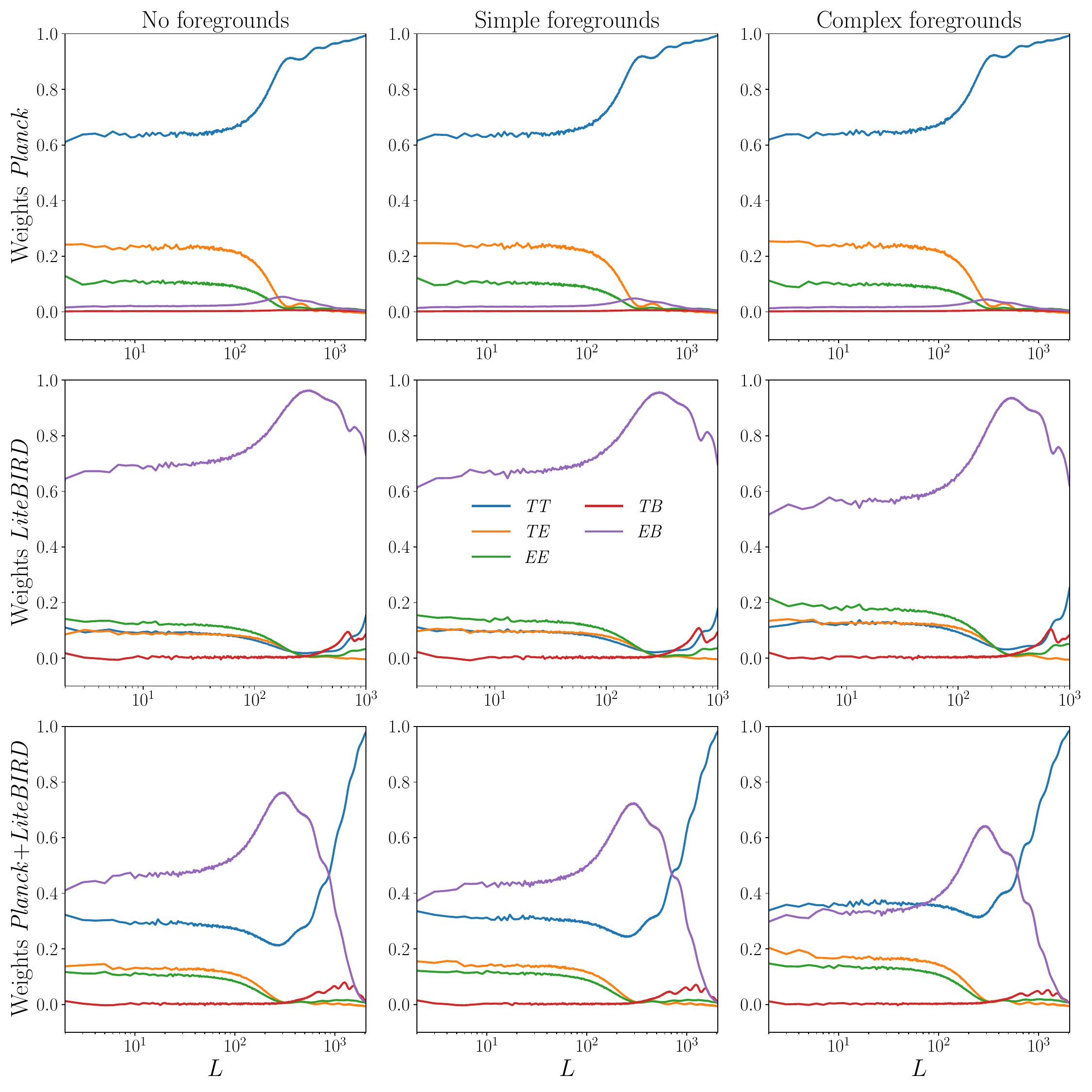}
    \caption{
    Weights used to calculate the minimum-variance (MV) estimator for all the experiments and simulations considered.
    }
    \label{fig:Weights_MV_all}
\end{figure} 

The weights of the MV estimator are calculated using the average biases, which do not vary with the simulation index. For that reason, the RDN0 bias, which averages to zero, is not included in the covariance matrix $N_L^{XYWZ}$. Although the MF is subtracted in section~\ref{sec: qe recons}, some residual MF is observed at very low multipoles. For instance, in ref.~\cite{P18:phi} the multipoles $L<8$ had to be excluded due to high sensitivity to the MF. In contrast, \textit{LiteBIRD}'s polarization-dominated reconstruction has a negligible MF compared to the lensing signal. Hence, by incorporating this into the weighting scheme, an unbiased recovery of the lensing signal at multipoles $L<8$ could be achieved in the future for the first time, with a minimal loss in the SNR.

To debias the MV estimator, we need to take into account the RDN0 term in the matrix $N_{L,i}^{XYWZ}$, 
\begin{equation}
    N_{L,i}^{XYWZ} = MC\text{-}N_{L}^{(0),XYWZ}+RD\text{-}N_{L,i}^{(0),XYWZ}+N_L^{(1),XYWZ},
\end{equation}
leading to the following expression for the debiased MV lensing power spectrum:
\begin{equation}\label{cl_pp_MV_debiased}
\hat{C}_{L,i}^{\phi^{\rm MV}\phi^{\rm MV}}=\frac{1}{R_{L}^{\rm MC}}\left (\frac{1}{f_{\rm sky,4}}C_L^{\hat{\phi}_i^{\rm MV} \hat{\phi}_i^{\rm MV}}-\sum_{XY}\sum_{WZ}w^{XY}_Lw^{WZ}_LN_{L,i}^{XYWZ}\right).
\end{equation}

Figure~\ref{fig:Weights_MV_all} shows the MV weights for all the experiments and simulations considered. For \textit{Planck}, the $TT$ estimator dominates, whereas for \textit{LiteBIRD} the $EB$ estimator prevails. Finally, for the combination of \textit{Planck} and \textit{LiteBIRD}, $EB$ dominates over $TT$ at large and intermediate scales, and $TT$ dominates over $EB$ for small scales. When increasing the foreground complexity, the $TT$ estimator coming from \textit{Planck} gains importance with respect \textit{LiteBIRD}'s $EB$, due to the increase of the foregrounds complexity in polarization.

\section{Results} \label{sec:results}

In this section, we present our results in terms of the lensing maps (section~\ref{sec: lensing maps}), lensing band powers (section~\ref{sec: band powers}), and the overall signal-to-noise of the reconstruction that we achieve (section~\ref{sec: SNR}). We show results for all simulation complexities and \textit{Planck}, \textit{LiteBIRD}, and \textit{Planck}+\textit{LiteBIRD} analyses, showcasing that in all situations the combination of \textit{Planck} and \textit{LiteBIRD} improves significantly over \textit{LiteBIRD} alone.

\subsection{Lensing maps}\label{sec: lensing maps}

Here we address the quality of the lensing reconstruction by inspecting the recovered maps for the different QEs and experiments. All the reconstructed maps shown in this section are filtered using the WF presented in section~\ref{sec: filtering}. Additionally, we plot the lensing-deflection field\footnote{The lensing deflection is defined as $\mathbf{\alpha}=\nabla\phi$, or $\alpha_{LM}=\sqrt{L(L+1)}\phi_{LM}$ in harmonic space.} instead of the lensing potential. The WF lensing-deflection map, $\hat{\alpha}_{LM}^{XY,\mathrm{WF}}$, is obtained from the lensing reconstruction as,
\begin{equation}
    \hat{\alpha}_{LM}^{XY,\mathrm{WF}} = \sqrt{L(L+1)}\frac{C_L^{\phi\phi, \mathrm{fid}}}{C_L^{\phi\phi, \mathrm{fid}} + N_L^{XYXY}}\hat{\phi}_{LM}^{XY},
\end{equation}
where $C_L^{\phi\phi, \mathrm{fid}}$ is the input lensing potential and $N_L^{XYXY}$ is the reconstruction noise from eq.~\eqref{biases MV} with $XY\in\{TT, TE, TB, EE, EB, \rm{MV}\}$.

\begin{figure}[t]
    \centering
    \begin{minipage}[t]{\textwidth}
        \centering
        \includegraphics[width=0.8\textwidth]{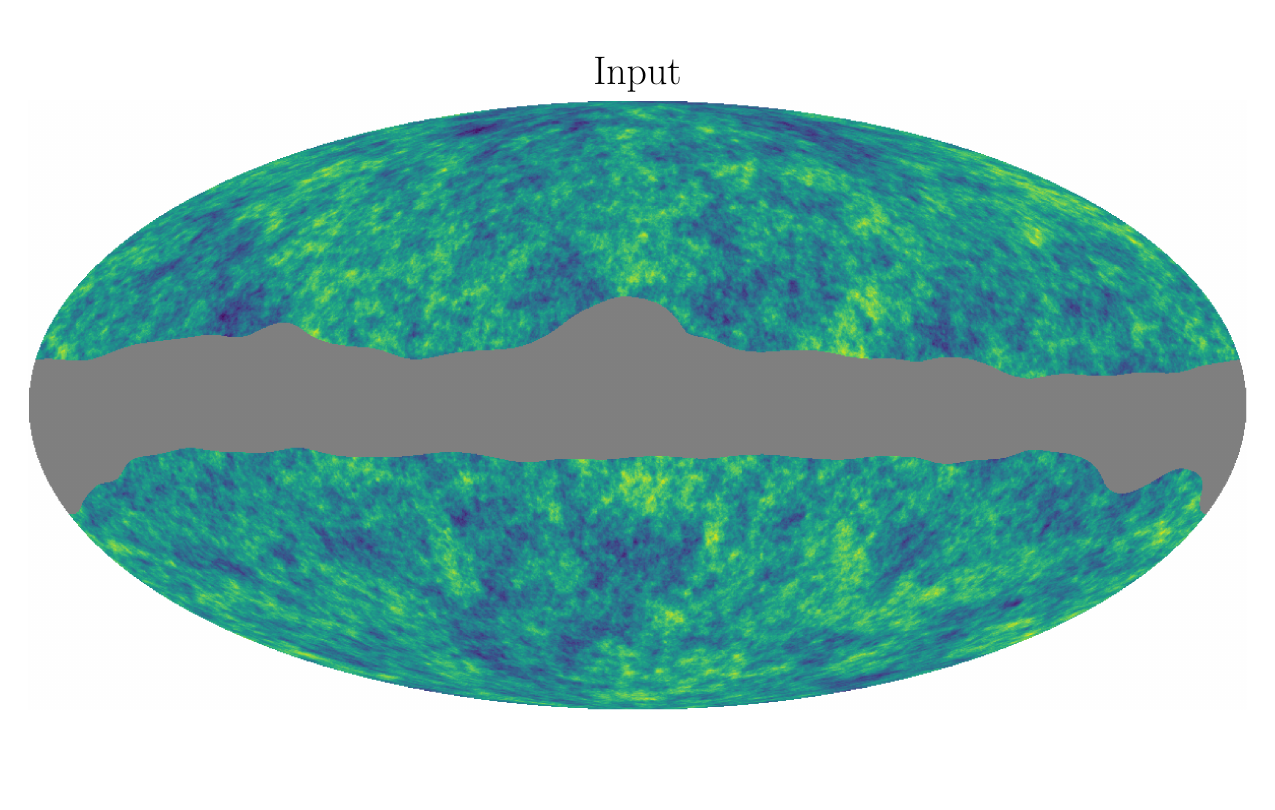} 
    \end{minipage}

    \vspace{-0.25cm} 

    \begin{minipage}[t]{\textwidth}
        \centering
        \includegraphics[width=0.8\textwidth]{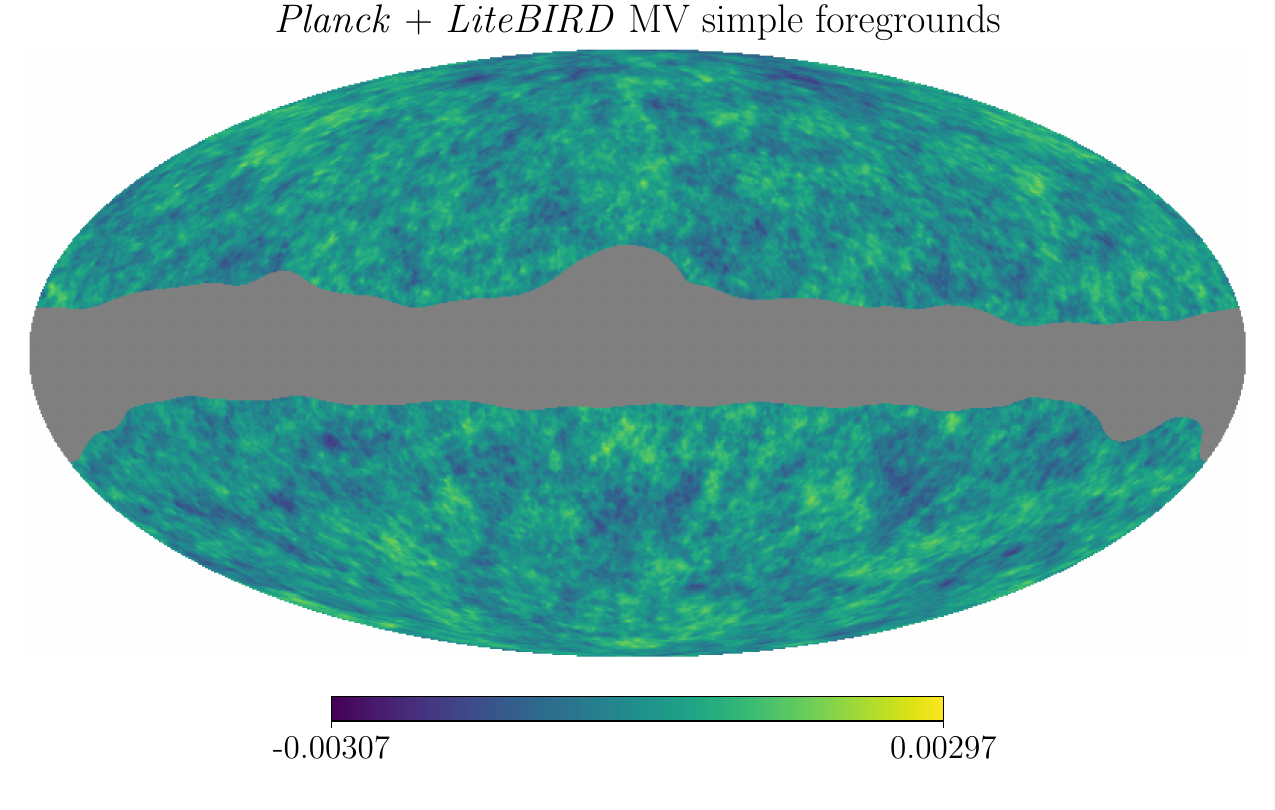}
    \end{minipage}
    \caption{
    Example of one of the 400 simple-foregrounds simulations of the lensing-deflection map, $\hat{\alpha}_{LM}^{\mathrm{MV,WF}} = \sqrt{L(L+1)}\hat{\phi}_{LM}^{\mathrm{MV,WF}}$, corresponding to the gradient mode or $E$ mode of the lensing deflection angle.  The upper panel shows the input lensing-deflection map, whereas the lower panel shows the Wiener-filtered lensing-deflection coming from the \textit{Planck} + \textit{LiteBIRD} MV reconstruction. The gray regions represent the apodized $80\,\%$ \textit{Planck} Galactic mask. For ease of visualization, the mask has been binarized after apodization to clearly delineate excluded and included regions.}
    \label{fig:full_sky_lens_recons}
\end{figure}

In the bottom panel of figure~\ref{fig:full_sky_lens_recons}, we show the WF MV lensing-deflection map from the combination of \textit{Planck} and \textit{LiteBIRD}, corresponding to the simple-foregrounds case. In comparison with the input lensing-deflection map shown in the top panel, we find very good agreement between the two maps.

\begin{figure}[t]
    \centering
    \includegraphics[width=0.9\textwidth]{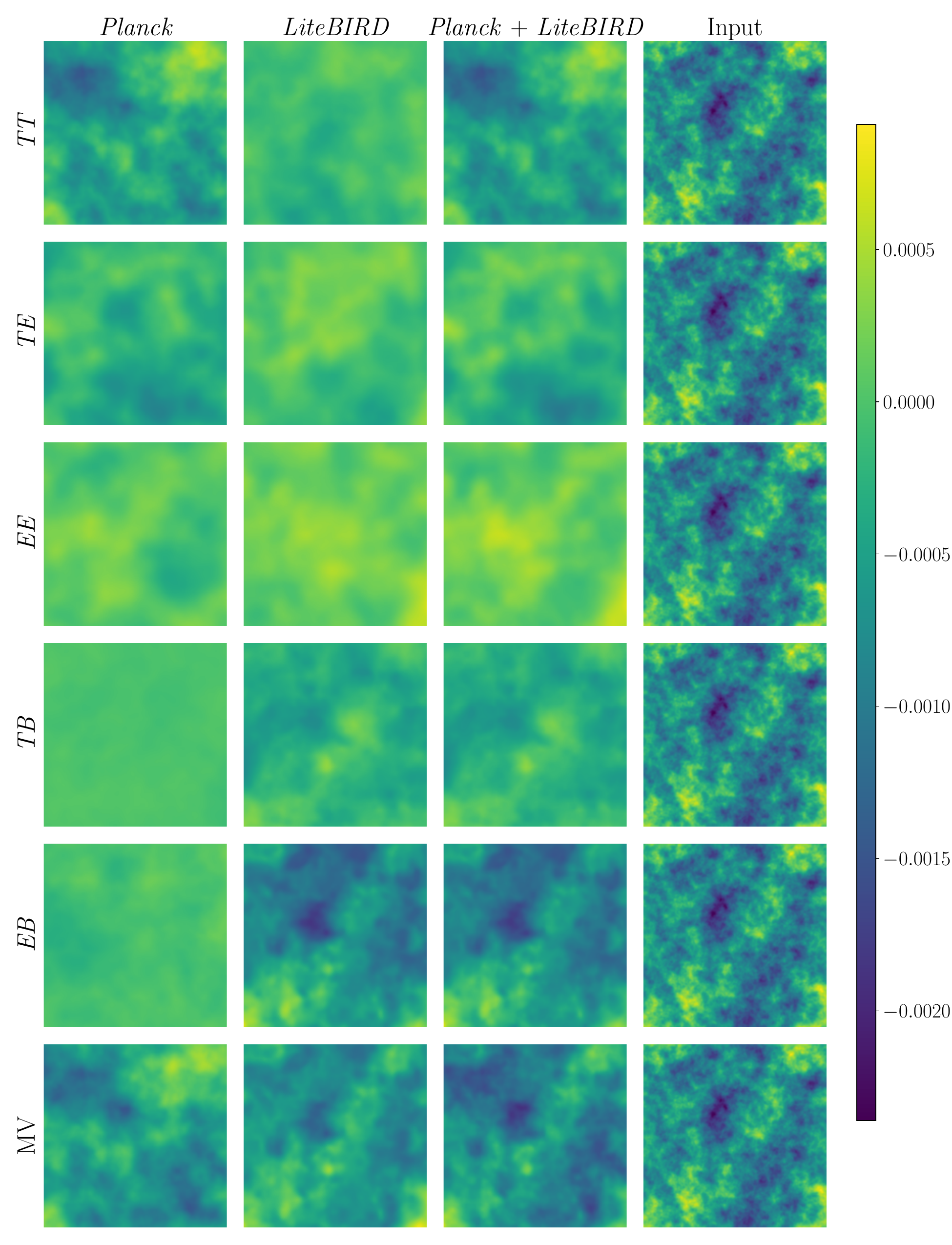}
    \caption{
    $12.5^\circ \times 12.5^\circ$ gnomonic projection of the lensing-deflection maps, $\hat{\alpha}_{LM}^{\mathrm{MV,WF}} = \sqrt{L(L+1)}\hat{\phi}_{LM}^{\mathrm{MV,WF}}$, corresponding to the gradient mode or $E$ mode of the lensing deflection angle. They are reconstructed from simple-foregrounds simulations and centered at Galactic coordinates $(l, b) = (270^\circ, -35^\circ)$. All the QEs for all the experiments are plotted alongside the true input lensing-deflection map. 
    }
    \label{fig:Lensing_patches_comparison}
\end{figure}

Figure~\ref{fig:Lensing_patches_comparison} shows the same sky patch for the different QEs and experiments considered in the case of simple foregrounds. From this plot, one can see which estimators contribute the most and at which angular scales to the final \textit{LiteBIRD}, \textit{Planck}, and their combined lensing estimate. It is particularly interesting to see how combining the different QEs to form the MV estimator leads to a clearly better result.  Focusing on \textit{Planck}, we can conclude that the main contributions to the MV estimator come from $TT$, $EE$, and $TE$. $TT$ is the estimator that contributes the most to the final shape of the MV map. This is consistent with the weights shown in figure~\ref{fig:Weights_MV_all}. Continuing with \textit{LiteBIRD}, two major differences with \textit{Planck} can be observed. First, \textit{LiteBIRD}’s MV reconstruction is better than \textit{Planck}’s, as the former is closer to the input lensing deflection. Secondly, no huge differences are observed between the $EB$ and MV maps, which agrees with $EB$ being the estimator with the greatest contribution; $TB$ is the second largest contributor to the MV map. 

Finally, let us focus on the reconstructed maps from the combination of \textit{Planck} and \textit{LiteBIRD}. As explained in section~\ref{sec: MV lensing} and shown in figure~\ref{fig:Lensing_patches_comparison}, \textit{Planck}'s signal comes mostly from the $TT$ estimator and \textit{LiteBIRD}'s most important contribution is the $EB$ estimator. This is what we see for the combination of both experiments in figure~\ref{fig:Lensing_patches_comparison}: the $TT$ map is equivalent to \textit{Planck}’s and the $EB$ map is equivalent to \textit{LiteBIRD}’s. It is also interesting to see that different structures present in the input map at different angular scales are observed in the $TT$ and $EB$ maps, with both contributing to the MV map. The MV map reconstructed from the combination of both experiments will be the best full-sky lensing-deflection measurement in the near future.

\subsection{Lensing band powers} \label{sec: band powers}

In this section, we show the lensing band powers recovered for the simple-foregrounds case and all the experiments and QE combinations. Finally, we also compare the performance of \textit{LiteBIRD} alone versus the combination of \textit{Planck} and \textit{LiteBIRD} for the $EB$ and MV estimators.

\begin{figure}[t]
    \centering
    \includegraphics[width=1\textwidth]{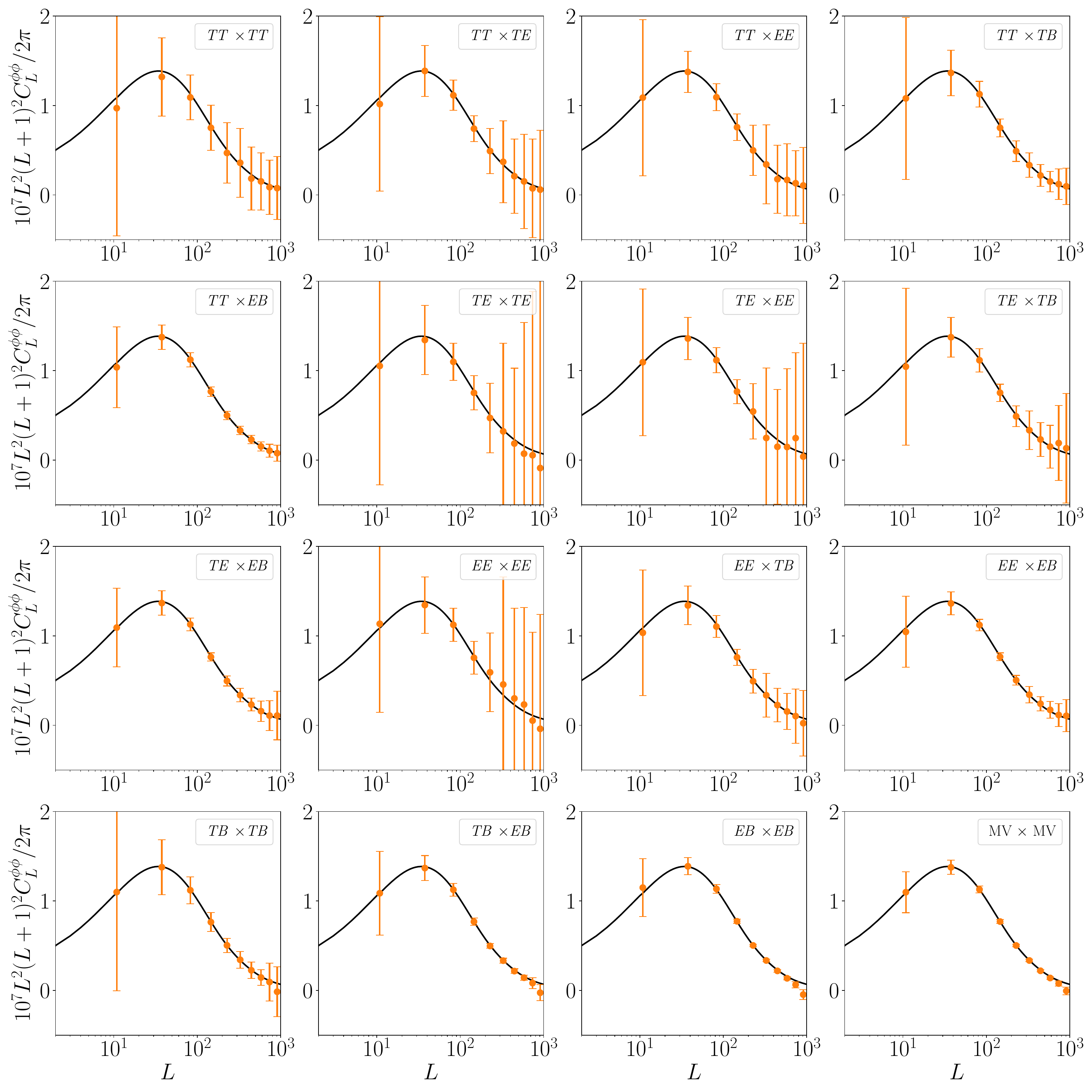}
    \caption{
    Lensing band powers for \textit{LiteBIRD}'s simple-foregrounds simulations. All the auto- and cross-spectra of the five QEs, and the minimum-variance (MV) estimator, are plotted. The orange points correspond to the average over all the simulations and the error bars correspond to the $1\,\sigma$ confidence intervals. The black solid line shows the input lensing power spectrum.
    }
    \label{fig:qcl_debiased_LiteBIRD_simple_foregrounds}
\end{figure} 

First, we define our binning configuration. The edges of the bins, $E_b$, are given by:
\begin{equation}
    E_b = \left (\sqrt{\ell_{\rm min}}+\frac{\sqrt{\ell_{\rm max}}-\sqrt{\ell_{\rm min}}}{N_\mathrm{bins}}\, b\right )^2,\ b\in \{0, 1,\ldots, N_\mathrm{bins}\},
\end{equation}
where $\ell_{\rm min}=2$, $\ell_{\rm max}=1000$ for \textit{LiteBIRD} and $\ell_{\rm max}=2048$ for \textit{Planck} and \textit{Planck} + \textit{LiteBIRD}, $b$ is the bin index, and $N_\mathrm{bins}$ is the number of bins. We decided to use $N_\mathrm{bins}=10$ bins for \textit{LiteBIRD}, and $N_\mathrm{bins}=14$ for \textit{Planck} and \textit{Planck} + \textit{LiteBIRD}. The bin centers are then
\begin{equation}
    L_b = \frac{E_{b-1}+E_{b}}{2}, \ b\in \{1,2,\ldots, N_\mathrm{bins}\}.
\end{equation}

Uniform binning is performed within the bin to obtain the binned power spectra $\hat{C}_{b,i}^{\phi^{XY}\phi^{WZ}}$ and $C_{b}^{\phi\phi, \mathrm{fid}}$. Similar to ref.~\cite{P18:phi}, the binning function is chosen to have unit response to the fiducial power spectrum,
\begin{equation} \label{eq band powers}
\hat{C}_{L_b,i}^{\phi^{XY}\phi^{WZ}}=\hat{A}_{b,i}^{\phi^{XY}\phi^{WZ}}C_{L_b}^{\phi\phi, \mathrm{fid}} = \frac{\hat{C}_{b,i}^{\phi^{XY}\phi^{WZ}}}{C_{b}^{\phi\phi, \mathrm{fid}}}C_{L_b}^{\phi\phi, \mathrm{fid}}\,,
\end{equation}
where $\hat{C}_{L_b,i}^{\phi^{XY}\phi^{WZ}}$ is the binned lensing power spectrum at $L=L_b$, and $\hat{A}_{b,i}^{\phi^{XY}\phi^{WZ}}$ is the lensing amplitude with respect to the input lensing power spectrum in the band power $b$ and the simulation $i$ ($\hat{A}_{b,i}^{\phi^{XY}\phi^{WZ}}=1$ for $\hat{C}_{b,i}^{\phi^{XY}\phi^{WZ}}=C_{b}^{\phi\phi, \mathrm{fid}}$). Finally, $C_b^{\phi\phi,\rm{fid}}$ and $C_{L_b}^{\phi\phi,\rm{fid}}$ are the input lensing power spectrum at bin $b$ and at multipole $L_b$, respectively.

\begin{figure}[t]
    \centering
    \includegraphics[width=1\textwidth]{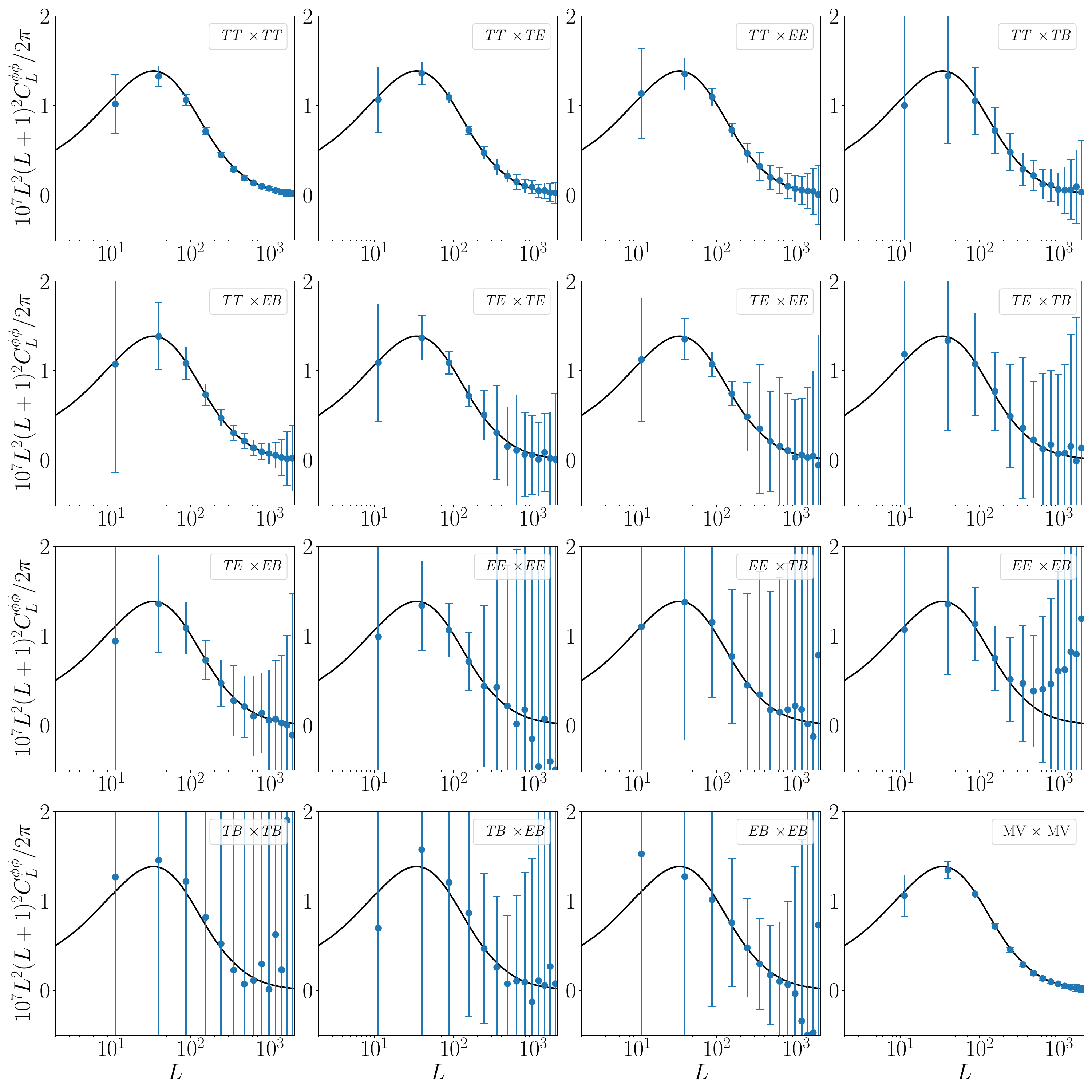}
    \caption{
    Lensing band powers for \textit{Planck}'s simple-foreground simulations. All the auto- and cross-spectra of the five QEs, and the minimum-variance (MV) estimator, are plotted. The blue points correspond to the average over all the simulations and the error bars correspond to the $1\,\sigma$ confidence intervals. The black solid line shows the input lensing power spectrum.
    }
    \label{fig:qcl_debiased_Planck_simple_foregrounds}
\end{figure} 

\begin{figure}[t]
    \centering
    \includegraphics[width=1\textwidth]{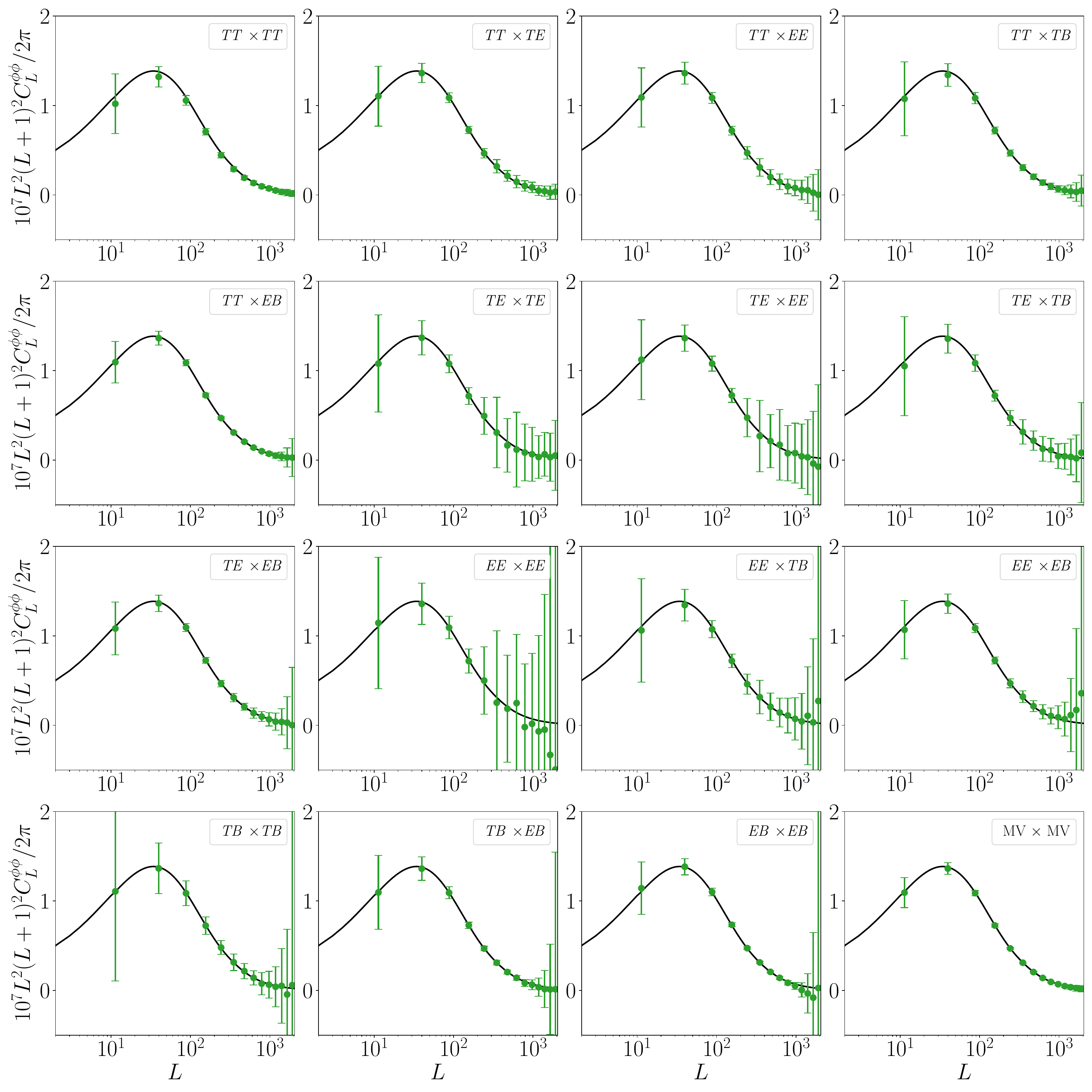}
    \caption{
    Lensing band powers for the combination of \textit{Planck} and \textit{LiteBIRD} in simple-foreground simulations. All the auto- and cross-spectra of the five QEs, and the minimum-variance (MV) estimator, are plotted. The green points correspond to the average over all the simulations and the error bars correspond to the $1\,\sigma$ confidence intervals. The black solid line shows the input lensing power spectrum.
    }
    \label{fig:qcl_debiased_Planck_LiteBIRD_simple_foregrounds}
\end{figure}

Figures \ref{fig:qcl_debiased_LiteBIRD_simple_foregrounds}, \ref{fig:qcl_debiased_Planck_simple_foregrounds}, and \ref{fig:qcl_debiased_Planck_LiteBIRD_simple_foregrounds}, show the band powers for the simple-foreground case for \textit{LiteBIRD}, \textit{Planck}, and the combination of \textit{Planck} and \textit{LiteBIRD}, respectively. For \textit{LiteBIRD}, we show that we achieve an unbiased reconstruction for all the different QEs, except $EB$. Only for the $EB$ estimator and at very small scales do we observe $>1\,\sigma$ deviations. Moreover, such biases do not impact either the SNR or the precision of cosmological parameter estimation. For that reason, we decide to leave the further investigation of this small bias to future work. The estimators with the smallest error bars are $TB$ and $EB$ but, since they are extremely correlated, $TB$ does not contribute much to \textit{LiteBIRD}'s MV estimator. For \textit{Planck}, the $TT$ and the $TE$ estimator are the best QEs measured. When compared with \textit{LiteBIRD}, we observe that \textit{LiteBIRD} has a better estimate of the large-scale lensing power spectrum than \textit{Planck}. Consequently, with the combination of \textit{Planck} and \textit{LiteBIRD} we get the best of both worlds: good measurements of the lensing potential through $TT$ thanks to \textit{Planck}, and through $EB$ thanks to \textit{LiteBIRD}. This is accompanied by an overall reduction of the error bars of all QEs.

\begin{figure}[t]
    \centering
    \includegraphics[width=1\textwidth]{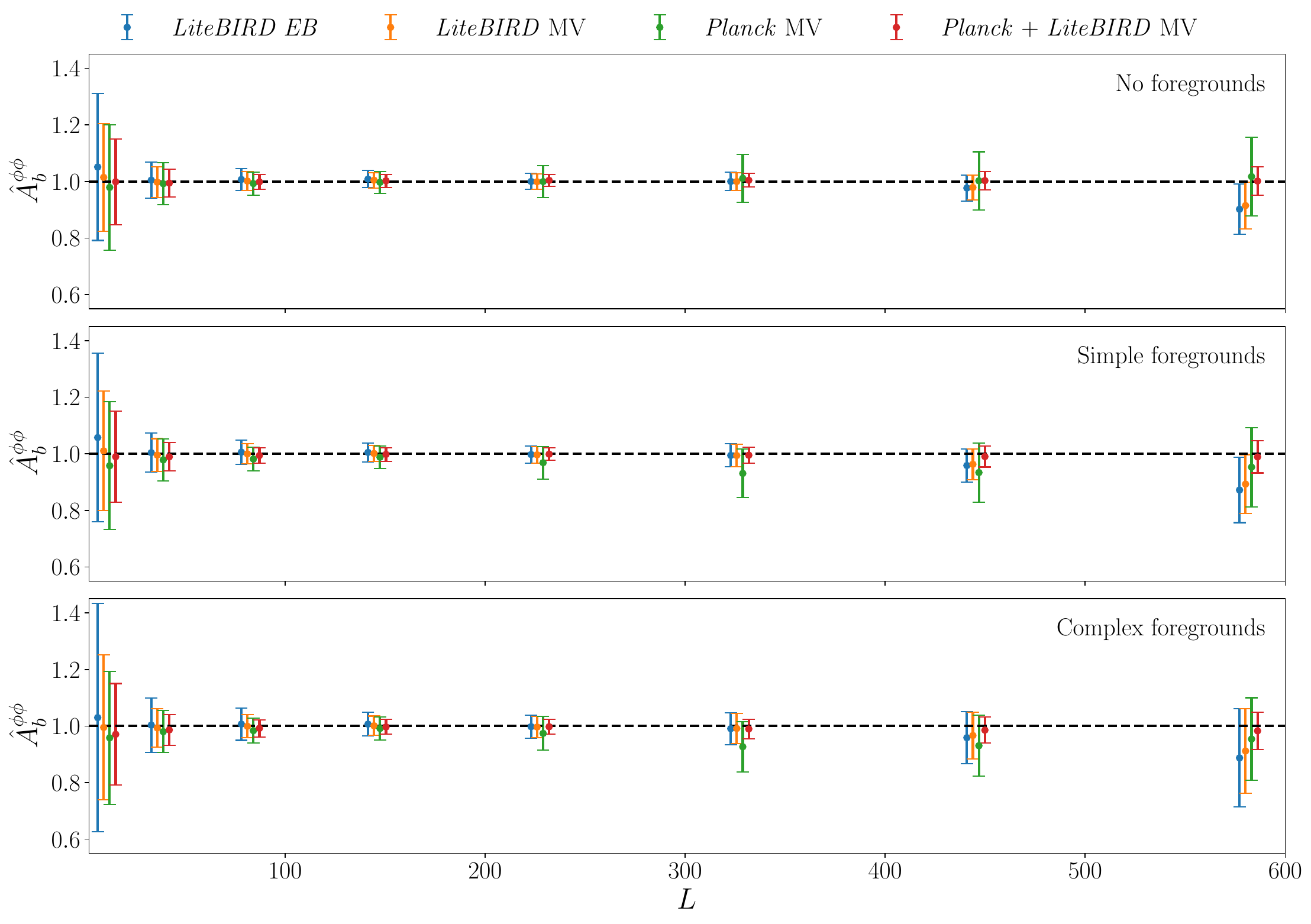}
    \caption{
    Lensing amplitude band powers as defined in equation~\eqref{eq band powers} for \textit{LiteBIRD} $EB$, \textit{LiteBIRD} MV, \textit{Planck} MV and \textit{Planck} + \textit{LiteBIRD} MV for all the foregrounds complexity scenarios. \textit{Planck} and \textit{Planck} + \textit{LiteBIRD} is binned the same way as \textit{LiteBIRD} for comparison purposes. The error bars correspond to the $1\,\sigma$ confidence intervals. The black dashed line correspond to the perfect reconstruction $\hat{A}_b^{\phi\phi}=1$.
    }
    \label{fig:bandpowers_A_L_b_all_fg}
\end{figure}

Finally, figure~\ref{fig:bandpowers_A_L_b_all_fg} shows the lensing amplitude per band power for the $EB$ and MV QEs of \textit{LiteBIRD} and for the MV QE of \textit{Planck} and \textit{Planck} + \textit{LiteBIRD}, for all the simulation complexities. We only include the band powers with $L<600$, because at $L=600$ the lensing SNR has already saturated, as will be explained in section~\ref{sec: SNR}. As expected, when the complexity of the foregrounds increases, the error bars increase as well. As previously mentioned, there is a small bias at small scales for \textit{LiteBIRD}. This bias disappears when including \textit{Planck} data, thanks to \textit{Planck}'s higher resolution. Another key aspect of the complementarity of the combination: \textit{LiteBIRD}'s lensing reconstruction allows us to access the large and intermediate scales of the lensing power spectrum, while \textit{Planck} grants access to information from the intermediate to small scales.

\subsection{Signal-to-noise ratio} \label{sec: SNR}

In this section, we estimate the SNR for the different experiments, QEs, and foreground complexities. For the SNR calculation, we want to estimate the overall lensing amplitude as a weighted sum of the amplitudes across band powers. The optimal weights, $w_b$, are derived by obtaining the MV lensing amplitude, $\hat{A}_i^{\phi^{XY}\phi^{WZ}}$, under the constraint that the sum of the weights is equal to one. For each simulation, the MV lensing amplitude is given by
\begin{equation}\label{lens amp MV}
\hat{A}_i^{\phi^{XY}\phi^{WZ}}=\sum_bw_b\hat{A}_{b,i}^{\phi^{XY}\phi^{WZ}}=\frac{\sum_{bb'}(\mathrm{Cov}_i^{-1})_{bb'}\hat{A}_{b,i}^{\phi^{XY}\phi^{WZ}}}{\sum_{bb'}(\mathrm{Cov}_i^{-1})_{bb'}}\, ,
\end{equation}
where $\mathrm{Cov}_i=\mathrm{Cov}_i(\hat{A}_{b}^{\phi^{XY}\phi^{WZ}}, \hat{A}_{b'}^{\phi^{XY}\phi^{WZ}})$ is the covariance matrix estimated using all the simulations except simulation $i$. Then, the final lensing amplitude is given by the mean and sample standard deviation,
\begin{equation}
    \hat{A}^{\phi^{XY}\phi^{WZ}} = \left \langle\hat{A}_i^{\phi^{XY}\phi^{WZ}}\right \rangle \pm \sigma\left(\hat{A}_i^{\phi^{XY}\phi^{WZ}}\right),
\end{equation}
%
both calculated using the 400 estimations of $\hat{A}_i^{\phi^{XY}\phi^{WZ}}$. The uncertainty includes the contribution from the cosmic variance of the lensing potential, thanks to being a simulation-based approach. From the lensing amplitude, we can compute the overall SNR, which is defined as,
\begin{equation}\label{SNR amp est}
    \mathrm{SNR}\left(\hat{A}_i^{\phi^{XY}\phi^{WZ}}\right)=\frac{1}{\sigma\left (\hat{A}_i^{\phi^{XY}\phi^{WZ}}\right )}.
\end{equation}

\begin{figure}[t]
    \centering
    \includegraphics[width=1\textwidth]{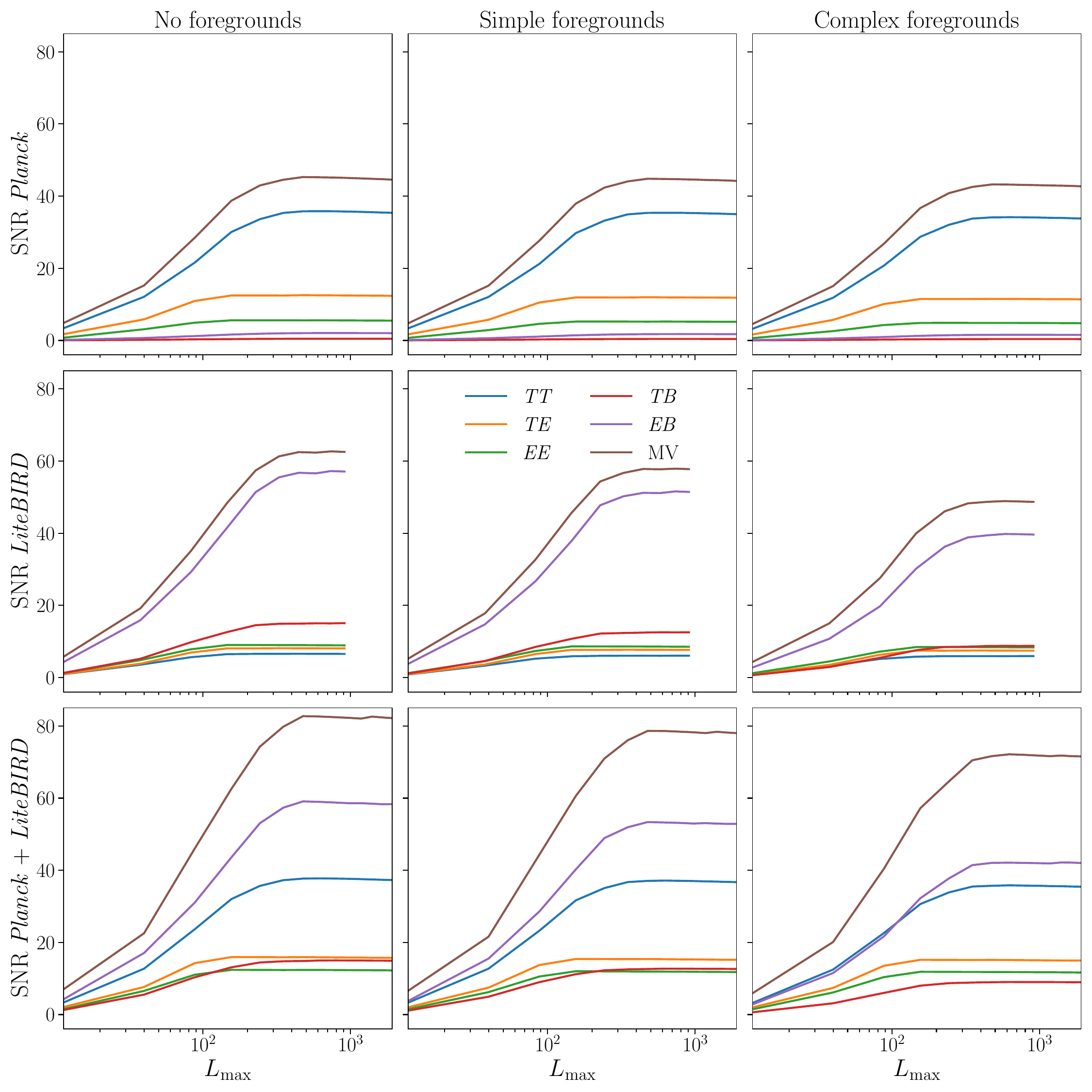}
    \caption{
    Cumulative signal-to-noise ratio (SNR) up to a maximum multipole, $L_{\rm max}$, for all the QEs. The SNR is plotted for \textit{LiteBIRD}, \textit{Planck}, and \textit{Planck} + \textit{LiteBIRD} for all the foreground complexities considered.
    }
    \label{fig:SNR_all_exp_fg}
\end{figure}

Figure \ref{fig:SNR_all_exp_fg} shows the cumulative SNR for the different experiments, foreground complexities, and QEs. The cumulative SNR is calculated as presented in eq.~\eqref{SNR amp est}, but with eq.~\eqref{lens amp MV} only evaluated up to the bin $b$, such as $L_b\leq L_{\rm max}$. From Figure \ref{fig:SNR_all_exp_fg}, we observe that \textit{Planck}'s highest-SNR QE is $TT$, followed by $TE$. For \textit{LiteBIRD}, the highest SNR QE is $EB$, which clearly dominates over the rest of the QEs. For the combination of \textit{Planck} and \textit{LiteBIRD}, the highest-SNR QEs are $EB$, followed by $TT$ for all the foreground complexities. This means that \textit{LiteBIRD}'s polarization contributes more to the MV estimator than \textit{Planck}'s temperature. For all experiments, the cumulative SNR saturates at multipoles $L\approx 400$ because at smaller scales the lensing signal is very low and the reconstruction noise blows up. Finally, while \textit{Planck}'s MV SNR does not vary much when increasing the foreground complexity, \textit{LiteBIRD}'s does. This might be related to a higher intensity of the CMB signal with respect to the foregrounds in temperature compared to polarization. Another reason could be the component-separation method chosen here, which might not be optimal for dealing with complex foregrounds. This last item will need to be studied in future work. 

\begin{figure}[t]
    \centering
    \includegraphics[width=0.7\textwidth]{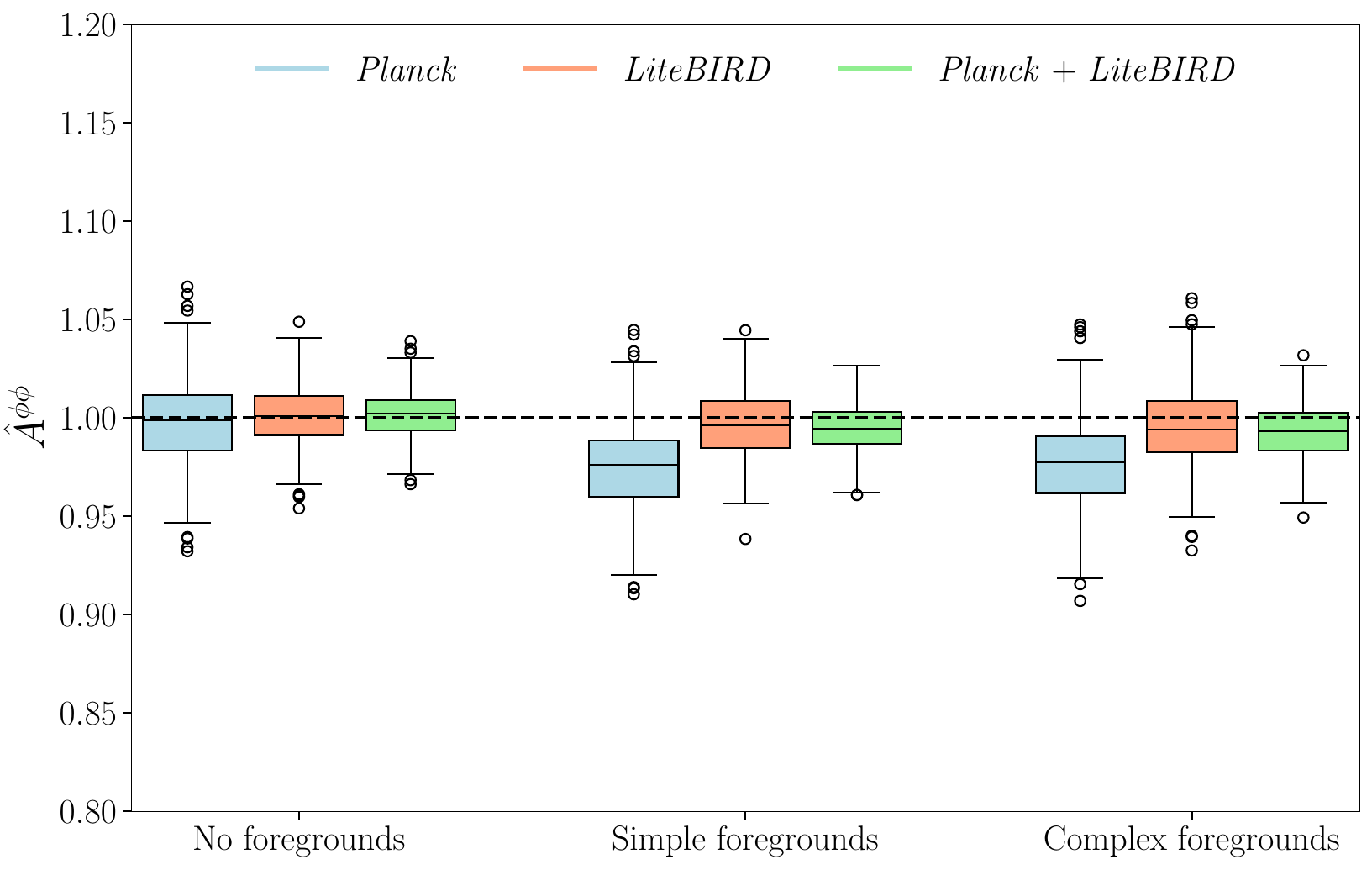}
    \caption{
    Box plot for the MV lensing amplitude, $\hat{A}^{\phi\phi}$, of \textit{Planck}, \textit{LiteBIRD} and \textit{Planck} + \textit{LiteBIRD} in all the simulations considered. The black dashed line represents the reference value for the perfect reconstruction, $\hat{A}^{\phi\phi}= 1$. Boxes span from the first to the third quartile, giving the $0.67\,\sigma$ intervals, with the horizontal black line inside the box showing the median. Lower and upper whiskers correspond to $2.7\,\sigma$ intervals. Individual points outside the whiskers are $\geq 2.7\,\sigma$ outliers.
    }
    \label{fig:A_lens_MV_all_exp_all_fg}
\end{figure}

Figure \ref{fig:A_lens_MV_all_exp_all_fg} shows box plots for the \textit{Planck}, \textit{LiteBIRD}, and \textit{Planck} + \textit{LiteBIRD} MV lensing amplitude for all the foreground complexities considered. In the no-foregrounds case, we obtain unbiased estimates for all the experiments. Once foregrounds are included, small deviations from the fiducial value appear. The most significant deviations correspond to \textit{Planck}'s simple- and complex-foregrounds simulations, where $1\,\sigma$ negative biases appear. Not all contributions from foregrounds are captured in this work, which potentially causes these small biases. More work needs to be done on foreground debiasing in the future.

In table~\ref{tab:SNR_estimation}, we present the SNR calculated for all experiments, foreground complexities, and QEs considered. Our results are compatible with published \textit{Planck} lensing analyses~\cite{P18:phi, Planck_PR4_lensing}. The SNR of \textit{LiteBIRD}'s MV estimator is between $14$ to $23\,\%$ higher than \textit{LiteBIRD}'s $EB$-only QE, which is a significant enough improvement to make the MV the default estimator for future analyses. Furthermore, \textit{LiteBIRD}'s MV SNR is higher than \textit{Planck}'s across all foreground complexities, which demonstrates that a richer lensing signal can be extracted from precise but low-resolution polarization measurements than from higher-resolution temperature data. Finally, the expected improvement on the SNR from the combination of \textit{Planck} and \textit{LiteBIRD} with respect to \textit{LiteBIRD} alone ranges between $34$ and $48\,\%$. This means that the future combination of \textit{Planck} and \textit{LiteBIRD} will improve the best full-sky lensing map to date (i.e. \textit{Planck}'s) by around $69$ to $76\,\%$, depending on how complex polarized Galactic emission turns out to be.

Since the SNR is estimated from simulations, there is an associated error which is not given in table~\ref{tab:SNR_estimation}. Assuming that the lensing amplitudes are Gaussian distributed, the error on our SNR calculation, $\Delta \text{SNR}$, can be estimated from the variance of the sample standard deviation, as demonstrated in appendix~\ref{Appendix Error SNR}. The relative error is then given by
\begin{equation} \label{eq error SNR}
    \frac{\Delta \mathrm{SNR}}{\mathrm{SNR}} = \frac{1}{\sqrt{2N_{\rm sims}}} = 0.035,
\end{equation}
where $N_{\rm sims}=400$ is the number of simulations we used. The relative SNR uncertainty is about $4\,\%$, which is a good trade-off between precision and the computational cost required to run a larger number of simulations. We found consistent results when applying naive bootstrapping to assess the uncertainty of the SNR calculation.

\begin{table}[t]
    \centering
    \begin{tabular}{|c|c|c|c|c|}
    \hline
    QEs & Experiment & SNR no fg. & SNR simple fg. & SNR complex fg. \\
    \hline 
    \multirow{4}{*}{ $TT$ } &  \textit{Planck} & 35.4 & 35.0 & 33.8 \\
    & \textit{LiteBIRD}  & 6.5 & 6.0 & 5.9 \\
    & \textit{Planck} + \textit{LiteBIRD}  & 37.3 & 36.7 & 35.5 \\
    \hhline{~----}
    & \% improvement  &  $+$470\phantom{$+$} & $+$510\phantom{$+$} & $+$497\phantom{$+$} \\
    \hline
    \multirow{4}{*}{ $EE$ } &  \textit{Planck} & 5.5 & 5.2 & 4.79 \\
    & \textit{LiteBIRD}  & 8.9 & 8.5 & 8.4 \\
    & \textit{Planck} + \textit{LiteBIRD}  & 12.2 & 11.8 & 11.7 \\
    \hhline{~----}
    & \% improvement  &  $+$37.3\phantom{$+$} & $+$38.3\phantom{$+$} & $+$39.7\phantom{$+$} \\
    \hline
    \multirow{4}{*}{ $TE$ } &  \textit{Planck} & 12.4 & 11.8 & 11.4 \\
    & \textit{LiteBIRD}  & 8.1 & 7.7 & 7.4 \\
    & \textit{Planck} + \textit{LiteBIRD}  & 15.7 & 15.2 & 15.0 \\
    \hhline{~----}
    & \% improvement  &  $+$95\phantom{$+$} & $+$98\phantom{$+$} & $+$101\phantom{$+$} \\
    \hline
    \multirow{4}{*}{ $TB$ } &  \textit{Planck} & 0.470 & 0.408 & 0.377 \\
    & \textit{LiteBIRD}  & 15.1 & 12.5 & 8.8 \\
    & \textit{Planck} + \textit{LiteBIRD}  & 14.9 & 12.6 & 9.0 \\
    \hhline{~----}
    & \% improvement  &  $-$0.90\phantom{$-$} & $+$0.77\phantom{$+$} & $+$2.00\phantom{$+$} \\
    \hline
    \multirow{4}{*}{ $EB$ } &  \textit{Planck} & 2.02 & 1.75 & 1.55 \\
    & \textit{LiteBIRD}  & 57 & 51 & 39.6 \\
    & \textit{Planck} + \textit{LiteBIRD}  & 58 & 53 & 42.0 \\
    \hhline{~----}
    & \% improvement  &  $+$2.20\phantom{$+$} & $+$2.79\phantom{$+$} & $+$6.1\phantom{$+$} \\
    \hline
    \multirow{4}{*}{MV} &  \textit{Planck} & 44.6 & 44.2 & 42.7 \\
    & \textit{LiteBIRD}  & 63 & 58 & 48.7 \\
    & \textit{Planck} + \textit{LiteBIRD}  & 82 & 78 & 72 \\
    \hhline{~----}
    & \% improvement  &  $+$31.4\phantom{$+$} & $+$35.1\phantom{$+$} & $+$47.0\phantom{$+$} \\
    \hline
    \end{tabular}
    \caption{Signal-to-noise ratio (SNR) for all the QEs, experiments, and foreground complexities considered. The ``\% improvement'' row corresponds to the percentage improvement of \textit{Planck}+\textit{LiteBIRD} with respect to \textit{LiteBIRD} alone. The relative error on the SNR calculation is $4\,\%$, which was considered for setting the number of significant figures.}
    \label{tab:SNR_estimation}
\end{table}

\section{Applications} \label{sec:applications}

The CMB lensing map contains a wealth of cosmological information and has many different science applications, ranging from cosmological parameter estimation (section \ref{sec: cosmo params}) to internal delensing (section \ref{sec: delensing}). Here we explore a handful of examples to contextualize the improvement that the combined \textit{Planck} + \textit{LiteBIRD} lensing map will represent.

\subsection{Cosmological parameter estimation} \label{sec: cosmo params}

We estimate the cosmological parameters -- cold dark matter and baryon densities, $\Omega_{\rm c}h^{2}$ and $\Omega_{\rm b}h^{2}$, the Hubble constant $H_{0}$, the optical depth to reionization $\tau$, and the amplitude and scalar spectral index of primordial fluctuations, $A_{\rm s}$ and $n_{\rm s}$ -- for the base $\Lambda$CDM model using only the power spectrum of the gravitational potential, $\hat{C}_{\ell}^{\phi\phi}$. This is done using the MCMC sampler~\cite{2002PhRvD..66j3511L, 2013PhRvD..87j3529L} implemented in \texttt{cobaya}\footnote{\url{https://cobaya.readthedocs.io/en/latest/}}~\cite{2021JCAP...05..057T}. We obtain the parameters using a Gaussian likelihood, 
\begin{equation}\label{eq: likelihood}
    -2\log{\mathcal{L}} = \sum_{bb'}[\hat{C}_{L_{b}}^{\phi\phi}-C_{L_{b}}^{\phi\phi} (\boldsymbol{\theta})] \bR{C}_{bb'}^{-1}[\hat{C}_{L_{b'}}^{\phi\phi}-C_{L_{b'}}^{\phi\phi} (\boldsymbol{\theta})],
\end{equation}
where $\hat{C}_{L_{b}}^{\phi\phi}$ is the measured mean lensing power spectrum using 100 simulations with the MV estimator, $C_{L_{b}}^{\phi\phi}(\boldsymbol{\theta})$ is the theory lensing power spectrum, calculated with the \texttt{CLASS} Boltzmann solver for $\boldsymbol{\theta}$ cosmological parameters, and $\bR{C}_{bb'}$ is the covariance matrix obtained from 300 simulations, which are independent from the 100 ones used to obtain the mean spectrum. Taking into account that the inverse of the above covariance matrix is not an unbiased estimator, we rescale it by the Hartlap factor~\cite{Hartlap}:
\begin{equation}
    \alpha_{\mathrm{cov}} = \frac{N_\mathrm{sims}-N_{\mathrm{bins}}-2}{N_\mathrm{sims}-1},
\end{equation}
where $N_\mathrm{sims}=300$ is the number of simulations used to estimate the covariance matrix, and $N_{\mathrm{bins}}$ is the number of band powers ($N_\mathrm{bins}=10$ for \textit{LiteBIRD}, and $N_\mathrm{bins}=14$ for \textit{Planck} and \textit{Planck} + \textit{LiteBIRD}).

The likelihood in eq.~(\ref{eq: likelihood}) assumes that the band powers are Gaussian distributed, which should be close to true, given the central limit theorem. Following ACT's analysis~\cite{ACT_DR6_lensing}, we performed a Kolmogorov–Smirnov test~\cite{Massey01031951} and found that all band powers are consistent with a Gaussian distribution.

We use the priors summarized in table~\ref{tabpriors}. These are similar to the priors assumed in the latest \textit{Planck} and ACT analyses~\cite{ACT_DR6_lensing, P18:phi, Planck_PR4_lensing}. The optical depth is fixed to the \textit{Planck}-2018  best-fit value~\cite{P18:cosmological-parameters}, which is also the input value for our simulations, since lensing alone is not sensitive to it. Additionally, we fix the total neutrino mass to 0.06 eV with a single massive neutrino.

\begin{table}[t]
    \centering
    \begin{tabular}{|l|c|}
        \hline
        Parameter & Prior \\
        \hline
        $H_0$ & $[40, 100]$ \\
        $\ln(10^{10}A_{\rm s})$ & $[2, 4]$ \\
        $n_{\rm s}$ & $\mathcal{N}(0.965, 0.02)$ \\
        $\Omega_{\rm b} h^2$ & $\mathcal{N}(0.0224, 0.0005)$ \\
        $\Omega_{\rm c} h^2$ & $[0.005, 0.99]$ \\
        $\tau$ & 0.0544 \\
        \hline
    \end{tabular}
    \caption{Priors used for cosmological parameter inference with the lensing spectrum. Square brackets indicate uniform priors. $\mathcal{N}(\mu, \sigma)$ denotes Gaussian priors with mean $\mu$ and standard deviation $\sigma$.}\label{tabpriors}
\end{table}

\begin{figure}[t]
    \begin{subfigure}[t]{.5\linewidth}
        \centering
        \includegraphics[width=1.05\linewidth]{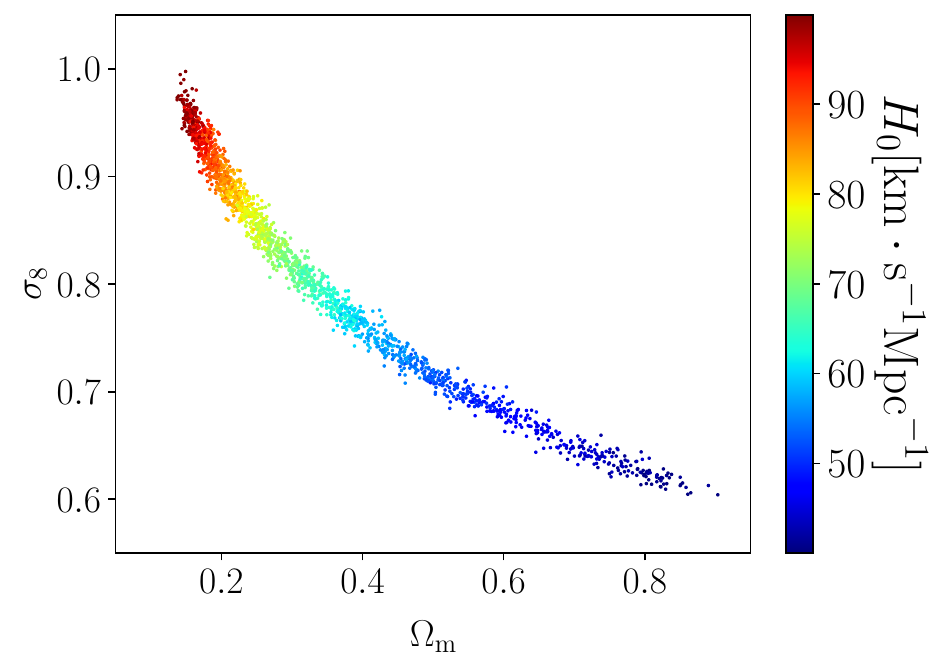}
        \label{fig:sub1}
    \end{subfigure}
     \begin{subfigure}[t]{.5\linewidth}
        \centering
        \includegraphics[width=0.935\linewidth]{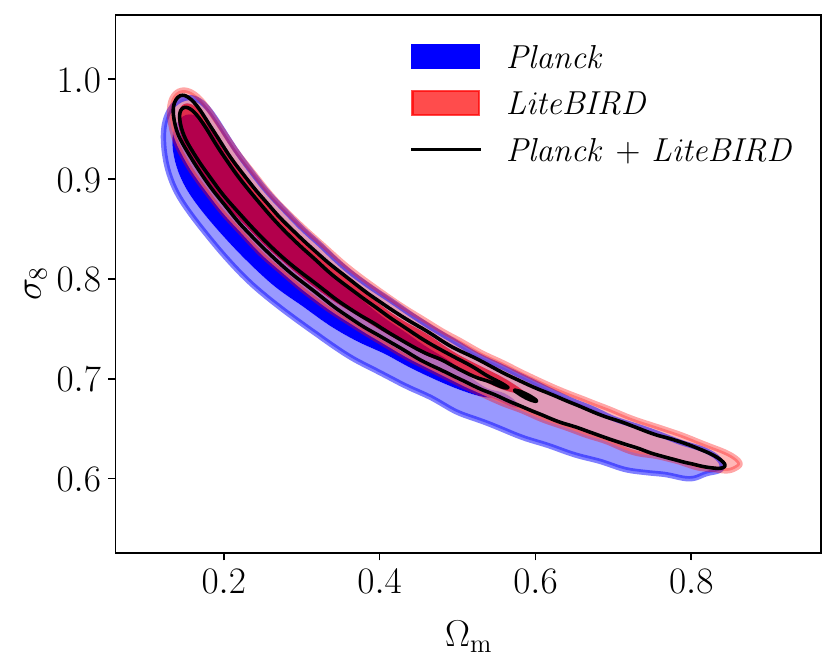}
        \label{fig:sub2}
    \end{subfigure}
    \caption{\textit{Left panel: } Posterior in $\sigma_{8}$--$H_{0}$--$\Omega_{\rm m}$ space from our lensing-only likelihood projected in a 2D space with the third dimension, $H_{0}$, represented by different colours. This is obtained for the combination of \textit{Planck} and \textit{LiteBIRD} experiments, and for the simple-foreground model. \textit{Right panel:} Constraints in the $\sigma_{8}$--$\Omega_{\rm m}$ plane obtained from the lensing-only likelihood, and the simple-foreground model. Contours of the $68\,\%$ and the $95\,\%$ marginalized confidence levels are given for the three experiments: \textit{Planck} only (blue); \textit{LiteBIRD} only (red); and \textit{Planck} + \textit{LiteBIRD} (solid black line).}
    \label{fig:degeneracy_H0_OmegaM_Sigma8}
\end{figure}

Cosmological weak-lensing observables are influenced by the amplitude of density fluctuations at late times, characterized by $\sigma_{8}$, and the matter density parameter $\Omega_{\rm m}$, with an additional dependence on the Hubble parameter $H_{0}$. The left panel of figure~\ref{fig:degeneracy_H0_OmegaM_Sigma8} shows the degeneracy line between these three parameters. As discussed in Appendix D of ref.~\cite{2024Madhavacheril}, small and large scales of CMB lensing are sensitive to two different combinations of these parameters, i.e.~two different surfaces in the $\sigma_{8}$--$H_0$--$\Omega_{\rm m}$ parameter space. In particular, the line in the left panel emerges from the intersection of these two planes. This degeneracy can be broken through a joint analysis of CMB lensing and BAO data, which constrains an additional surface in the parameter space that intersects the CMB lensing line at a single point (see, e.g., refs.~\cite{2024Madhavacheril,Planck_PR4_lensing}). However, this goes beyond the scope of this work.

The right panel of figure~\ref{fig:degeneracy_H0_OmegaM_Sigma8} shows again constraints in the $\sigma_8$--$\Omega_{\rm m}$ plane, but this time presenting the difference between the three configurations we have considered for the simple foreground model: \textit{Planck}-only; \textit{LiteBIRD}-only; and \textit{Planck} + \textit{LiteBIRD}. In particular, uncertainties are reduced when \textit{Planck} and \textit{LiteBIRD} data are combined compared to the $\textit{LiteBIRD}$-only case, and specially with respect to the $\textit{Planck}$-only scenario. This is because the error bars in the power spectrum reconstruction are smaller, as can be seen in Figures \ref{fig:bandpowers_A_L_b_all_fg} and \ref{fig:A_lens_MV_all_exp_all_fg}.

\begin{figure}[t]
    \centering
    \includegraphics[width=0.7\textwidth]{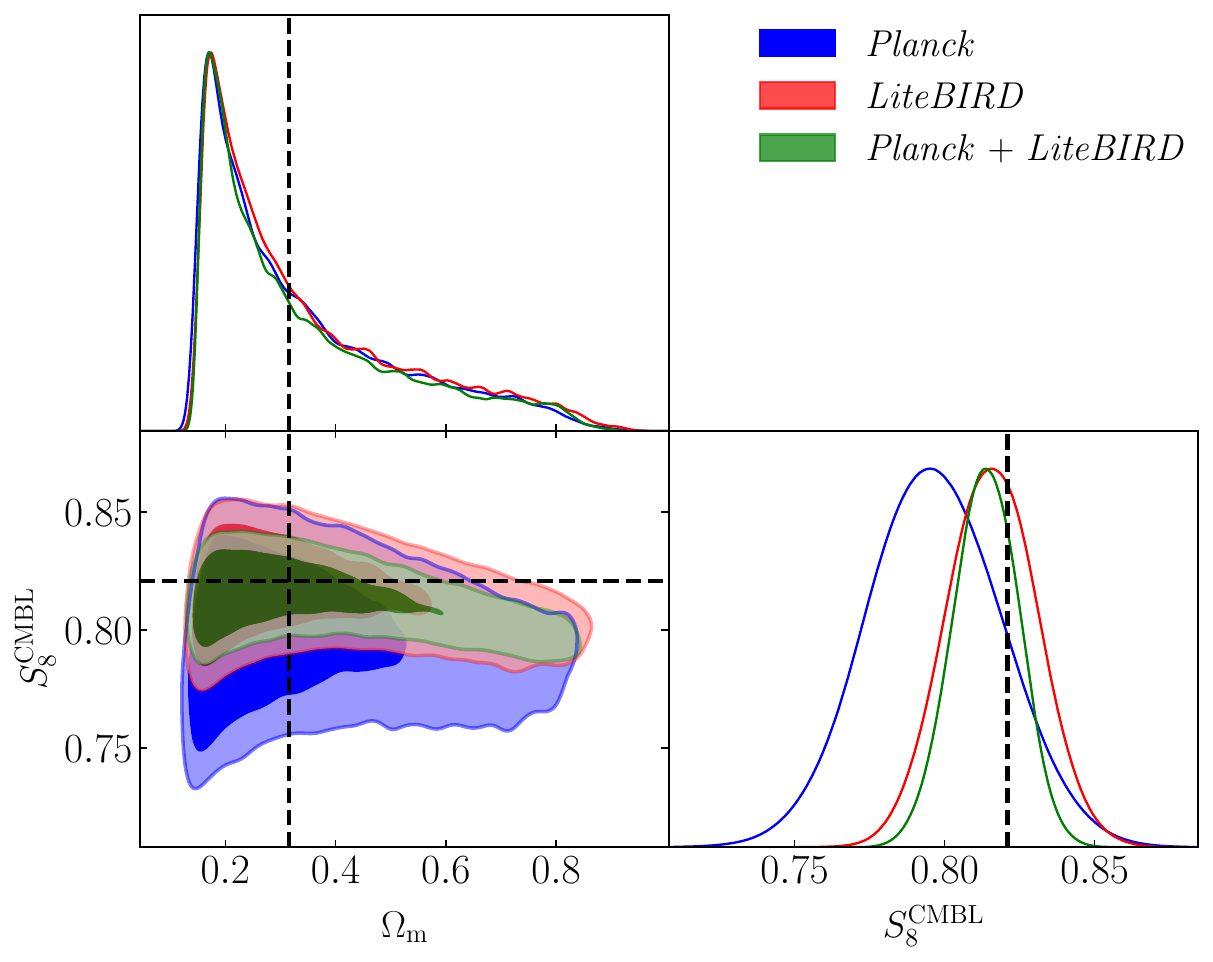}
    \caption{Marginalized constraints on $S_{8}^{\mathrm{CMBL}}$ and $\Omega_{\rm m}$ from our lensing-only likelihood and simple-foreground simulations of the \textit{Planck}-only (blue), \textit{LiteBIRD}-only (red), and \textit{Planck} + \textit{LiteBIRD} (green) analyses. Contours show the $68\,\%$ and the $95\,\%$ marginalized confidence levels. Diagonal panels show the corresponding $68\,\%$ 1D marginalized posteriors.}
    \label{fig:Omega_M_vs_Sigma_8}
\end{figure}

Following the ACT DR6 analysis~\cite{ACT_DR6_lensing}, we define the CMB-lensing-equivalent of the usual $S_8$ parameter as 
\begin{equation}
    S_{8}^{\mathrm{CMBL}} = \sigma_{8}\left(\frac{\Omega_{\rm m}}{0.3}\right)^{0.25},
\end{equation}
which approximately corresponds to the best-constrained direction in the $\sigma_{8}$--$\Omega_{\rm m}$ plane. Figure~\ref{fig:Omega_M_vs_Sigma_8} shows the posteriors of $\Omega_{\rm m}$ and $S_{8}^{\mathrm{CMBL}}$ for the three configurations assuming again our simple-foreground model. The slight bias towards lower values of $S_8^{\rm CMBL}$ is mainly due to the small biases on the overall lensing amplitude seen in figure~\ref{fig:A_lens_MV_all_exp_all_fg}. Posteriors from the no-foreground simulations also exhibit a small bias, which may be attributed to the masking procedure. We use the no-foreground case to calibrate a transfer function for the lensing power spectrum and correct for it in the simple- and complex-foreground scenarios. This approach ensures that we remove all the possible biases arising from everything that is not residual foreground contamination, which should be common for all the foreground models. Furthermore, for the \textit{Planck} + \textit{LiteBIRD} complex-foreground case, we tested that, if we calibrate the residuals with the fiducial spectrum for each band power and subtract it previous to the MCMC sampling, we are able to recover an unbiased posterior with a negligible impact on the uncertainty. Hence, we believe that leftover foreground residuals in the lensing band powers are the cause of the bias in $S_8^\mathrm{CMBL}$. A more detailed treatment of foreground contaminants is left for future work.

Finally, table~\ref{tab_summary_parameters} summarizes the 1$\,\sigma$ uncertainties on $S_{8}^{\mathrm{CMBL}}$  for the three configurations and foreground scenarios considered, highlighting the relative improvement between a \textit{LiteBIRD}-only and \textit{Planck}+\textit{LiteBIRD} analysis. We obtain $S_8^\mathrm{CMBL}$ uncertainties that are roughly compatible with the $\sigma_{S_8^\mathrm{CMBL}}=0.0270$~\cite{P18:phi} and $\sigma_{S_8^\mathrm{CMBL}}=0.0216$~\cite{Planck_PR4_lensing} lensing-only analyses of \textit{Planck} PR3 and PR4 data,\footnote{Note the factor of $0.3^{-0.25}$ needed to convert the $\sigma_8\Omega_{\rm m}^{0.25}$ \textit{Planck} constraints into our $\sigma_8(\Omega_{\rm m}/0.3)^{0.25}$ $S_8$ measurements.} respectively, thus validating our forecasting framework. The combination of \textit{Planck} + \textit{LiteBIRD} provides a factor of 2 improvement in the $S_{8}^{\mathrm{CMBL}}$ uncertainty compared to \textit{Planck}-only constraints. A similar improvement is obtained compared to the current ACT constraints~\cite{ACT_DR6_lensing}. The improvement with respect to the \textit{LiteBIRD}-only analysis increases with the complexity of the foregrounds. This is because \textit{LiteBIRD}'s MV estimator is mostly based on $EB$, which suffers from the increase of foreground complexity between our different \texttt{psym3} models, while \textit{Planck's} MV is dominated by $TT$ and remains mostly stable.

\begin{table}[t]
    \centering
    \begin{tabular}{|c|c|c|c|}
    \cline{2-3}
    \multicolumn{1}{c|}{} & \multicolumn{2}{c|}{$\sigma_{S_{8}^{\mathrm{CMBL}}}$} \\
    \hline
    Lensing map & Simple fg. & Complex fg. \\
    \hline 
    \textit{Planck} &  0.0226 & 0.0233 \\
     \textit{LiteBIRD} & 0.0152 & 0.0192 \\
     \textit{Planck} + \textit{LiteBIRD} & 0.0111 & 0.0124 \\
     \hline
     \% improvement & $+$27\phantom{$+$} & $+$36\phantom{$+$}\\
    \hline
    \end{tabular}
    \caption{Uncertainty on the $S_{8}^{\mathrm{CMBL}}$ parameter ($\sigma_{S_{8}^{\mathrm{CMBL}}}$) for the MV lensing reconstruction from the considered experiments and foreground models. The ``\% improvement'' row corresponds to the porcentage improvement of \textit{Planck} + \textit{LiteBIRD} with respect to the \textit{LiteBIRD} only case.}
    \label{tab_summary_parameters}
\end{table}


\subsection{Improving tensor-to-scalar ratio constraints through internal delensing}\label{sec: delensing}

In this section, we address the delensing of \textit{LiteBIRD}'s $B$-mode polarization that is achievable with the combination of \textit{Planck} and \textit{LiteBIRD} data. We also compare the delensing efficiency obtained with \textit{LiteBIRD}'s $EB$ and MV QEs to quantify the improvement that using all \textit{LiteBIRD} temperature and polarization channels allows compared to our previous forecast~\cite{Lonappan_lensing, Namikawa_delensing_2024}. Although delensing can also improve constraints on cosmological parameters~\cite{2017Green, 2022Hotinli, 2023Ange} and primordial non-Gaussianity~\cite{2020Coulton}, here we focus on the improvement it represents to tensor-to-scalar ratio measurements, since that is the prime science objective of \textit{LiteBIRD}.

In harmonic space, lensed $B$ modes can be described as the convolution of $E$-mode anisotropies with the lensing potential power spectrum~\cite{Lewis:2006:review}. 
Hence, we can \textit{delens} $B$-mode measurements by subtracting a $B$-mode template built from the convolution of the observed $E$ modes and the reconstructed lensing potential. In this work, we decide to stay at the power spectrum level, calculating the delensing efficiency per multipole, $A_\ell^{\rm lens}$, as
\begin{equation} \label{eq: Alens_delensing}
    A_\ell^{\rm lens} = \frac{C_\ell^{BB}-C_\ell^{BB,\mathrm{temp}}}{C_\ell^{BB}} = \frac{\sum_{\ell' L}\Xi_{\ell\ell' L}C_{\ell'}^{EE}C_{L}^{\phi\phi}\left[1-W_{\ell'}^{EE}W_{L}^{\phi\phi}\right]}{\sum_{\ell' L}\Xi_{\ell\ell' L}C_{\ell'}^{EE}C_{L}^{\phi\phi}},
\end{equation}
where $C_\ell^{BB,\mathrm{temp}}$ is the $B$-mode template power spectrum, the kernel $\Xi_{\ell\ell' L}$ is defined in eq.~(9) of ref.~\cite{Namikawa_delens_eq_2015}, and $C_{\ell}^{EE}$ and $C_{L}^{\phi\phi}$ are the fiducial lensed $E$-mode and lensing potential power spectra. $W_{\ell}^{EE}$ and $W_L^{\phi\phi}$ are the Wiener-filtered $E$ mode and lensing potential, respectively defined as
\begin{align}
    W_{\ell}^{EE} = & \frac{C_{\ell}^{EE}}{\langle\bar{C}_{\ell}^{EE}\rangle}\,,\\
    W_{L}^{\phi\phi} = &\frac{C_{L}^{\phi\phi}}{C_{L}^{\phi\phi}+AN\text{-}N_L^{(0)}}\,,
\end{align}
where $\langle\bar{C}_{\ell}^{EE}\rangle$ is the mean observed $E$-mode power spectrum (including both noise and foreground residuals) calculated using eq.~\eqref{mean_observed_PS}, and $AN\text{-}N_L^{(0)}$ is the analytical estimation of the N0 noise bias, for the $EB$ or MV QEs. Note that the analytic formula in eq. \eqref{eq: Alens_delensing} is derived under the assumption that higher-order lensing contributions are negligible. This assumption is valid if we use the gradient-level template for delensing~\cite{gradient_delensing_Anton_2021}.

When performing internal delensing (i.e., using exclusively the information contained in CMB data), the $B$-mode field to be delensed is the same field used to estimate the lensing potential. If the same multipoles are used for both lensing estimation and delensing, additional biases appear, leading to a reduction of the delensing power~\cite{Teng_delensing_2011, BaleatoLizancos_delensing, Namikawa_internal_delensing_2017}. In order to avoid such biases, we adopt the same strategy as in ref.~\cite{Namikawa_delensing_2024} and estimate the lensing potential excluding the information from $B$-mode polarization at multipoles $\ell\leq 190$, which are the angular scales targeted for \textit{LiteBIRD}'s $B$-mode science. This will increase by 15 to 30\,\% the N0 bias for the $TB$ and $EB$ QEs, and therefore the N0 bias for the MV estimator.

\begin{figure}[t]
    \centering
    \includegraphics[width=0.65\textwidth]{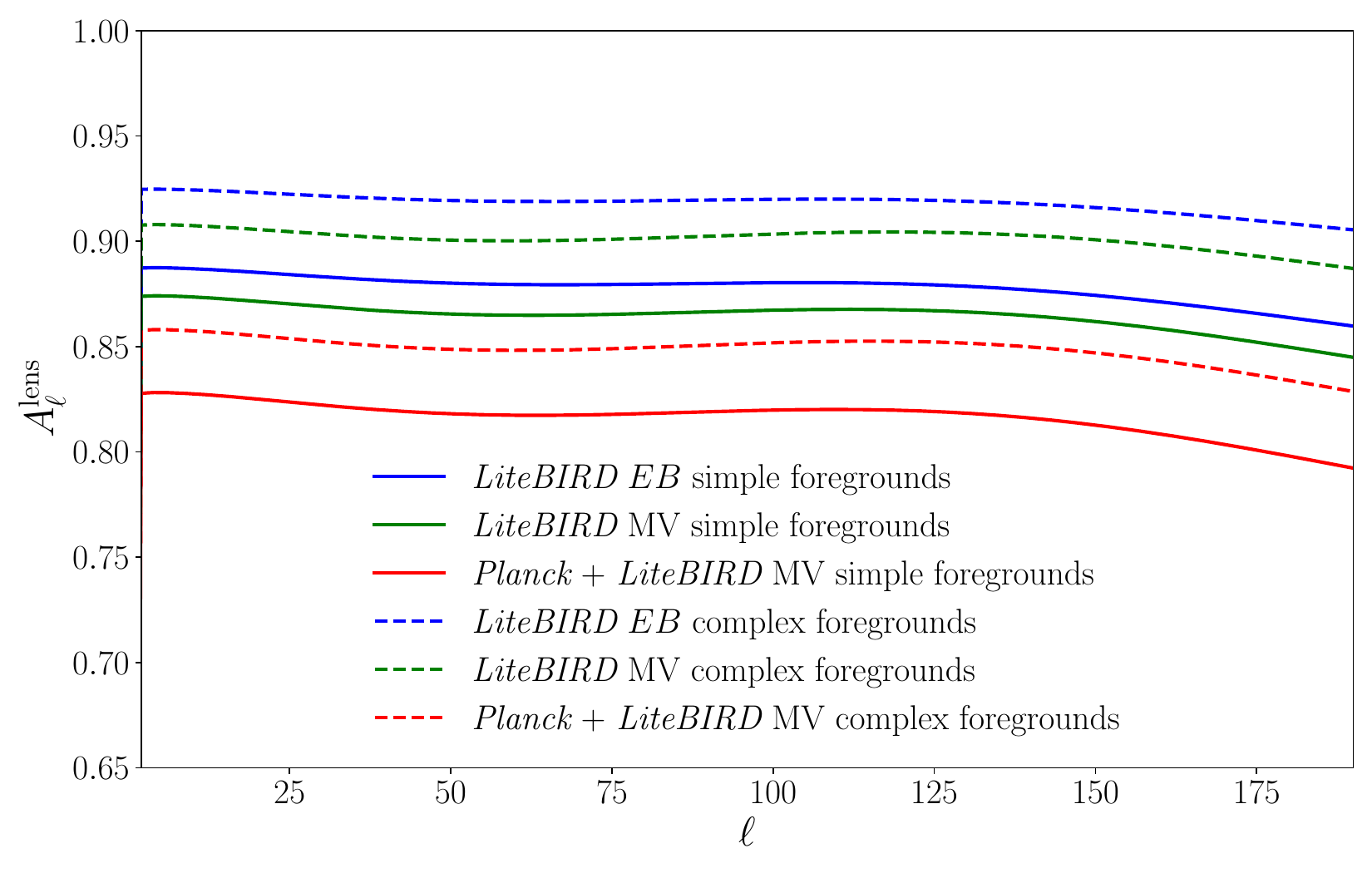}
    \caption{
    Delensing efficiency for \textit{LiteBIRD} and \textit{Planck} + \textit{LiteBIRD} for two different foreground complexities. Solid lines correspond to the simple-foregrounds case, whereas dashed lines show the complex-foregrounds case. We have considered three QEs: \textit{LiteBIRD}'s $EB$ in blue; \textit{LiteBIRD}'s MV in green; and the Planck + \textit{LiteBIRD} MV in red.
    }
    \label{fig:delensing_efficiency}
\end{figure}

In figure~\ref{fig:delensing_efficiency}, we calculate the delensing efficiency for the simple and complex foregrounds and for \textit{LiteBIRD} $EB$, \textit{LiteBIRD} MV, and \textit{Planck} + \textit{LiteBIRD} MV estimators. The results show that the delensing efficiency is around $0.81$ in the best-case scenario corresponding to \textit{Planck} + \textit{LiteBIRD} MV and simple-foreground simulations. In the worst case, i.e., \textit{LiteBIRD}'s $EB$ QE with complex foregrounds, the delensing efficiency is on average $0.91$. Compared to our previous analysis~\cite{Namikawa_delensing_2024}, we find that the delensing efficiency can be improved by $1.5\,\%$ by considering the \textit{LiteBIRD} MV estimator, and by $7\,\%$ with the \textit{Planck} + \textit{LiteBIRD} MV estimator. For consistency with previous forecasts~\cite{Namikawa_delensing_2024,LiteBIRD_PTEP_2023}, we will focus on the simple-foregrounds case to calculate our $r$ constraints. 

The likelihood we adopt to draw our $r$ constraints is given by
\begin{equation}
    -2\log \mathcal{L}\, (r) = f_{\rm sky}\sum_{\ell=2}^{190} (2\ell +1 )\left [\frac{\hat{C}_\ell^{BB}}{C_\ell^{BB}(r)}-\log\left(\frac{\hat{C}_\ell^{BB}}{C_\ell^{BB}(r)}\right)-1 \right],
\end{equation}
where $\hat{C}_\ell^{\rm BB}$ is the measured $BB$ spectrum and $C_\ell^{\rm BB}$ is the fiducial $BB$ spectrum as observed by \textit{LiteBIRD}. This likelihood captures the non-Gaussianity of the angular power spectrum and is exact in the full-sky limit~\cite{Hamimeche_Lewis_2008, Likelihood_r_2011, LiteBIRD_PTEP_2023}. Although it does not consider the multipole couplings due to the presence of the mask, we have included an $f_{\rm sky}$ factor to account for the sky coverage. The observed and fiducial $BB$ are calculated as
\begin{align}
    \hat{C}_\ell^{\rm BB}\phantom{r)} &=   A_\ell^{\rm lens} C_\ell^{BB, \mathrm{scalar}} + \left (\frac{r^{\rm fid}}{10^{-3}}\right) C_\ell^{BB, \mathrm{tensor}}+\langle\hat{N}_\ell\rangle,\\
    C_\ell^{\rm BB}\,(r) & = A_\ell^{\rm lens} C_\ell^{BB, \mathrm{scalar}} + \left (\frac{r}{10^{-3}}\right) C_\ell^{BB, \mathrm{tensor}}+\langle\hat{N}_\ell\rangle,
\end{align}
where $A_\ell^{\rm lens}$ is the delensing efficiency, $C_\ell^{\rm BB, scalar}$ and $C_\ell^{\rm BB, tensor}$ are the scalar and tensor (calculated for $r=10^{-3}$ with \texttt{CLASS}) $B$-mode power spectra respectively, $r^{\rm fid}$ is the fiducial input value assumed for the observation, and $\langle\hat{N}_\ell\rangle$ is the average noise and foreground residual power spectrum from the simple-foregrounds simulations after masking with \textit{Planck}'s $80\,\%$ Galactic mask (i.e., $f_{\rm sky}=0.8$).

\begin{figure}[t]
    \centering
    \includegraphics[width=0.9\textwidth]{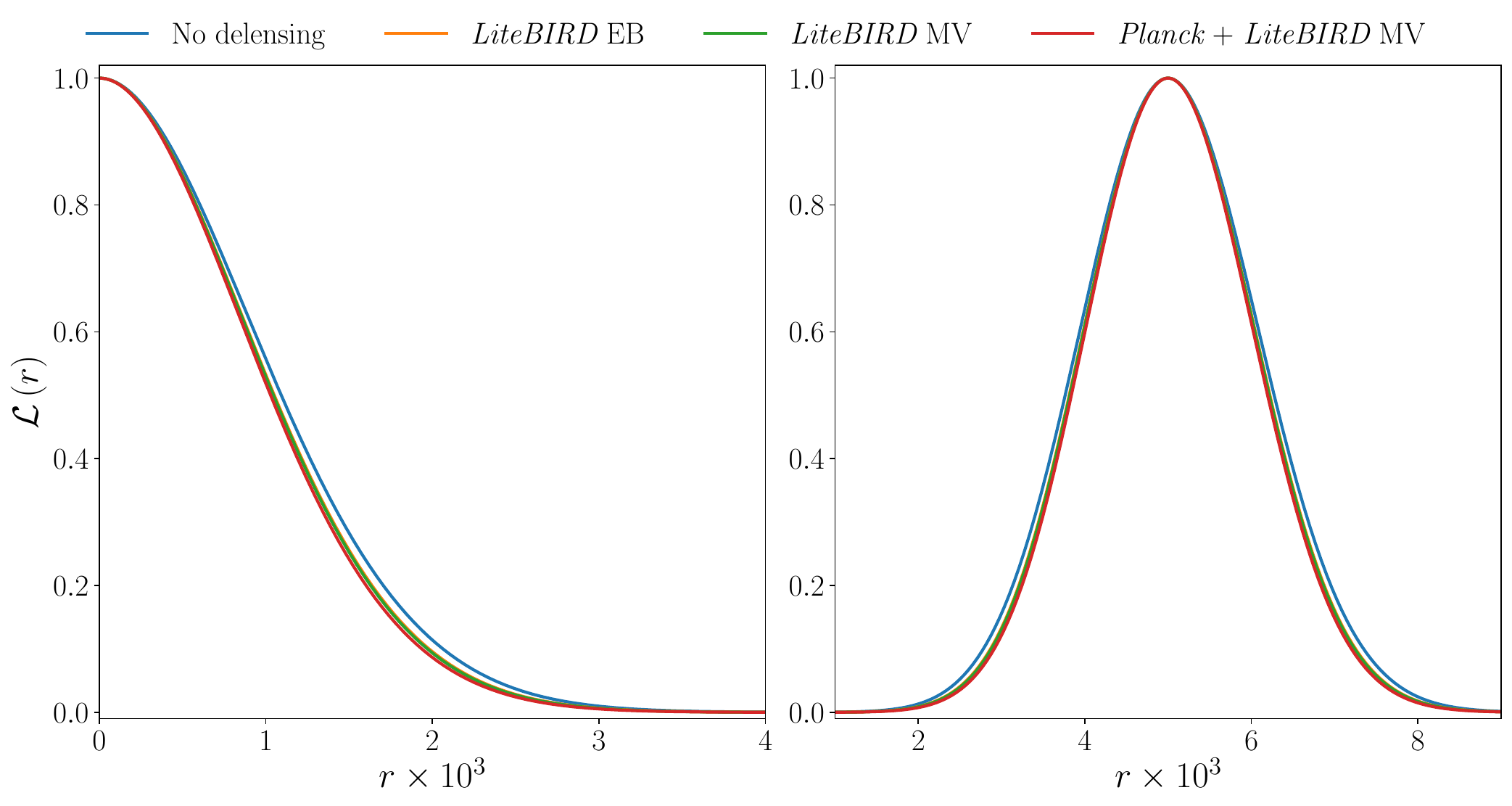}
    \caption{
    Likelihood distribution of $r$ for two different $r$ input values, $r^{\rm inp}=0$ (left) and $r^{\rm inp}=5\times 10^{-3}$ (right). We have plotted the case without delensing (no delensing), and with delensing using \textit{LiteBIRD} EB (orange), \textit{LiteBIRD} MV (green), and \textit{Planck} + \textit{LiteBIRD} MV (red). We have used the low complexity foregrounds delensing efficiency and the simple-foreground and noise residuals.
    }
    \label{fig:delensing_likelihood}
\end{figure}

\begin{table}[t]
    \centering
    \begin{tabular}{|c|c|c|c|c|c|}
        \hline
        \multirow{2}{*}{Cases} & \multicolumn{2}{c|}{$r^{\rm fid}=0$} & \multicolumn{3}{c|}{$r^{\rm fid}=5\times 10^{-3}$} \\
        \cline{2-6}
        & $\sigma_r\times 10^3$ & $\%$ & $\sigma_r^-\times 10^3$ & $\sigma_r^+\times 10^3$ & $\%$ \\
        \hline
        No delensing & 0.96 & -- & $-1.02$ & 1.11 & -- \\
        \textit{LiteBIRD} $EB$ & 0.92 & $-4.1$ & $-0.98$ & 1.06 & $-4.0$ \\
        \textit{LiteBIRD} MV & 0.91 & $-4.5$ & $-0.98$ & 1.06 & $-4.4$ \\
        \textit{Planck} + \textit{LiteBIRD} MV & 0.90 & $-6.1$ & $-0.96$ & 1.04 & $-6.1$ \\
        \hline
    \end{tabular}
    \caption{Comparison of \textit{LiteBIRD} constraints on $r$ without delensing, and when delensing with \textit{LiteBIRD}'s $EB$, \textit{LiteBIRD}'s MV, and the \textit{Planck} + \textit{LiteBIRD} MV QEs. We show constraints for two fiducial valures of $r$. $\sigma_r$, $\sigma_r^-$, and $\sigma_r^+$ are the asymmetric $1\,\sigma$ confidence intervals defined in eq.~\eqref{confidence_interval}. The ``$\%$'' column highlights the reduction in error bars compared to the base no-delensing case.}
    \label{tab:constraints_r}
\end{table}

We define asymmetric $1\,\sigma$ error bars on $r$ as the values verifying the following equation:
\begin{equation}\label{confidence_interval}
    \frac{\int_{r_\mathrm{ML}-\sigma_r^-}^{r_\mathrm{ML}+\sigma_r^+}\mathcal{L}\, (r)\mathrm{d}r}{\int_0^{\infty}\mathcal{L}\, (r)\mathrm{d}r}=0.683,
\end{equation}
where $r_\mathrm{ML}$ gives the maximum of the likelihood function, and $\sigma_r^-$ and $\sigma_r^+$ are the upper and lower error bars, respectively. We consider two fiducial values of $r$: the case of no primordial $B$ modes, i.e., $r^{\rm fid}=0$; and a primordial $B$-mode spectrum of $r^{\rm fid}=5\times 10^{-3}$. In the $r^{\rm fid}=0$ case, we fix $\sigma_r^-= 0$ and calculate $\sigma_r\equiv\sigma_r^+$. For the $r^{\rm fid}=5\times 10^{-3}$ case, $\sigma_r^-$ and $\sigma_r^+$ are determined so that $(r_\mathrm{ML}-\sigma_r^-, r_\mathrm{ML})$ and $(r_\mathrm{ML}, r_\mathrm{ML}+\sigma_r^+)$ are equiprobable intervals capturing the slight asymmetry of the distribution.

Figure~\ref{fig:delensing_likelihood} shows the likelihood distribution for the two different $r^{\rm fid}$ values, $r^{\rm fid}=0$ and $r^{\rm fid}=5\times 10^{-3}$. As one might expect from the low delensing efficiencies obtained in figure~\ref{fig:delensing_efficiency}, internally delensing \textit{LiteBIRD}'s $B$ modes only amounts to a minor improvement on $r$ constraints. This improvement is driven by the reduction of the $B$-mode variance around the recombination bump. Note that, due to our choice of component-separation algorithm, large-scale $B$ modes are not optimally cleaned (see section~\ref{subsec:apply HILC}). Hence, all our $r$ constraints would improve if a parametric foreground-cleaning method had been used to allow access to the information contained in the reionization bump. However, this will only affect the constraints on $r$, not the effectiveness of delensing.

In table~\ref{tab:constraints_r}, we quantify the $1\,\sigma$ confidence intervals obtained for the different cases. We find that, irrespective of the $r^{\rm fid}$ value considered, constraints on $r$ improve by $6\,\%$ when delensing with \textit{Planck} + \textit{LiteBIRD} MV estimator, and by around $4\%$ when delensing with \textit{LiteBIRD}-only lensing estimates. Still, internal delensing makes a marginal contribution. Instead, a multitracer delensing approach must be used to significantly improve \textit{LiteBIRD}'s sensitivity to primordial $B$ modes. As shown in ref.~\cite{Namikawa_delensing_2024}, delensing with a combined lensing map derived from \textit{LiteBIRD} CMB data, measurements of the CIB, and galaxy surveys like \textit{Euclid} or LSST, can reduce $\sigma_r$ by about $15~\%$.
\section{Summary and conclusions} \label{sec:summary}

In this paper, we have performed a comprehensive study of the gravitational lensing reconstruction that is possible through the combination of \textit{Planck}'s and \textit{LiteBIRD}'s full-sky temperature and polarization data. We covered the complete data analysis pipeline: from component separation using the multi-frequency observations of \textit{Planck} and \textit{LiteBIRD} to the final inference of cosmological parameters using our lensing reconstruction. Throughout the paper, we have tested the robustness of our results against two different Galactic foreground models of increasing complexity, which include synchrotron, dust, AME, free-free, and CO emissions. Increasing the complexity of diffuse Galactic emission has a higher impact on \textit{LiteBIRD} than on \textit{Planck}, although a significant improvement in SNR is found regardless of the foreground model considered.

The main results of our study can be summarized as follows,
\begin{itemize}
    \item Compared to previous studies that were limited to the $EB$ QE~\cite{Lonappan_lensing}, we have demonstrated that \textit{LiteBIRD} can obtain an enhanced lensing reconstruction thanks to the combination of temperature and polarization information and the MV estimator. Compared to ref.~\cite{Lonappan_lensing}, our lensing estimate also benefits from the use of small-scale information up to $\ell_\mathrm{max}=1000$ and the inclusion of \textit{LiteBIRD}'s 22 (redundant) frequency channels. We predict that \textit{LiteBIRD}'s MV lensing reconstruction will reach a $49$ to $58\,\sigma$ detection over $80\,\%$ of the sky, depending on the final complexity of polarized Galactic emission. 
    \item \textit{Planck} and \textit{LiteBIRD} are highly complementary experiments for lensing reconstruction. While \textit{Planck} offers access to the small-scale temperature anisotropies, \textit{LiteBIRD} provides precision measurements of the large-scale CMB polarization. Therefore, the next best full-sky lensing map will come from the combination of \textit{Planck} and \textit{LiteBIRD} data, reaching a $72$ to $78\,\sigma$ detection on $80\,\%$ of the sky, depending on the final complexity of polarized Galactic emission. Moreover, the lack of MF in polarization will grant access to the large scales of the lensing estimate for the first time.

    \item The \textit{Planck}+\textit{LiteBIRD} lensing estimate will help shed more light into the current tensions between local and cosmological measurements of the Hubble constant and the growth of structucte by probing the $\sigma_8$--$H_0$--$\Omega_{\rm m}$ plane.  In particular, the \textit{Planck}+ \textit{LiteBIRD} MV lensing reconstruction will allow a 0.011 to 0.013 uncertainty in $S_8^\mathrm{CMBL}$. This constitutes a factor of 2 improvement compared to \textit{Planck} constraints.
    \item Delensing \textit{LiteBIRD} with the \textit{Planck}+\textit{LiteBIRD} MV lensing estimate will reduce by $15$ to $19\,\%$ the amplitude of lensed $B$ modes (i.e., $A^\mathrm{lens}$ of $81$ to $85\,\%$), which improves the $9$ to $13\,\%$ reduction ($A^\mathrm{lens}$ of $87$ to $91\,\%$) possible with \textit{LiteBIRD}'s MV lensing reconstruction. However, this modest delensing fraction only allows a $6\,\%$ improvement of constraints on $r$. A multi-tracer delensing approach is needed to achieve a competitive delensing of \textit{LiteBIRD} data~\cite{Namikawa_delensing_2024}.
\end{itemize}

The main goal of the present paper is to showcase the potential of a combined \textit{Planck} + \textit{LiteBIRD} lensing analysis, using simulations of a simplified instrument model and leaving to future work the study of a more realistic scenario. Throughout this paper, we have assumed homogeneous white noise, but \textit{Planck} and \textit{LiteBIRD} have inhomogeneous noise due to the scanning strategy. This effect could introduce a larger mean-field bias~\cite{Hanson:2009:noise} and render our diagonal harmonic filtering suboptimal, thus making pixel-space filtering mandatory for an optimal lensing reconstruction~\cite{Eriksen:2004:wiener}. Other instrumental systematics like $1/f$ noise, miscalibration of the polarization angle, pointing offsets, and beam imperfections can impact the lensing reconstruction~\cite{Beam_systematics_2009, Hanson:2010:beam}. Finally, other extragalactic foregrounds that were not included in this paper such as the CIB, the tSZ, and point sources also play an important role in the lensing reconstruction, biasing the results if not taken into account correctly~\cite{2021Fabbian, 2022Lembo, extragalactic_Anton_2022}. For all these cases, other component-separation methods (see, e.g., refs.~\cite{NILC,2023Carones,2023Galloway,2009Stompor,2020delaHoz}) or an optimal lensing analysis, including spatial variation in the estimator normalization \cite{Optimal_filtering_2019}, are needed for an accurate lensing reconstruction.

\section*{Acknowledgments}
%
This work is supported in Japan by ISAS/JAXA for Pre-Phase A2 studies, by the acceleration program of JAXA research and development directorate, by the World Premier International Research Center Initiative (WPI) of MEXT, by the JSPS Core-to-Core Program of A. Advanced Research Networks, and by JSPS KAKENHI Grant Numbers JP15H05891, JP17H01115, and JP17H01125.
The Canadian contribution is supported by the Canadian Space Agency.
The French \textit{LiteBIRD} phase A contribution is supported by the Centre National d’Etudes Spatiale (CNES), by the Centre National de la Recherche Scientifique (CNRS), and by the Commissariat à l’Energie Atomique (CEA).
The German participation in \textit{LiteBIRD} is supported in part by the Excellence Cluster ORIGINS, which is funded by the Deutsche Forschungsgemeinschaft (DFG, German Research Foundation) under Germany’s Excellence Strategy (Grant No. EXC-2094 - 390783311).
The Italian \textit{LiteBIRD} phase A contribution is supported by the Italian Space Agency (ASI Grants No. 2020-9-HH.0 and 2016-24-H.1-2018), the National Institute for Nuclear Physics (INFN) and the National Institute for Astrophysics (INAF).
Norwegian participation in \textit{LiteBIRD} is supported by the Research Council of Norway (Grant No. 263011) and has received funding from the European Research Council (ERC) under the Horizon 2020 Research and Innovation Programme (Grant agreement No. 772253 and 819478).
The Spanish \textit{LiteBIRD} phase A contribution is supported by MCIN/AEI/10.13039/501100011033, project refs. PID2019-110610RB-C21, PID2020-120514GB-I00, PID2022-139223OB-C21, PID2023-150398NB-I00 (funded also by European Union NextGenerationEU/PRTR), and by MCIN/CDTI ICTP20210008 (funded also by EU FEDER funds).
Funds that support contributions from Sweden come from the Swedish National Space Agency (SNSA/Rymdstyrelsen) and the Swedish Research Council (Reg.~no.~2019-03959).
The UK  \textit{LiteBIRD} contribution is supported by the UK Space Agency under grant reference ST/Y006003/1 -- ``LiteBIRD UK: A major UK contribution to the LiteBIRD mission -- Phase1 (March 25).''
The US contribution is supported by NASA grant no. 80NSSC18K0132.
%

MRG acknowledges financial support from the Formación del Profesorado Universitario program of the Spanish Ministerio de Ciencia, Innovación y Universidades. CGA thanks the funding from the Formación de Personal Investigador (FPI, Ref. PRE2020-096429) program of the Spanish Ministerio de Ciencia, Innovación y Universidades.

MRG and MR have been supported by MICIU/AEI/10.13039/501100011033 and by FEDER, UE under the project with reference PID2022-140670NA-I00. MRG, CGA, PV and MR have been supported from Universidad de Cantabria and Consejería de Educación, Formación Profesional y Universidades del Gobierno de Cantabria, via the ``Actividad estructural para el desarrollo de la investigación del Instituto de Física de Cantabria''. MRG, CGA, PV, MR, RBB, FJC, EMG and GPC acknowledges financial support from the Complementary Plan in Astrophysics and High-Energy Physics (CA25944), project C17.I02.P02.S01.S03 CSIC, supported by the Next Generation EU funds, RRF and PRTR mechanisms, and the Government of the Autonomous Community of Cantabria.
We acknowledge Santander Supercomputacion support group at the University of Cantabria who provided access to the supercomputer Altamira Supercomputer at the Institute of Physics of Cantabria (IFCA-CSIC), member of the Spanish Supercomputing Network, for performing simulations/analyses.

We acknowledge the use of the \texttt{healpy}~\cite{healpy}, \texttt{mpi4py}~\cite{mpi4py}, \texttt{pysm3}~\cite{Thorne_2017, Zonca_2021, pysm3_2025}, \texttt{NaMaster}~\cite{pymaster}, \texttt{cobaya}~\cite{2021JCAP...05..057T}, \texttt{scipy}~\cite{scipy_2020}, \texttt{lensQUEST}~\cite{Beck_lensquest_2018, Beck_Thesis}, \texttt{lenspyx}~\cite{Lenspyx_citation}, \texttt{cmblensplus}~\cite{cmblensplus}, \texttt{GetDist} \cite{GetDist_Lewis_2019}, \texttt{numpy}~\cite{numpy_harris2020array}, and \texttt{matplotlib}~\cite{Matplotlib_2007} software packages.

\bibliographystyle{Class/JHEP}
\bibliography{Misc/cite}

\appendix
\section{HILC biases}  \label{Appendix Bias HILC}

In this appendix, we discuss the biases present in the cleaned power spectrum. The HILC's objective is to minimize the power spectrum of the cleaned map under the constraint that the CMB power is preserved. However, it will not completely eliminate the noise nor the foregrounds, always leaving a positive foreground and noise residual. Interestingly, there is also an additional negative bias. This bias, known as the ILC bias \citep{HILC_derivation_2008,NILC}, originates from the chance correlations between signal, foregrounds, and noise that can appear on large angular scales when calculating the frequency-frequency covariance matrix from the observed maps.

The mean cleaned power spectrum is given by
\begin{equation}\label{chance corr eq}
    \langle C_\ell^\mathrm{clean}\rangle = \langle C_\ell^\mathrm{s}\rangle + \left \langle \frac{1}{\bm{1}^\text{T}{(\bR{C}_\ell^\mathrm{fg+n})}^{-1}\bm{1}}\right \rangle + (1-n_{\rm c})\frac{\langle C_\ell^\mathrm{s}\rangle}{2\ell+1},
\end{equation}
where $\langle C_\ell^\mathrm{s}\rangle$ is the mean signal power spectrum, $\bR{C}_\ell^\mathrm{fg+n}$ is the sum of the covariance matrices of foregrounds and noise, and $n_{\rm c}$ is the number of frequency channels. See refs.~\cite{HILC_derivation_2008, Delabrouille_HILC_derivation} for the derivation of this equation. The last term in eq.~\eqref{chance corr eq} corresponds to the ILC bias, which is negative and depends only on the number of frequency channels and the signal power spectrum. The $2\ell +1$ factor in the denominator confines the ILC bias to large scales, for which less modes are available to estimate the covariance matrix.

\begin{figure}[t]
    \centering
    \includegraphics[width=0.65\textwidth]{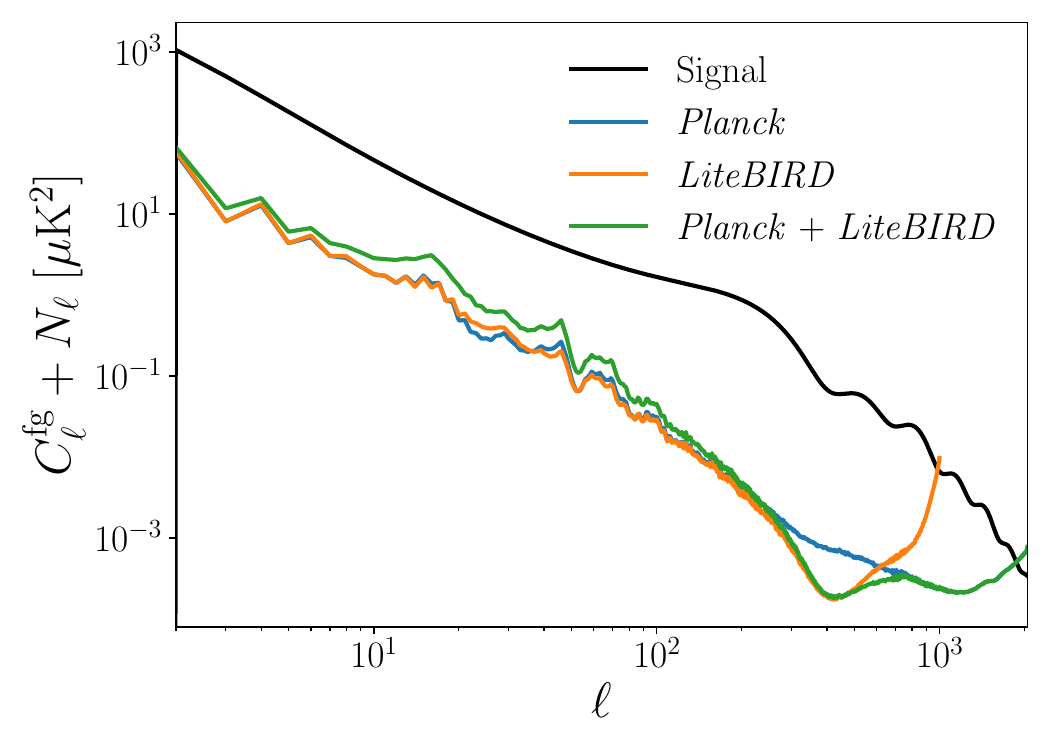}
    \caption{
    Foreground and noise residuals from our component-separated temperature map for the complex foregrounds. The black line corresponds to the temperature input CMB signal. We show residuals for \textit{LiteBIRD} (blue), \textit{Planck} (orange), and \textit{Planck} + \textit{LiteBIRD} (green) computed from the average of $400$ full-sky simulations. Temperature power spectra are calculated with the $97\,\%$ \textit{Planck} Galactic mask.
    }
    \label{fig:res_T_log_s5_d10}
\end{figure} 

The ILC bias is more relevant in situations in which the noise and foreground residuals are very low compared to the signal. This is especially the case for temperature, where it produces an increase of the foreground and noise residual power at large scales. This increase results from a partial loss of the reconstructed CMB signal, due to the correlated component of the signal leaking into the foreground and noise residuals. Figure~\ref{fig:res_T_log_s5_d10} shows the impact of the ILC bias, with residuals at multipoles $\ell \lesssim 100$ that are higher for \textit{Planck} + \textit{LiteBIRD} ($n_{\rm c}=31$) than \textit{LiteBIRD} ($n_{\rm c}=22$), which in turn are higher than \textit{Planck}'s ($n_{\rm c}=9$) due to the larger number of maps combined.

ILC-bias mitigation strategies have been incorporated in our HILC implementation. The frequency-frequency covariance matrix is binned across multipoles using a weighted average. To do that, each bin is computed using at least 1000 $a_{\ell m}$, mitigating the chance correlations between different modes. Additionally, the weights are smoothed using a uniform filter that consists of a moving average with a window size of $\Delta \ell=30$. With this approach, the ILC bias has a minor impact only in the $TT$ power spectrum at largest scales ($\ell \leq 10$), introducing a small negative bias in the cleaned power spectrum compared to the fiducial spectrum. There are more advanced methods which can potentially reduce the ILC bias, such as removing a small subset of $a_{\ell m}$ to calculate the weights using the remaining ones. This allows the weights to be uncorrelated to the data, ideally removing the ILC bias \cite{HILC_mitigation_bias_2024}. Because the lensing signal is mostly dominant at small scales, we believe that the mitigation strategy implemented in this paper is sufficient.

\section{Mean-field cross-spectra estimator}  \label{Appendix MF cross}

In this appendix, we derive eq.~\eqref{MF free PS}, which is used to calculate the lensing power spectra with the MF subtracted at the map level. The purpose of this equation is to avoid introducing the additional biases in the MF subtraction that appear due to cross-correlations when the same simulation set is used for the estimation of both the lensing power spectrum and the MF. To prevent this, we must guarantee that the $i$-th simulation is not used to calculate the MF estimate that will be subtracted from simulation number $i$.

To calculate our MF-free lensing power spectra, we start by dividing our 400 simulations into two splits, named $S_1=\{1,\ldots, 200\}$ and $S_2=\{201,\ldots, 400\}$. For each split, we compute the MF as the average over simulations: $\hat{\phi}_{LM}^{\mathrm{MF}1,XY} = \langle \hat{\phi}_{LM}^{i,XY}\rangle_{S_{1}}$ and $\hat{\phi}_{LM}^{\mathrm{MF}2,XY} = \langle \hat{\phi}_{LM}^{i,XY}\rangle_{S_{2}}$. Let us also define the MF estimate where the $i$-th simulation has been excluded from the average, $\phi_{i,LM}^{\mathrm{MF1}, XY}$ and $\phi_{i,LM}^{\mathrm{MF2}, XY}$. Then, the cross-spectra estimator for simulation $i$ is given by
\begin{equation} \label{MF exc}
    C_L^{\hat{\phi}_i^{XY} \hat{\phi}_i^{WZ}} = \frac{1}{2L+1}\sum_{M=-L}^{L}\left (\phi_{i,LM}^{XY}-\phi_{i,LM}^{\mathrm{MF1}, XY}\right) \left (\phi_{i,LM}^{WZ}-\phi_{i,LM}^{\mathrm{MF2}, XY}\right )^\ast.
\end{equation}

Equation~\eqref{MF exc} provides an MF-free lensing estimate without additional biases since the $i$-th simulation has been excluded either from MF1 or MF2. However, for computational efficiency, we do not want to recompute the mean excluding the $i$-th simulation for every realization, nor do we want to generate new simulations specifically for the MF estimation. Instead, let us assume without loss of generality that $i\in S_1$. Hence, $i\notin S_2$ and $\phi_{i,LM}^{\mathrm{MF2}, XY}=\phi_{LM}^{\mathrm{MF2}, XY}$. In that case, we find that
\begin{equation}
\begin{aligned}
\phi_{i,LM}^{XY}-\phi_{i,LM}^{\mathrm{MF1}, XY} &= \phi_{i,LM}^{XY}-\frac{1}{\frac{N_{\rm sims}}{2}-1}\sum_{\substack{j=1\\ j\neq i}}^{N_{\rm sims}/2}\phi_{j,LM}^{XY}\\
&=\phi_{i,LM}^{XY}\left ( 1+\frac{1}{\frac{N_{\rm sims}}{2}-1}\right)-\frac{1}{\frac{N_{\rm sims}}{2}-1}\sum_{j=1}^{N_{\rm sims}/2}\phi_{j,LM}^{XY}\\
&=\frac{\frac{N_{\rm sims}}{2}}{\frac{N_{\rm sims}}{2}-1}\phi_{i,LM}^{XY}-\frac{\frac{N_{\rm sims}}{2}}{\frac{N_{\rm sims}}{2}-1}\phi_{LM}^{\mathrm{MF1}, XY}\\
&=\frac{N_{\rm sims}}{N_{\rm sims}-2}\left(\phi_{i,LM}^{XY}-\phi_{LM}^{\mathrm{MF1}, XY} \right).
\end{aligned}
\end{equation}

Then, eq.~\eqref{MF exc} is equal to:
\begin{equation}
    C_L^{\hat{\phi}_i^{XY} \hat{\phi}_i^{WZ}} = \frac{1}{2L+1}\sum_{M=-L}^{L}\frac{N_{\rm sims}}{N_{\rm sims}-2}\left (\phi_{i,LM}^{XY}-\phi_{LM}^{\mathrm{MF1}, XY}\right) \left (\phi_{i,LM}^{WZ}-\phi_{LM}^{\mathrm{MF2}, XY}\right )^\ast.
\end{equation}

\section{Error on the SNR}  \label{Appendix Error SNR}

In this appendix, we derive the error on the signal-to-noise ratio (SNR) appearing in eq.~\eqref{eq error SNR}. For that, we first need to calculate the variance of the sample standard deviation. The sample standard deviation $S$ is defined by
\begin{equation}
    S = \sqrt{\frac{1}{n-1}\sum_{i=1}^n(A_i-\bar{A})^2},
\end{equation}
where $n=N_{\rm sims}$, $A_1,\ldots, A_n$ are the lensing amplitudes estimated from the simulations, and $\bar{A}$ is the mean lensing amplitude.

Under the assumption that $A_i$ are independent samples from a normal distribution, the sample standard deviation follows a scaled $\chi$ distribution, 
\begin{equation}\label{eq scaled chi}
    \frac{\sqrt{n-1}}{\sigma}S \sim \chi_{n-1},
\end{equation}
where $\sigma$ is the population standard deviation, and $\chi_{n-1}$ is the $\chi$ distribution with $n-1$ degrees of freedom.

From eq.~\eqref{eq scaled chi}, we can calculate the variance of the sample standard deviation:
\begin{equation}
    \mathrm{Var}(S) = \mathrm{Var}\left (\frac{\sigma}{\sqrt{n-1}}\chi_{n-1}\right ) = \frac{\sigma^2}{n-1}\mathrm{Var}(\chi_{n-1})\cong \frac{\sigma^2}{2n},
\end{equation}
where the variance of $\chi_{n-1}$ was calculated using Stirling's approximation,
\begin{equation}
    \mathrm{Var}(\chi_{n-1}) = \frac{n-1}{2n}\left [1+\mathcal{O}\left(\frac{1}{n}\right)\right].
\end{equation}
Finally, the relative uncertainty on the SNR is given by the following expression,
\begin{equation}
    \frac{\Delta \mathrm{SNR}}{\mathrm{SNR}} = \left|\frac{\partial\mathrm{SNR}}{\partial\sigma}\right| \frac{\Delta\sigma}{\mathrm{SNR}} = \frac{\Delta\sigma}{\sigma} = \frac{1}{\sqrt{2n}}.
\end{equation}

\end{document}